%% file: main.tex
\documentclass[twocolumn,tight]{aastex631}
\graphicspath{{./}{}}
\usepackage[T1]{fontenc}
\usepackage{amsmath}

\begin{document}

\newcommand{\cp}[1]{\textcolor{red}{{CP:} #1}}

\title{exoALMA. XVIII. Interpreting large scale kinematic structures as moderate warping}

\input{authors.tex}


\begin{abstract}
The exoALMA program gave an unprecedented view of the complex kinematics of protoplanetary disks, revealing diverse structures that remain poorly understood. We show that moderate disk warps ($\sim 0.5-2^\circ$) can naturally explain many of the observed large-scale velocity features with azimuthal wavenumber $m = 1$. Using a simple model, we interpret line-of-sight velocity variations as changes in the projected Keplerian rotation caused by warping of the disk. While not a unique explanation, this interpretation aligns with growing observational evidence that warps are common. We demonstrate that such warps can also produce spiral structures in scattered light and CO brightness temperature, with $\sim 10$~K variations in MWC~758. Within the exoALMA sample, warp properties correlate with stellar accretion rates, suggesting a link between the inner disc and outer disc kinematics. If warps cause large-scale kinematic structure, this has far reaching implications for turbulence, angular momentum transport, and planet formation.
\end{abstract}


\section{Introduction}

The ALMA Large Program exoALMA has provided an unprecedented view of the outer kinematic structure of protoplanetary discs \citep{Teague_exoALMA}. Through high-resolution observations of the $^{12}$CO and $^{13}$CO $J=3-2$ emission lines in particular \citep{Loomis_exoALMA, Zawadzki_exoALMA}, the survey has revealed that disc kinematics are often asymmetric, exhibiting large-scale deviations from simple Keplerian rotation \citep[][Fukagawa et al., in prep.]{Izquierdo_exoALMA, Stadler_exoALMA}. While many of these features can plausibly be linked to local perturbations—such as planets \citep{Pinte_exoALMA, Gardener_exoALMA}, instabilities \citep{Barraza_exoALMA}, or laminar flows \citep{Zulete_ea_2024}, explaining the largest-scale structure remains an important open challenge.

A recurring pattern in these data is the presence of ${m = 1}$-like azimuthal asymmetries in the line-of-sight (LOS) velocity fields. These features often extend across large portions of the disc and appear in most exoALMA targets to varying degrees, suggesting a global, disc-scale origin. Intriguingly, simulations have shown that disc warping can produce similar kinematic behaviour \citep{Young_ea_2022}. In this paper, we explore the hypothesis that such features can arise from moderate warping of the disc plane. Specifically, we consider smooth radial variations in inclination and position angle of a few degrees. In doing so, we demonstrate that such warps may account not only for the widespread, coherent, low-amplitude kinematic asymmetries observed in the outer disc, but also for corresponding features in scattered light and brightness temperature.

The theory of disc warping has a long and well-developed history, with foundational predictions dating back to early studies of viscous accretion flows with misaligned angular momentum \citep{Papaloizou_ea_1983, Pringle96}.  In the low-viscosity regime typical of protoplanetary discs ($H/R \gtrsim \alpha$, where $\alpha$ is the canonical turbulence parameter, $H$ is the disc pressure scale height and $R$ is cylindrical radius in disc coordinates) , warps are expected to propagate as bending waves. These travel at the local sound speed and are damped over a timescale $\tau_\mathrm{damp} \sim 1/\alpha \Omega_\mathrm{K}$, where $\Omega_\mathrm{K}$ is the Keplerian frequency \citep{Lubow_ea_2000, Ogilvie_ea_2013}. This means that $\tau_\mathrm{damp} \sim 1$~Myr for $\alpha \sim 10^{-4}$ at $100$~au around a solar mass star.  Internal torques between neighbouring disc rings—driven by pressure gradients and resonant radial, azimuthal, and vertical motions \citep["sloshing" and "breathing" modes; e.g.][]{Lodato_ea_2007, Dullemond_Kimmig_ea_2022, Held_ea_2024}—further shape the global disc structure. In addition, parametric instability can be excited via resonance between vertical shear and inertial waves \citep{Ogilvie_ea_2013, Paardekooper_Ogilvie_2019}, potentially generating strong turbulence and accelerating warp decay \citep{Deng_ea_2022}. Counterintuitively, warping may not only produce dust substructures \citep{Longarini_ea_2021a}, but also promote rapid dust settling into the midplane \citep{Aly_ea_2024}.

{In cases of extreme warping,} large misalignments can lead to disc breaking, in which the disc separates into discrete planes \citep[e.g.][]{Lodato_ea_2010,Facchini_ea_2013, Facchini_ea_2018, Nixon_ea_2013, Dogan_ea_2023, Young_ea_2023}, resulting in precessing shadows \citep{Nealon_ea_2020}. These shadows may in turn have dramatic impact on disc structure \citep{Zhang_Zhu_2024, Ziampras_ea_2025}. { The narrow shadows \citep[e.g. in the sample of][]{Bohn2022} require large inner disk misalignments that imply a torn disc \citep[$\gg H/R$; e.g.][]{Price_ea_2018, Nealon_ea_2020}. Meanwhile, both geometric \citep{Muro-Arena_ea_2020, Debes_ea_2023} and dynamical \citep{Nealon_ea_2018, Nealon_ea_2019} models show that modest warps give rise to broad shadows in scattered light, which extend over $\sim 180$~degrees in azimuth. This work pertains to these more moderate warp structures, although misaligned (torn) inner disks are plausibly related phenomena.}

Observational evidence for {widespread warping and misaligned inner discs} is growing. High-contrast imaging in the near-infrared has revealed shadows in numerous systems \citep{Benisty_ea_2023}, for example TW Hydra \citep{Debes_ea_2016, Debes_ea_2023}, HD 142527 and DoAr 44 \citep[e.g]{Casassus_ea_2018}. `Dipper' optical or near infrared (NIR) light curves are common and often interpreted as occultation by warped or misaligned inner discs \citep{Cody_ea_2014, Stauffer_ea_2015, Ansdell_ea_2016b, Ansdell_ea_2016a}. At mm-wavelength, molecular line observations have indicated kinematic misalignments in systems such as HD 100546 and HD 142527 \citep{Pineda_ea_2014, Casassus_ea_2015}, although misalignments can be difficult to distinguish from radial flows \citep{Rosenfeld_ea_2014, Zulete_ea_2024}. {VLTI/GRAVITY} observations of discs with shadows show that several of the cases have unambiguous misalignments between inner and outer discs \citep{Bohn2022}.
HST \citep[e.g.][]{Watson_ea_2007} and more recently JWST scattered light observations have shown asymmetric lobes above and below the midplane of edge-on discs, and appearing among $\sim75$~percent in the sample of \citet{Villenave_ea_2023}. These lobes vary in relative brightness with wavelength, suggestive of an inner disc misalignment or moderate warp \citep{Juhasz_Facchini_2017,Nealon_ea_2019, Kimmig_ea_2025}.

The origin of disc warping remains an open question. While a rotation axis of the star tilted with respect to the magnetic field or inner disc may cause inner disc misalignment \citep[e.g.][]{Lai_1999, Foucart_Lai_2011, Romanova_ea_2021}, it is not clear whether this applies to the {moderate warping} at larger spatial scales we explore in this work. Large scale {warps} in the outer disc may still be caused by magnetic fields, or they may be self-induced due to radiation driven instability \citep{Pringle96, Armitage_Pringle_1997}, driven by perturbations from companions or flybys \citep{Kraus_ea_2020,Nealon_ea_2020, Cuello_ea_2023} or late infall of material \citep{Kuffmeier_ea_2023}. Misaligned stellar or substellar companions can torque the disc and induce warps or even disc breaking \citep[e.g.][]{Nealon_ea_2018, Zhu_ea_2019}. However, for systems not in stellar multiples, flybys are not expected to be common \citep{Rosotti_ea_2014, Winter_ea_2024b, Shuai_ea_2022}. Alternatively, continued infall of misaligned material from the surrounding envelope can reorient the outer disc while the inner disc remains aligned with the stellar spin \citep[e.g.][]{Bate_ea_2010, Kuffmeier_ea_2024}, with an increasing number of observational case studies \citep{Ginski_ea_2021, Garufi_ea_2024}. These mechanisms may help to explain the growing body of observational evidence pointing to misalignment between inner and outer disc structures.

In this work, we systematically apply a simple warp model to the exoALMA sample, using residual velocity maps derived from Keplerian fitting. Our goal is to determine whether coherent warps can explain the $m = 1$ structure seen in many discs, and to explore possible physical correlations with other system properties such as accretion rate and non-axisymmetry in the dust. We show that moderate warps can account for the large-scale structure in many systems, with implications for our understanding of angular momentum transport and the physical state of protoplanetary discs.


\section{Methodology}
\label{sec:methodology}

\subsection{Data}

We aim to explore whether radially-dependent perturbations to inclination and position angle (i.e. a warped disc) can explain the non-axisymmetric structures in the LOS velocity in the exoALMA dataset \citep[for an overview, see][]{Teague_exoALMA}. We restrict ourselves to the fiducial resolution $^{12}$CO LOS residual velocity maps (\( \delta v_{\rm los} \)) obtained through the \textsc{Discminer} fitting procedure \citep{Izquierdo_exoALMA}. Unless otherwise stated, we always use the outcome of the analysis pipeline performed on continuum subtracted cubes, with a nominal beam size of $0.15''$ and channel spacing $100$~m~s$^{-1}$, clipped at $3\sigma$. We apply our procedure on residuals obtained from the \textsc{Discminer} analysis, which models the channel maps in terms of a fixed inclination and position angle, with a parameterised emission surface height and intensity, and a Keplerian rotation curve. The residuals are deprojected and defined with the azimuthal coordinate $\phi=0$ along the red-shifted major axis.

\subsection{Linear approximation}

While we note that tools exist in the literature to model the kinematics of warped discs \citep{Casassus_Perez_2019, Casassus_2022}, we aim to achieve a very simple, flexible model that is easy to apply without fitting numerous parameters. Our method is similar to the `tilted ring' approach that has been applied historically to modeling galaxy rotation curves \citep[e.g.][]{Begeman_ea_1989}. We do not aim to fully fit radial, vertical and azimuthal velocity variations, but to interpret all line of sight variations as far as possible as due to the projection of the Keplerian azimuthal component. This allows us to efficiently fit for radially dependent {profiles} in the warp structure, but our results should be interpreted as a `maximal' tilt amplitude that could be inferred from the data.

In order to model the observed perturbations as warped discs, we start by assuming a circular ring of material orbiting with azimuthal velocity:
\begin{equation}
\vec{v} = v_\phi(R) \, \hat{e}_\phi, \quad \text{with } v_\phi(R) = \pm \sqrt{\frac{G M_*}{R}},
\end{equation}where we will assume hereafter that $v_\phi$ is positive. The unperturbed LOS velocity (with fixed inclination \( i_0 \) and position angle \( \mathrm{PA}_0 \)) is then:
\begin{equation}
    v_{\rm los}^{\rm flat}(R, \phi) = v_\phi(R) \, \sin i_0 \, \cos(\phi - \mathrm{PA}_0).
\end{equation}Without loss of generality, we will take PA$_0 = 0$ to simplify the following expressions, and we rotate all discs to conform to this definition in figures ($\phi=0$ corresponding to the red-shifted semi-major axis).

Now, if we allow the disc orientation (inclination and position angle) to vary with radius:
\begin{align}
    i = i_0 + \delta i(R), \\
    \mathrm{PA} = \mathrm{PA}_0 + \delta \mathrm{PA}(R).
\end{align}If we assume small perturbations (\( \delta i, \delta \mathrm{PA} \ll 1 \) in radians) then:
\begin{align}
    \sin(i_0 + \delta i) &\approx \sin i_0 + \delta i \, \cos i_0, \\
    \cos(\phi - \delta \mathrm{PA}) &\approx \cos\phi + \delta \mathrm{PA} \, \sin\phi.
\end{align}Substituting into the projected velocity:
\begin{align}
    v_{\rm los}^{\rm warp} &\approx v_\phi(R) \left[ (\sin i_0 + \delta i \, \cos i_0)(\cos \phi + \delta \mathrm{PA} \, \sin \phi) \right] \\
    &\approx v_\phi(R) \left[ \sin i_0 \, \cos \phi + \delta i \, \cos i_0 \, \cos \phi + \delta \mathrm{PA} \, \sin i_0 \, \sin \phi \right].
\end{align}Thus, the residual in the LOS velocity is:
\begin{align}
\label{eq:delta_vlos}
    \delta v_{\rm los} &= v_{\rm los}^{\rm warp} - v_{\rm los}^{\rm flat}\\& \nonumber = v_\phi(R) \left[ \delta i(R) \, \cos i_0 \, \cos \phi + \delta \mathrm{PA}(R) \, \sin i_0 \, \sin \phi \right].
\end{align}We can rewrite equation~\ref{eq:delta_vlos} as:
\begin{equation}
\label{eq:delta_vlos_2}
\delta v_{\rm los}(R, \phi) = A(R) \cos\phi + B(R) \sin\phi.
\end{equation} We can then derive the inclination and position angle perturbations from the coefficients \( A(R) \) and \( B(R) \) as:
\begin{align}
\delta i(R) &= \frac{A(R)}{v_\phi(R) \cos i_0}, \\
\delta \mathrm{PA}(R) &= \frac{B(R)}{v_\phi(R) \sin i_0}.
\end{align}The coefficients \( A(R) \) and \( B(R) \) are obtained by least-squares fitting\footnote{We use the \texttt{linalg.lstsq} method from \textsc{Numpy} \citep{Numpy_citation}.} to the azimuthal slice of the residual field in each annulus.\footnote{The annulus radius $R$ is always understood to be the radial location in the deprojected coordinate system, assumed the same as the radius in disc coordinates. Strictly we should deproject differently at each radius and refit the \textsc{Discminer} iteratively. However, \textit{a posteriori} the warp angles are typically $\lesssim 3^\circ$, so expected offsets are much smaller than the beam size. We experimented by refitting the disc in newly deprojected coordinates to find very minor differences in the inferred warp structure.} We then assume an uncertainty equivalent to the square root of the residual root mean square sum divided by the number of beams that fit within $2\pi R$. This should be interpreted as a statistical uncertainty, not one that necessarily accounts for all the possible systematics inherent in the complexity of the exoALMA pipeline. In addition, as we discuss in Section~\ref{sec:uniqueness}, $A(R)$ and $B(R)$ also absorb any axisymmetric azimuthal deviations from Keplerian and radial velocities respectively. This means that the warp interpretation is not unique.

A necessary condition for warps to explain velocity structures is evident from equation~\ref{eq:delta_vlos_2}: \textit{$\delta v_{\rm los}$ must have azimuthal wavenumber of $m=\pm 1$ along a given annulus.} In Section~\ref{sec:results} we fit indiscriminately for warp structures, but the success of this fit can be understood as the degree to which a disc conforms to this criterion.

\subsection{Physical coordinates}

The above is derived entirely in the plane of the sky. It is useful to understand how these observed perturbations connect to physical warping in the disc. We therefore consider the literature definitions of three angles, which are the tilt $\beta$, the twist $\gamma$, and what we will call in this work the warp amplitude $\psi$. In Appendix~\ref{app:phys_coords} we define these angles formally and relate them to the perturbations $\delta i$ and $\delta$PA. The convenient small angle approximations are:
\begin{align}
\label{eq:beta_small_main}
\beta(R) &\approx \sqrt{ \delta i(R)^2 + \delta \mathrm{PA}(R)^2 \sin^2 i_0 }, \\
\label{eq:gamma_small_main}
\gamma(R) &\approx \arctan2\left( { \delta i(R) },{ -\sin i_0 \, \delta \mathrm{PA}(R) } \right)\\
\label{eq:psi_def_main}
    {\psi}(R) &= R \sqrt{ \left( \frac{\partial i}{\partial  R} \right)^2 + \sin^2 i \left( \frac{\partial \mathrm{PA}}{\partial R} \right)^2 }.
\end{align}In order to be able to compare discs, hereafter we will refer to a `tilt amplitude', by which we mean $ \beta_\mathrm{max}$ as defined by the maximum value of $\beta(R)$ for a given disc (for a specific molecular tracer and observational beam size). In the literature, $\psi$ is referred to as a `warp amplitude', and we will follow this nomenclature although it is strictly a gradient. However, as discussed in Appendix~\ref{app:phys_coords}, the warp coordinates are dependent on the reference coordinate system \citep[see also][]{Juhasz_Facchini_2017}. {The most physically relevant reference frame is that aligned with the total angular momentum of the system, as all warped and misaligned structures precess around this axis. However, we note that some numerical studies define the reference frame to be that of a perturbing binary, and, more pertinently in our context, for observed discs we do not know the total angular momentum vector. Care must therefore be taken when comparing the physical coordinates we infer in this work to numerical or analytic models.}

\section{Results and discussion}
\label{sec:results}
\subsection{Case study: MWC~758}

In order to understand how the warp model manifests on the observational properties of discs, it is instructive to explore in detail a single case study before looking at the properties of the broader exoALMA sample. We consider the case of MWC~758, which has famous spirals in scattered light \citep{Benisty_es_2015}. Comparing the inclination and position angle in the inner disc inferred with H-band VLTI/PIONIER observations \citep[$48^\circ$ and $100^\circ$ respectively --][]{Lazareff_ea_2017} to the continuum-derived values \citep[$7^\circ$ and $76^\circ$ respectively --][]{Curone_exoALMA} also suggests substantial misalignment. It also has a clear $m=1$ spiral in the LOS kinematic residuals ({this structure will be discussed further by} Fukagawa et al. in prep.). We consider here how the warping model may explain this spiral, and the consequences for other observational diagnostics.

\begin{figure*}[ht!]
    \centering
    \includegraphics[width=\textwidth]{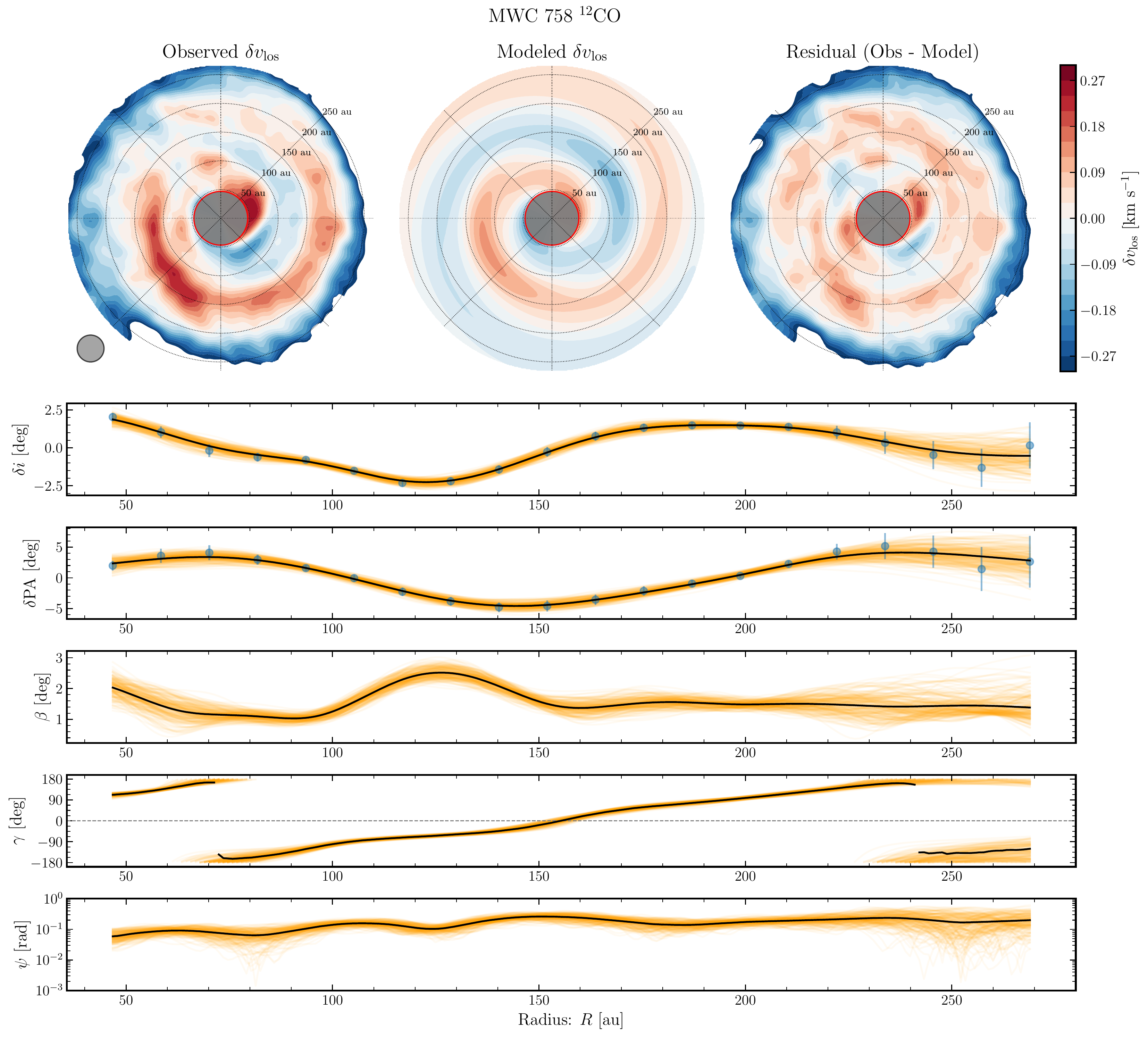}
    \caption{Top panels show residuals from the observed (left) vs modeled (right) $\delta v_{\rm los}$ fields for MWC~758 after fitting Keplerian velocity profiles. The flexible model is for a simple warped disc geometry, with perturbation in inclination and position angle. The colour scale is the LOS velocity  in km/s. Grey circles mask two times the central beam size. The beam size is also shown on the left hand side, assumed circular for visualisation. The bottom panels show radial profiles of \( \delta i \), \( \delta \mathrm{PA} \) and the physical warp properties tilt $\beta$, twist $\gamma$ and the warp amplitude $\psi$ for MWC~758 from our fitting procedure. Blue points and errorbars in \( \delta i \) and \( \delta \mathrm{PA} \) come from the least squares fitting procedure, Faint orange lines show posterior distributions from the GP model.}
    \label{fig:inc_pa_profiles}
\end{figure*}

\begin{figure}
    \centering
    \includegraphics[width=\linewidth]{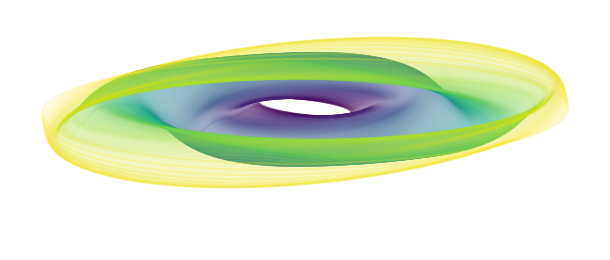}
    \caption{Visualisation of the warp structure via concentric rings with the profile for MWC~758 shown in Figure~\ref{fig:inc_pa_profiles}. The coordinate system is chosen such that the disc lies in the $x-y$ plane on average, and the $z$-axis is stretched by a factor four to emphasise the warp. The disc is then viewed slightly from above. }
    \label{fig:warp_structure}
\end{figure}

\subsubsection{Kinematic warp model outcome}

We show the radial profiles for perturbations in inclination $\delta i$ and position angle $\delta$PA in Figure~\ref{fig:inc_pa_profiles}. The data points and errorbars are from the annuli fitting procedure discussed in Section~\ref{sec:methodology}, while the black line shows the mean of a Gaussian process (GP) fit\footnote{To fit we use the \texttt{GaussianProcessRegressor} of \textsc{Scikit-Learn} \citep{scikit-learn_citation}, available from \url{https://scikit-learn.org/stable/modules/generated/sklearn.gaussian_process.GaussianProcessRegressor.html}.} used to interpolate between these points and estimate uncertainties. We adopt a Matern kernel with smoothness parameter $\nu=2.5$, and initiate the length scale at $2$ beam sizes. We sample LOS velocities every half beam size. Fitting the profiles with a Gaussian process has the advantage of (a) establishing uncertainties in our warp metrics by drawing samples from posteriors and (b) calculating the numerical derivatives required to quantify the warp amplitude $\psi$.

Although annuli are fit independently, we find that a coherent, near sinusoidal structure emerges in perturbation space, with a similar structure in both inclination and position angle. We also visualise the LOS velocity structure of the $^{12}$CO surface for MWC~758 (left), compared to the warped model from our GP fit (right). The total tilt amplitude is $ \beta_\mathrm{max} \approx 1.6 \pm 0.2^\circ$. The fits become uncertain in the outer disc region ($\sim 200-250$~au), where the data cannot be reproduced by a warp (i.e. $\delta i$ and $\delta$PA become consistent with zero with large uncertainties). In terms of our twisting parameter $\gamma$, the twist is a continuous, almost linear (periodic) function of radius. This is what gives rise to clear spiral structure, which becomes apparent when a substantial twist is present. Without a twist, the velocity pattern may alternate between red and blue with increasing radius at fixed azimuth, as with several examples discussed in Section~\ref{sec:full_sample}. {The warp amplitude for MWC~758 is approximately constant with radius, with $\psi\sim 0.1$. While this comes close to the analytic criterion for disc tearing \citep{Dogan_ea_2018}, most numerical studies do not find tearing for the small tilt amplitudes ($\beta_\mathrm{max}\ll H/R$) that we report here.}

As discussed in Section~\ref{sec:uniqueness}, the warp can only produce features that have $m=1$ periodicity on the annulus. In the outer region the velocities around the annuli are offset from the systemic velocity, rather than $m=1$ symmetric. In this case, the velocity field {could be the result of a wind (this will be assessed in future exoALMA publications -- Benisty et al. in prep.),} while variations in emission height (which we do not estimate here) and/or sloshing motions may also contribute. However, within $\sim 200$~au the warp model does reproduce a spiral pattern strikingly similar to the residuals from the Keplerian rotation curve observed in MWC~758. The coherence of the inferred warp structure is further circumstantial evidence in support of this interpretation.

\subsubsection{Scattered light spirals}
\label{sec:scattered_light_mwc758}
Given that warping can evidently generate spiral structure, we can further ask if it might play a role in generating the spiral arms seen in scattered light \citep{Benisty_es_2015, Ren_ea_2023, Orihara_ea_2025}. We therefore run a \textsc{Radmc3d}\footnote{\url{https://github.com/dullemond/radmc3d-2.0}} \citep{Dullemond_ea_2012} radiative transfer simulation to explore this.  We impose a midplane density $\rho_\mathrm{mid} = \rho_0 (R/R_0)^{-1}$, where $R_0 = 10$~au and our inner edge is $5$~au. Since we do not consider gas opacity or self-gravity, $\rho_0$ is unimportant except for the dust, for which we adopt $\rho_0 =2\times 10^{-15}$~g~cm$^{-3}$. We assume well-coupled dust in a vertically isothermal disc in hydrostatic equilibrium with a scale height $H(R) = 0.05 (R/R_0)^{1.03}$, corresponding to mild flaring. To be comparable to the extent of the scattered light observations, we truncate at an outer radius of $160$~au.

 In the range of radii for which we have kinematic constraints, we perturb the orientation of the disc at each inclination to match our profiles, without any other change to the density at a fixed radius. Unfortunately, at the distance of MWC~758 the velocity map resolution is prohibitive at radii within the scattered light spirals, where the structure is particularly important for producing shadows outside. However, we can in a general sense assess whether inclinations and position angle perturbations similar to those in MWC~758 may produce comparable structures. We do this by extrapolating a reasonable but arbitrary profile inside the region for which we have kinematic constraints (see Appendix~\ref{app:phys_warp}). 
 Here we do not aim to reproduce every constraint, but apply a simple model for the outer disc, which is not meant as a `fit'. We also miss physics, such as deviations from vertical hydrostatic equilibrium during warp propagation. Overall, the aim to reproduce the entire system would be a considerable effort beyond the scope of this work. Through our experiment we simply aim to answer the question: `can warping produce spiral arm structures in scattered light observations?'

\begin{figure*}
    \centering
    \includegraphics[width=\linewidth]{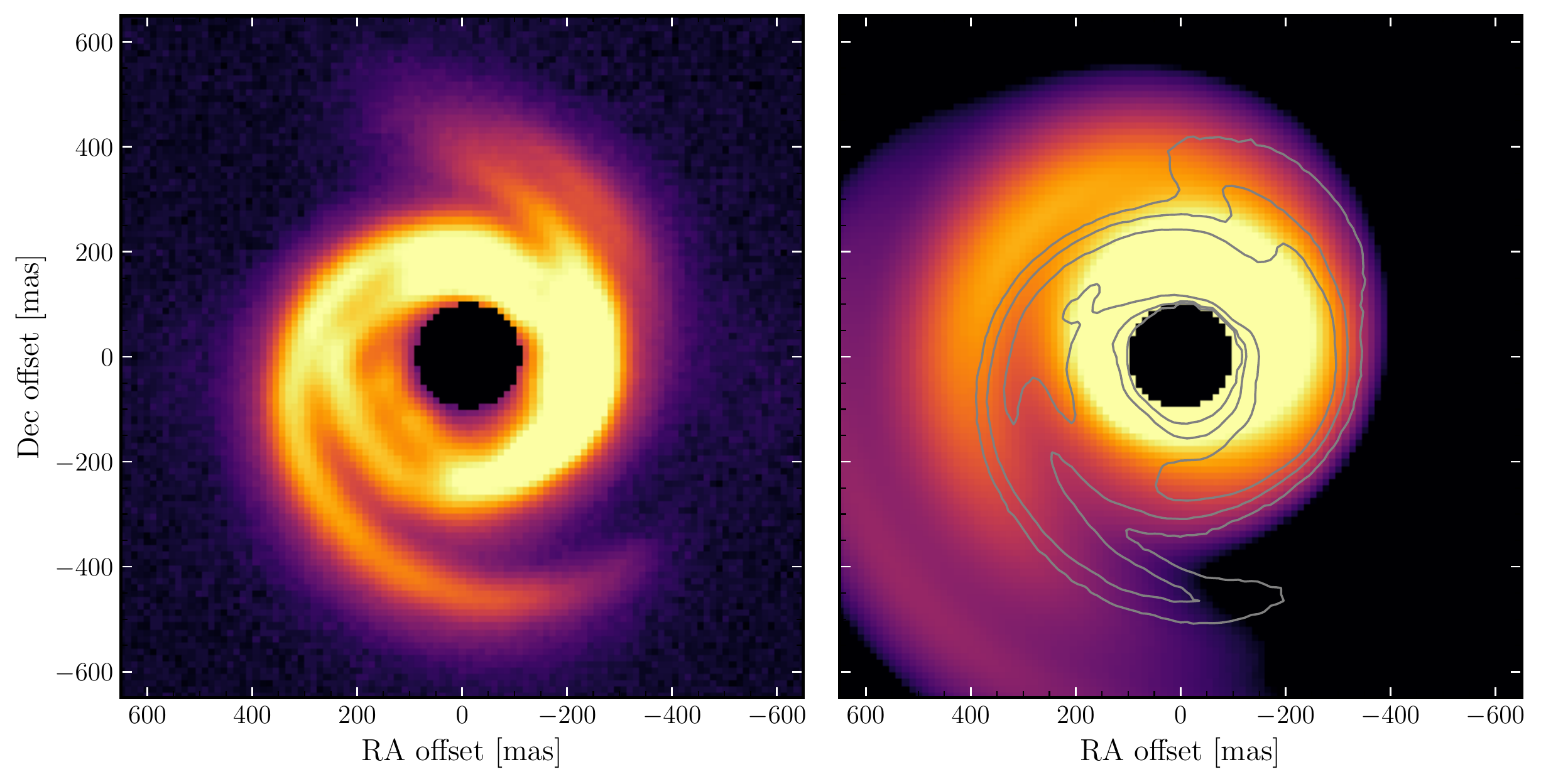}
    \caption{The total K-band polarised intensity map of MWC~758 \citep[left,][]{Ren_ea_2023} compared to the total intensity at $2.2$~$\mu$m from our \textsc{Radmc3d} model (right). Both are masked inside $100$~mas, which is the size of the coronograph. The contours of the right hand panel are $5$, $10$, $20$ and $50\, \sigma$ from the observed structure. We highlight that we do not have good constraints on warp structure inside of $150$~mas, where we have assumed a profile with comparable amplitude to the warp further out (see Appendix~\ref{app:phys_warp}).  }
    \label{fig:radmc_vs_MWC}
\end{figure*}

For scattering opacity, we adopt amorphous olivine with equal parts Mg and Fe and optical constants from \citet{Jaeger_ea_1994} and \citet{Dorschner_ea_1995} assuming $0.1\, \mu$m dust. The stellar spectrum is black body, with stellar radius $2R_\odot$ (although geometrically we assume a point source) and effective temperature $7600$~K. 
The total intensity from the radiative transfer calculation at $2.2$~$\mu$m is shown in Figure~\ref{fig:radmc_vs_MWC}, compared to the total polarised intensity as observed with VLT/SPHERE in the K-band \citep{Ren_ea_2023}. 
The structure is somewhat larger scale and less sharp than observed. We do not clearly obtain a spiral arm stretching north; it is possible that the visibility of this spiral arm is strongly dependent on flaring angle, breathing motion or inner disc structure. In Section~\ref{sec:TB_spirals}, we also discuss how emission from close to the mid-plane may plausibly produce $m=2$ structures.
However, given the simplicity of our model, it is remarkable that we qualitatively reproduce several aspects of the observed structure. 
Both a long spiral arm stretching south, and an overbrightness in the east are visible. In the future, combining tailored reconstruction of the shadowing in scattered light, as performed by \citet{Orihara_ea_2025}, in combination with kinematic modeling may produce a global picture of the disc geometry. 
Here we simply conclude that the warps required to give the observed signatures in kinematic residuals for MWC~758 are also capable of producing spirals in scattered light.

\subsubsection{Brightness temperature spirals}
\label{sec:TB_spirals}
MWC~758 not only shows a spiral structure in the LOS kinematic residuals, but also in the CO brightness temperature. Based on the thermal structure we infer from our \textsc{Radmc3D} model, we can also explore whether we expect warping to produce similar spirals in brightness temperatures. We might expect some variation in this temperature for the same reason as for the scattered light. Namely, in a warped disc, the surface at the same radius and height above the warped midplane may be irradiated differently depending on the azimuth.

\begin{figure*}
    \centering
    \includegraphics[width=\linewidth]{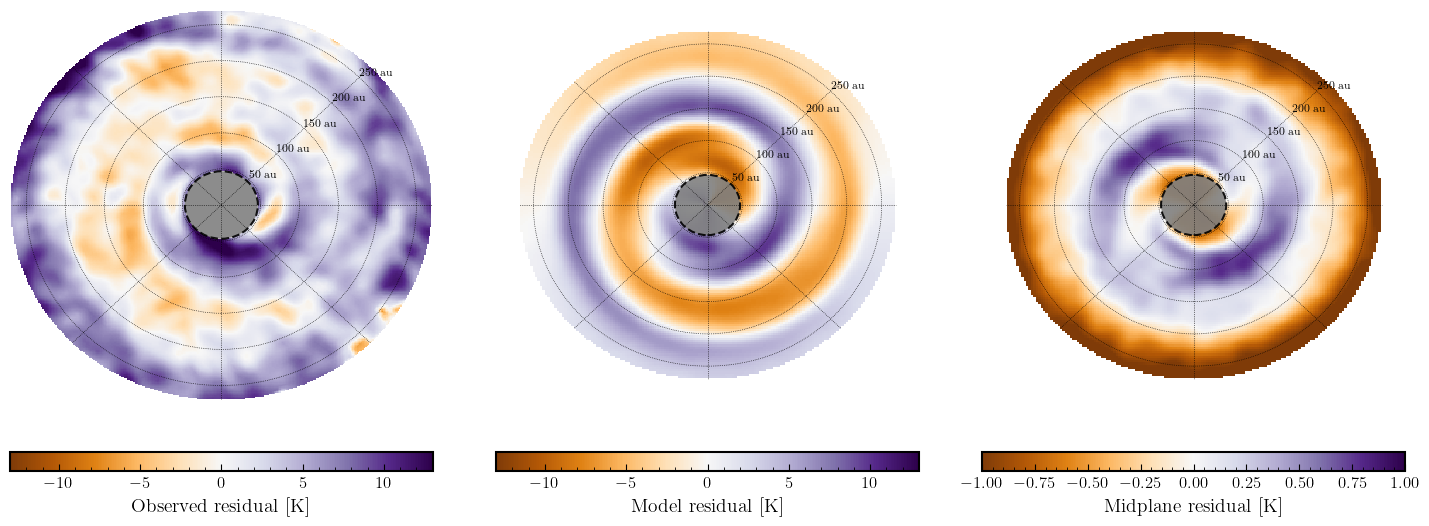}
    \caption{Residual brightness temperature of $^{12}$CO after subtracting an axisymmetric powerlaw as observed in MWC~758 (left) and our warped disc model at an estimated CO emission height (middle) and midplane from the model (right). The temperature quoted in each case is the difference from a fitted radial axisymmetric power law model. The model has been smoothed over a circular beam for visualisation purposes, assuming zero temperature residual outside of the simulated range. The CO surface temperature from the model (middle) broadly matches the observed structure (left), but the midplane pattern (right) resembles the $m=2$ symmetry in the inner region. }
    \label{fig:MWC_temperature}
\end{figure*}

To compare our model to the observed brightness temperature structure, we simply assume that the emission comes from where CO becomes self-shielding (we assume this vertical column is $N_\mathrm{ss} =10^{15}$~cm$^{-2}$, with depletion with respect to hydrogen $10^{-6}$ and dust-to-gas ratio of $10^{-2}$ -- but our results do not strongly depend on these parameters). We show the result compared to the observations in Figure~\ref{fig:MWC_temperature}. As for the scattered light model, there are some differences in the tightness of the spirals and the intensity of the structures. However, again, our model does well given its simplicity and lack of detailed physics. Numerous observational effects as well as physical effects, such as varying flaring angles and photochemistry may influence the structure \citep{Young_ea_2021}. Indeed, if the vertical motions in the outer disc are tracing a (thermal) wind, then this material could be expected to experience some temperature enhancement. Overall it is clear that warping is capable of producing structures comparable to what we observe.

Finally, we note that it is simple to understand that brightness temperature structure in the midplane of a warped disc must have $m=2$ symmetry if there are no other factors. We confirm this on the right hand panel of Figure~\ref{fig:MWC_temperature}. In this case, the temperature deviations for the axisymmetric powerlaw are much smaller, but the morphology somewhat better resembles the inner regions of MWC~758. This might hint that the $^{12}$CO emission is coming from deeper in the disc than we assume, for example due to greater photodissociation in the surface layers.  This may be the subject of focused experimentation in future work.

\subsection{Full exoALMA sample}
\label{sec:full_sample}

We now consider more briefly the remainder of the exoALMA sample. The full sample is discussed in Appendix~\ref{app:models_all}. We will discuss in Section~\ref{sec:backside} that cases where the backside of the disc is visible in the emission line profile are problematic. In these cases, the residuals have been extracted from a Keplerian model fitting the double bell line profile using \textsc{Discminer} \citep{Izquierdo_exoALMA}, and we highlight these cases below. We still report the warp structures for these discs (summarised in Table~\ref{tab:main}), but exclude them from our population level analysis in Section~\ref{sec:correlations}. {We highlight that, even without our analysis, many of the kinematic structures seen in the LOS velocity residuals show arcs and asymmetric features qualitatively similar to those found by \citet{Young_ea_2022} in their numerical simulations of warped discs.}

\subsubsection{Strong warping candidates}
\label{sec:strongwarp_candidates}

 Alongside MWC~758, the LOS residuals for the disc around CQ Tau is one of the most striking examples of a spiral structure that can be described well by a coherent, twisted warp with non-constant $\gamma(R)$. Intriguingly, these two cases are also those for which SO, a putative shock tracer \citep{Sakai_ea_2014}, has been detected \citep{Zagaria_ea_2025}. Speculatively, rapid `sloshing' motions produced by the warp may drive shock heating in these cases \citep[e.g.][and Section~\ref{sec:sloshing}]{Kimmig_ea_2024}.

 In addition to MWC~758 and CQ~Tau, the discs around HD~135344B, HD~143006 and J1604 all exhibit a similarly spiral-like structure in at least part of the disc. They also have substantial variations in position angle, which would require potentially contrived combinations of radial and azimuthal velocity variations to produce similar LOS signatures by axisymmetric perturbations. {Most of these discs also have strong evidence of warping uncovered by previous studies, as discussed in Appendix~\ref{app:models_all}.} We therefore suggest that these cases represent some of the clearest warped disc candidates, although this precludes neither alternative explanations in these cases, nor the warping hypothesis in the remaining exoALMA sample.

\subsubsection{Ambiguous structure}
\label{sec:ambiguous_cases}

While the warp model has success in reproducing LOS residuals in several discs, some of the structures we attribute to the warp may be more readily explained in alternative ways. In particular, cases for which we see a systematic blue-red trend across the major axis can be interpreted as slower than Keplerian rotation due to radial pressure gradients \citep{Longarini_exoALMA, Stadler_exoALMA}.

It is not possible to unambiguously distinguish between pressure gradients and warping. However, regions of the outer disc where the apparent $\delta$PA is consistent with zero would require a coincidence of viewing angle if they are warped, making pressure gradients a more compelling explanation. Most of the discs have at least some outer structure that can be interpreted in this way. In fact, physically we expect all discs to exhibit this feature; it is even possible that a disc warp has obscured sub-Keplerian rotation in cases where we do not see it clearly.

We do not attempt to distinguish directly which parts of the assumed warp structure may be due to pressure gradients. However, from visual inspection some examples where substantial pressure-related residuals may lead to warp amplitude overestimation include: HD~34282, LkCa15, PDS 66, SY Cha and V4046 Sgr.

\subsubsection{Interpreting residuals}
\label{sec:residual_interp}

Residuals include some $m=2$ structures, some of which switch signs along major and minor axes. Such structures might be readily explained by small errors in the geometric center or the emission surface height ({this will be assessed by} Fukagawa et al. in prep.). Since \textsc{Discminer} fits these parameters for a non-warped disc model, future modifications that incorporate the warp might improve these residual features. In the context of this work, the expected errors in geometric fitting parameters should not greatly influence the warp structure we infer. The $m=2$ structures are not fit by the $m=1$ warp, and uncertainties in a global PA and inclination would just result in approximately constant $\beta$ and $\gamma$ values, with small $\psi$ throughout the disc.\footnote{We have validated this by offsetting parameters for synthetic \textsc{Discminer} models.} We therefore expect our warp metrics to be robust against these uncertainties, while accounting for the putative warp may improve residual extraction in future.

We also find that for some of the discs, including J1604, SY Cha and MWC~758, once extracting off the warp we find $m=0$ structures in the form of systematically red-shifted residuals in the inner region and blue-shifted residuals in the outer region, suggestive of a wind. It is possible in these cases that small uncertainties in the systemic velocity mean that the inner disc is actually at the systematic velocity, while the outer disc is somewhat more blue shifted than assumed. {Whether or not this can be interpreted as a wind will be discussed} in a coming exoALMA paper (Benisty et al. in prep.).

\subsubsection{Highly inclined discs and backside emission}

\label{sec:backside}
\begin{figure}
    \centering
    \includegraphics[width=\linewidth]{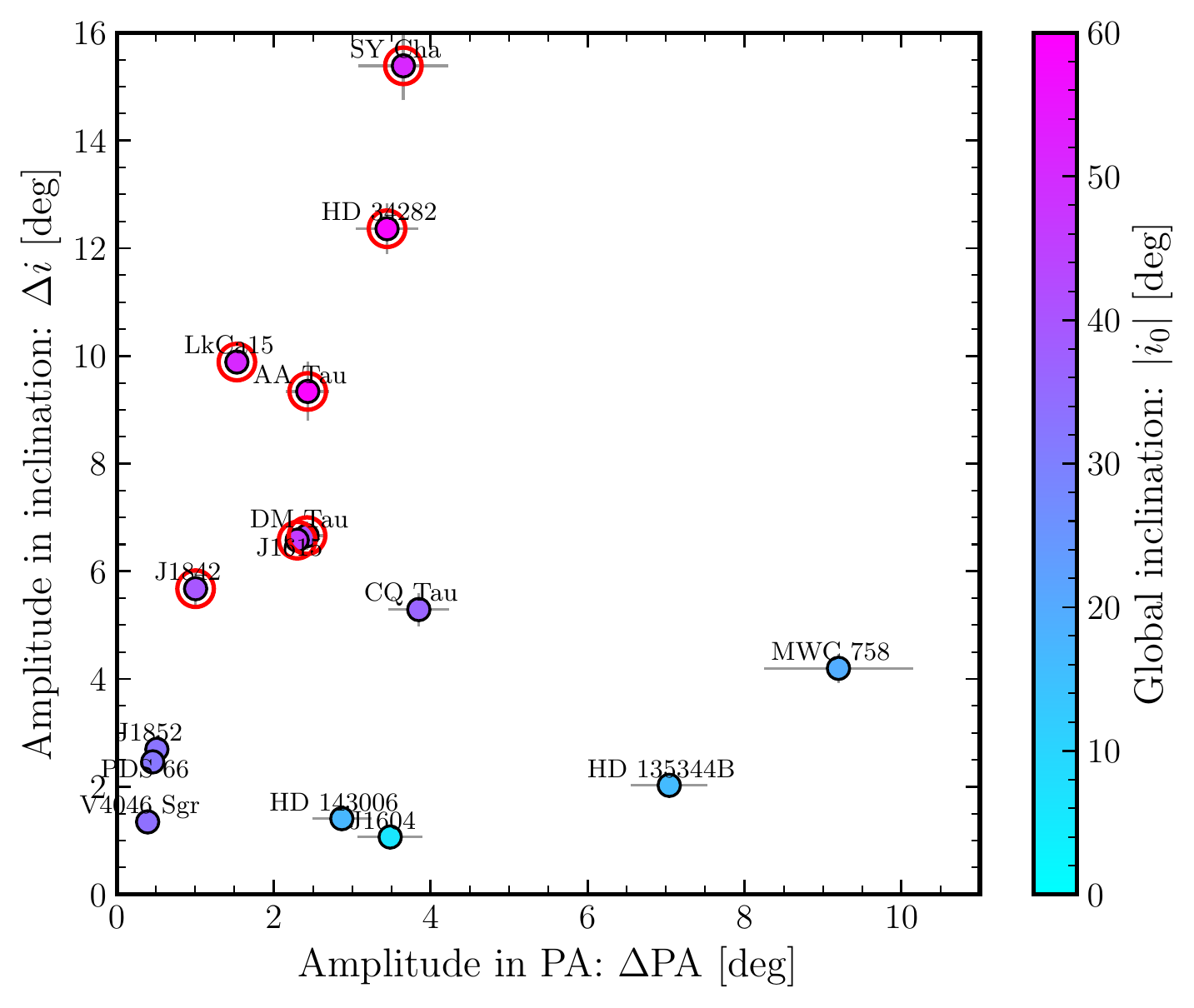}
    \caption{The relationship between inclination and PA tilt amplitude, with points being coloured by overall inclination of the system. The points circled in red are those for which the backside of the disc is visible in the $^{12}$CO line profiles. }
    \label{fig:inclination-amplitude}
\end{figure}

HD 34282, AA Tau, DM Tau, J1615, J1842, LkCa15 and SY Cha are high inclination discs, where the backside is visible \citep[for which double bell line profiles were applied in tomographic analysis --][]{Izquierdo_exoALMA}. These discs have systematically larger tilt amplitudes. We show the distribution of amplitudes in PA and inclination in Figure~\ref{fig:inclination-amplitude}, where this is evident. There are two obvious possible explanations for this finding. One explanation is that the noisy structure could simply trick the fitting procedure into adopting different inclinations (although in principle our Gaussian process modeling should automatically account for noise in fitting an $m=1$ structure).

A second plausible scenario is that there are substantial radial motions that have a greater LOS component in these high inclination discs. We have already discussed in Section~\ref{sec:ambiguous_cases} how are fitting procedure might attribute radial winds to warping. Alternatively, if the structures really are due to warps, then the high inclination discs may catch more of the radial sloshing motions that can contribute significantly to the LOS velocity residuals at high inclination, as discussed in Section~\ref{sec:sloshing}, although we do not find a strong emission height dependence (see Section~\ref{sec:13CO_all}). Radial sloshing combined with the backside contribution could provide an explanation for the difficulties in fitting the simple Keplerian models in these cases. This phenomenon warrants future exploration, but for now we note the warp properties are probably not reliably inferred in these high inclination cases. We therefore exclude them from our analysis in Section~\ref{sec:correlations}.

\subsubsection{Emission height dependence}
\label{sec:13CO_MWC758}
\label{sec:13CO_all}

\begin{figure}
    \centering
    \includegraphics[width=\linewidth]{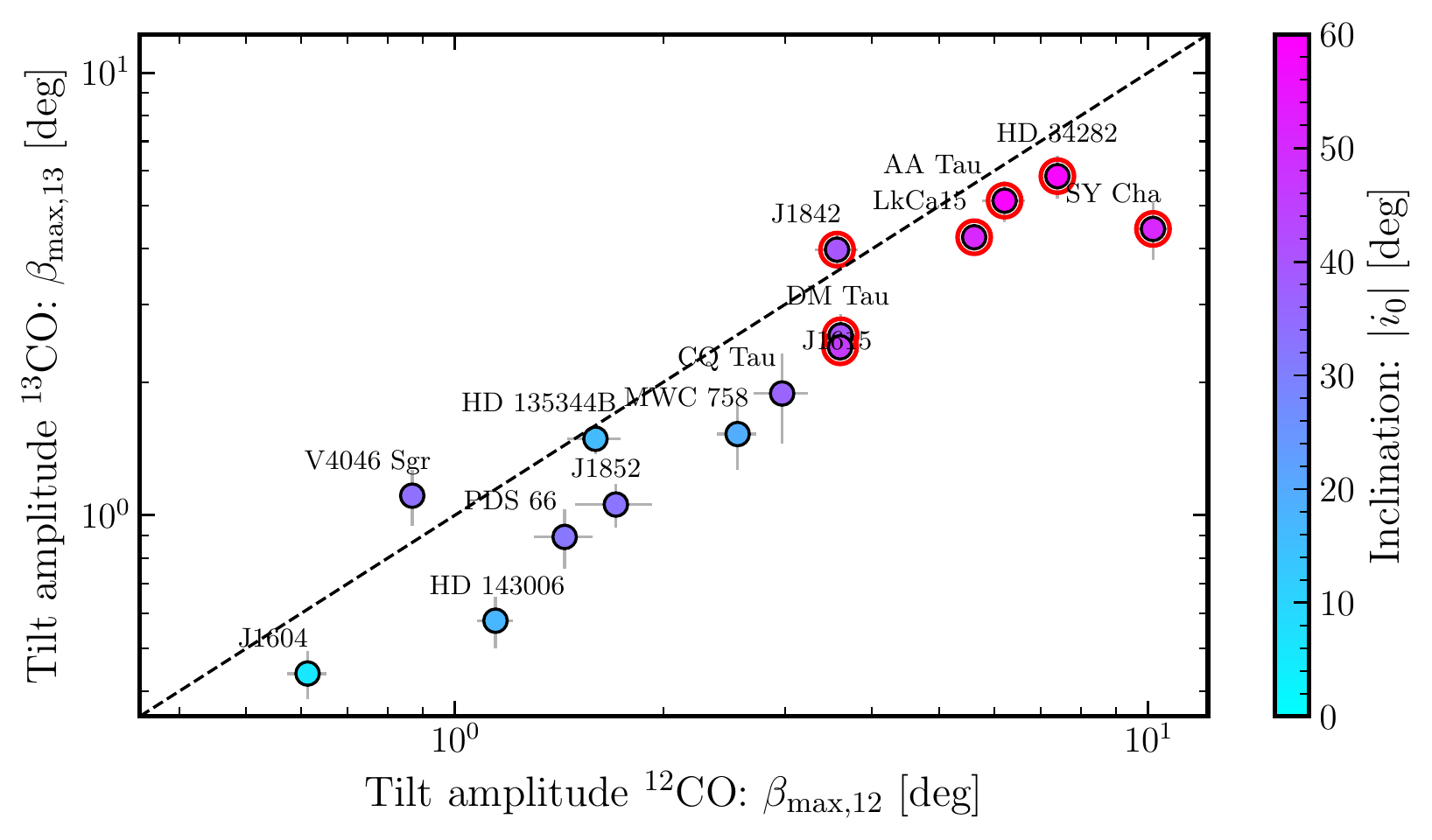}
    \caption{Comparison between the tilt amplitudes inferred using $^{12}$CO and $^{13}$CO isotopologues. The black dashed line shows $1:1$ agreement. Points are coloured by the global inclination of the disc. The discs for which the backside is visible in the line profiles are circled in red.}
    \label{fig:12vs13CO_warp_amplitude}
\end{figure}

One of our assumptions in applying the warp model is that the dominant contribution to the LOS velocity perturbations is azimuthal velocity in the natural coordinate system of a given annulus. However, the molecular emission surface for $^{12}$CO is in fact substantially above the midplane \citep{Galloway_exoALMA}, where we might expect more complex kinematics of the warp structure. We can estimate the sensitivity of our results to the finite emission height by repeating our experiment for isotopologues with a lower optical depth, emitting from lower down in the disc.

In Figure~\ref{fig:12vs13CO_warp_amplitude}, we show the dependence of $\beta_\mathrm{max}$ on the choice of isotopologue. We recover very similar tilt amplitudes for both $^{12}$CO and $^{13}$CO. The $^{13}$CO systematically exhibits a slightly lower $\beta_\mathrm{max}$, which might be the result of lower sensitivity in the outer disc, truncating the region over which we can fit the warp model. For discs that have been imaged at the larger beamsize $0.3''$, we checked the dependence of our results on resolution. We found in these cases good agreement in $^{12}$CO, and agreement in $^{13}$CO except for the cases DM Tau, SY Cha and HD 34282, which show discrepancies for $\beta_{\mathrm{max}}$ of $2-3^\circ$. The latter are all discs with a visible backside, which we exclude from our statistical analysis in Section~\ref{sec:correlations}. We conclude that our results, particularly for low inclination discs, are robust to the observational tracer and resolution. We emphasise that while this constitutes evidence that we are not biasing our warp fit by motions that are localised vertically or radially, this does not necessarily validate the warp interpretation.

\begin{table*}[t]
\centering
\caption{\label{tab:main}Properties of the discs in the exoALMA sample that we adopt or derive in this work. For the $\delta i$ and $\delta$PA values we report the total range resulting from our fitting procedure. For $\beta_\mathrm{max}$ we quote the $\pm 1 \,\sigma$ uncertainty. Stellar masses $M_*$ and outer radii $R_\mathrm{out}$ are from the \textsc{Discminer} fitting procedure \citep{Izquierdo_exoALMA}, and the non-axisymetric indices (NAIs) for the continuum are from \citet{Curone_exoALMA}. Stellar accretion rates $\dot{M}_\mathrm{acc}$ are mostly the same as those adopted by \citet{Curone_exoALMA}, with the following references: AA Tau -- \citet{Bouvier_ea_2013}, CQ Tau -- \citet{Donehew_Brittain_2011}, DM Tau, J1615, J1842, J1852, LkCa 15 -- \citet{Manara_ea_2014}, J1604 -- \citet{Sicilia-Aguilar_ea_2020}, HD 135344B -- \citet{Sitko_ea_2012}, HD 143006 -- \citet{Rigliaco_ea_2015}, HD 34282 -- \citet{Fairlamb_ea_2015}, MWC~758 -- \citet{Huelamo_ea_2018}, PDS 66 -- \citet{Ingleby_ea_2013}, SY Cha -- \citet{Manara_ea_2023}, V4046 Sgr -- \citet{Donati_ea_2011}.
}
\label{tab:warp_comparison}
\begin{tabular}{lcccccccccc}
\toprule
Source & $M_\star$ [$M_\odot$] & $i_0$ [$^\circ$] & $R_\mathrm{out}$ [au] & $\delta i$ [$^\circ$] & $\delta$PA  [$^\circ$] & $\beta_\mathrm{max}$ [$^\circ$] &  $\langle \log \psi \rangle_R$ & NAI & Acc. R.\footnote{Accretion rate: $\log \dot{M}_\mathrm{acc}$ [$M_\odot$~yr$^{-1}$]} & DB?\footnote{Double bell used to fit line profiles -- i.e. visible backside.}\\
\hline
MWC~758 & 1.40 & 19.4 & 266.6 & -2.3--2.0 & -4.8--5.2 & 2.56$\pm$0.17 &-0.85$\pm$0.03 & 0.43 & -7.15 & N\\
V4046 Sgr & 1.73 & -33.6 & 358.2 & -0.5--0.9 & -0.2--0.2 & 0.87$\pm$0.02 &-1.71$\pm$0.03 & 0.03 & -9.30 & N\\
HD 34282 & 1.62 & -58.3 & 741.6 & -5.2--8.3 & -2.8--1.5 & 7.40$\pm$0.33 &-0.67$\pm$0.04 & 0.11 & -7.70 & Y\\
AA Tau & 0.79 & -58.7 & 497.0 & -6.6--3.4 & -1.7--1.2 & 6.21$\pm$0.44 &-0.79$\pm$0.04 & 0.12 & -8.10 & Y\\
CQ Tau & 1.40 & -36.2 & 152.0 & -2.8--2.4 & -1.8--2.2 & 2.97$\pm$0.27 &-0.84$\pm$0.03 & 0.11 & -7.00 & N\\
DM Tau & 0.45 & 40.3 & 535.7 & -3.1--3.7 & -2.1--1.3 & 3.60$\pm$0.19 &-0.97$\pm$0.04 & 0.09 & -8.20 & Y\\
HD 135344B & 1.61 & -16.1 & 222.8 & -1.1--1.2 & -2.7--4.2 & 1.60$\pm$0.14 &-1.26$\pm$0.04 & 0.41 & -8.00 & N\\
HD 143006 & 1.56 & -16.9 & 170.6 & -0.3--1.1 & -1.6--1.4 & 1.14$\pm$0.07 &-1.40$\pm$0.05 & 0.21 & -8.10 & N\\
J1604 & 1.29 & 6.0 & 251.6 & -0.6--0.4 & -1.3--2.2 & 0.61$\pm$0.04 &-1.62$\pm$0.04 & 0.06 & -10.50 & N\\
J1615 & 1.14 & 46.1 & 538.2 & -3.7--3.7 & -1.6--1.0 & 3.60$\pm$0.13 &-1.13$\pm$0.03 & 0.04 & -8.50 & Y\\
J1842 & 1.07 & 39.4 & 317.1 & -2.2--3.7 & -1.2--1.0 & 3.56$\pm$0.25 &-1.33$\pm$0.04 & 0.07 & -8.80 & Y\\
J1852 & 1.03 & -32.7 & 247.0 & -2.3--1.0 & -0.2--0.5 & 1.71$\pm$0.22 &-1.55$\pm$0.04 & 0.02 & -8.70 & N\\
LkCa15 & 1.17 & 50.4 & 698.0 & -5.7--4.4 & -0.9--0.8 & 5.61$\pm$0.14 &-0.95$\pm$0.03 & 0.05 & -8.40 & Y\\
PDS 66 & 1.28 & -31.9 & 132.2 & -1.1--1.1 & -1.6--0.3 & 1.44$\pm$0.14 &-1.51$\pm$0.03 & 0.01 & -9.90 & N\\
SY Cha & 0.81 & -50.7 & 535.1 & -6.2--10.5 & -5.2--0.9 & 10.17$\pm$0.52 &-0.73$\pm$0.03 & 0.07 & -9.20 & Y\\
\hline
\end{tabular}
\end{table*}

\subsection{Correlations with system properties}
\label{sec:correlations}

\begin{figure}
    \centering
    \includegraphics[width=\linewidth]{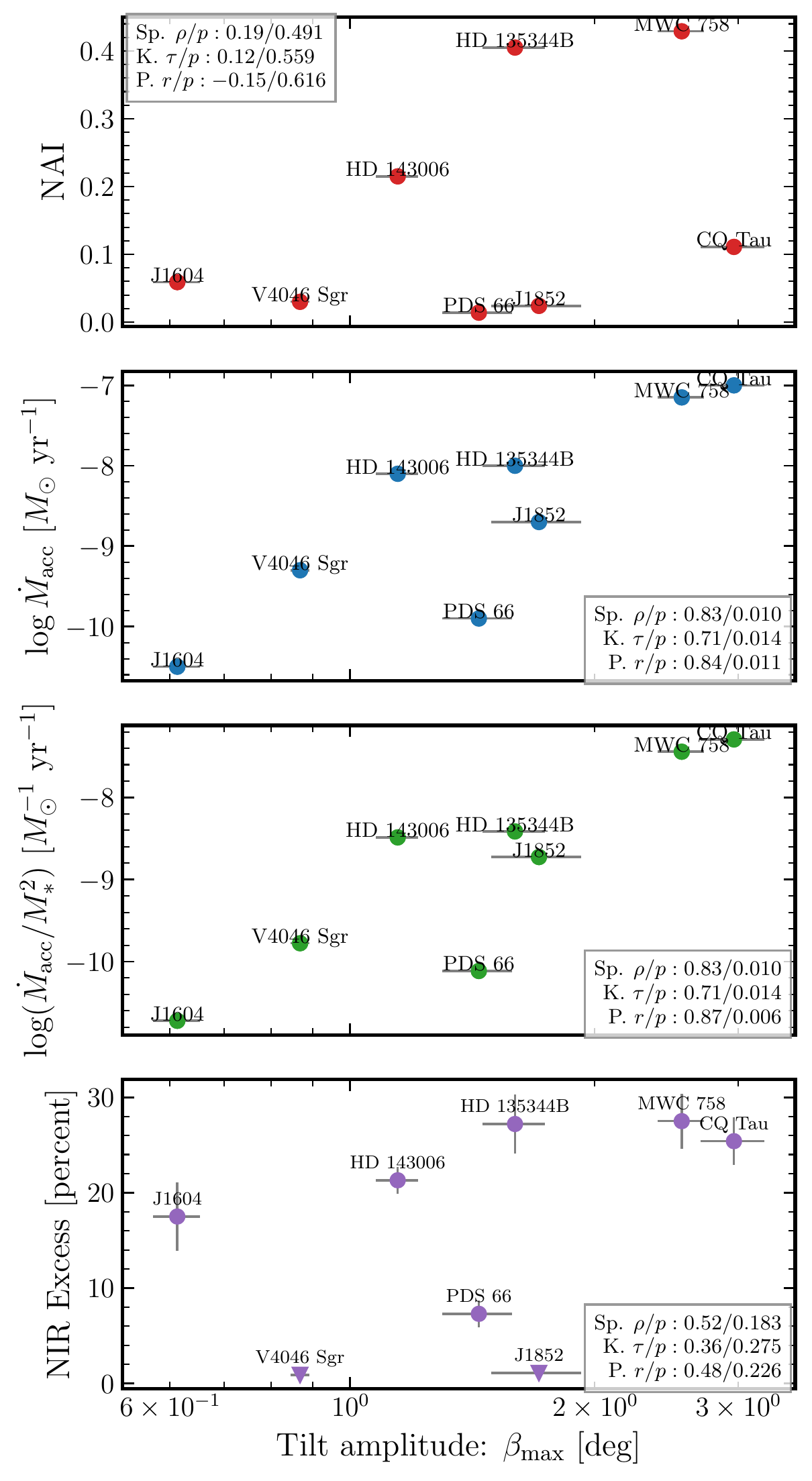}
    \caption{From top to bottom, we show how the the continuum non-axisymmetric index \citep[NAI, top --][]{Curone_exoALMA}, stellar accretion rates (middle top), normalised stellar accretion rates (to the square of the stellar mass, bottom middle) and NIR excess \citep[][bottom]{Garufi_ea_2018} depend on the range of tilt amplitudes we infer from our model.
    The outcome of Spearman rank, Kendall $\tau$ and permutation correlation tests are shown in terms of correlation statistic and $p$-value.
    We exclude cases where inclination is unfavourable due to the appearance of the back-side of the disc.}
    \label{fig:correlations}
\end{figure}

\begin{figure}
    \centering
    \includegraphics[width=\linewidth]{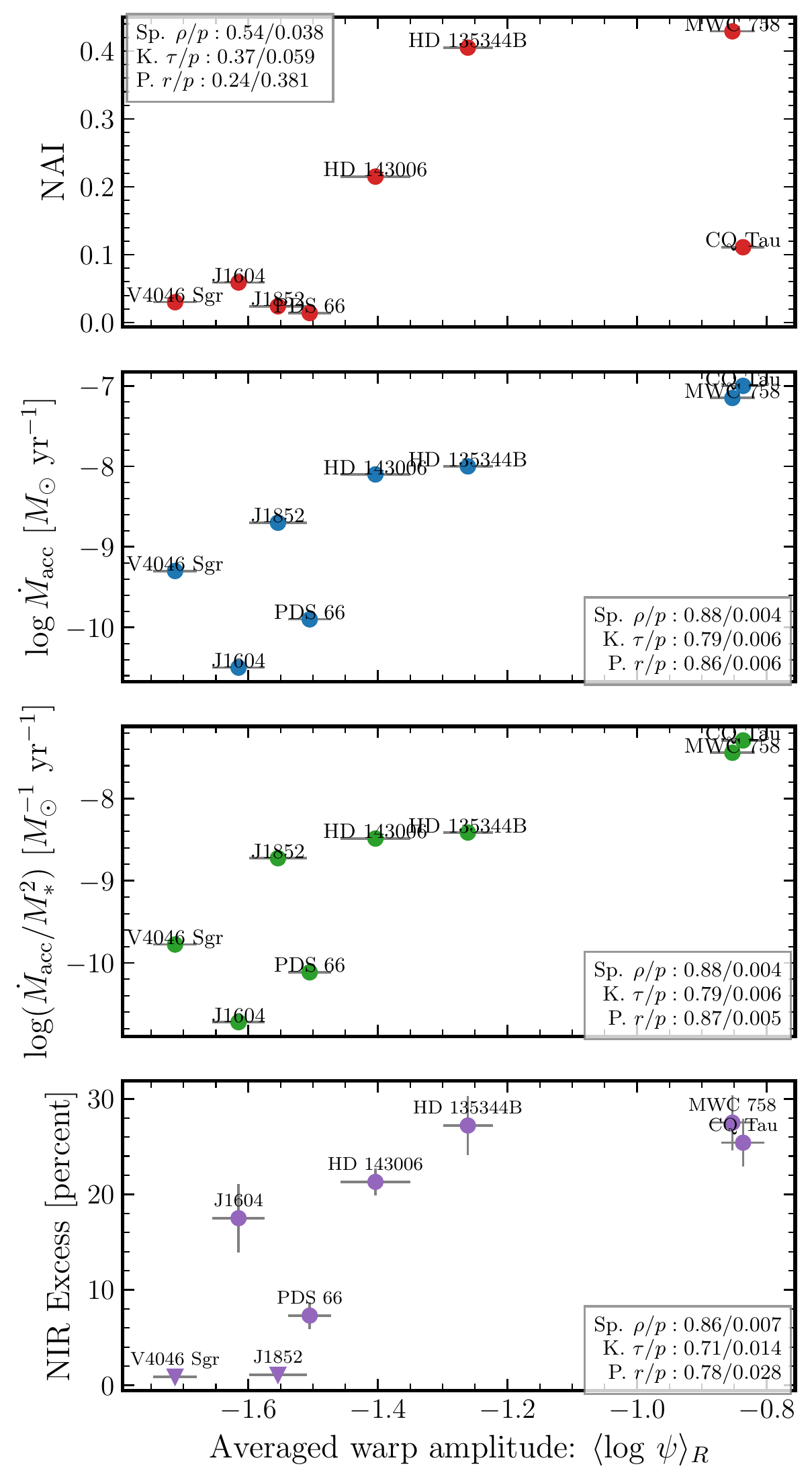}
    \caption{As for Figure~\ref{fig:correlations} but with the averaged logarithmic warp amplitude for the sample of exoALMA discs on the $x$-axis. Results of statistical correlation tests are shown as in Figure~\ref{fig:correlations}. }
    \label{fig:logpsi_mdotacc}
\end{figure}

We now restrict our consideration to  exoALMA discs with moderate inclinations, for which the backside emission is not visible, in order to search for correlations between the warp and system properties. In Figure~\ref{fig:correlations} we show how the amplitude of the warp $ \beta_\mathrm{max}$ correlates with non-axisymmetric structure in the dust \citep[measured via the non-axisymmetric index, or NAI --][]{Curone_exoALMA}, stellar accretion rate and stellar accretion rate normalised by the square of the stellar mass, the latter being the approximate observed scaling \citep[e.g.][]{Manara_ea_2017, Almendros-Abad_ea_2024, Delfini_ea_2025}, and the NIR excess \citep{Garufi_ea_2018}. We perform Spearman, Kendall $\tau$, and permutation statistical tests in each case. We implement the permutation test by comparing the correlation coefficient $r_{xy} = \sigma_{xy}/\sqrt{\sigma_{xx}\sigma_{yy}}$, where $\sigma_{xy}$ is the covariance between $x$ and $y$, over $10^4$ permutations.

We do not find any correlation between $\beta_\mathrm{max}$ and the NAI. Nor do we find a correlation with the NIR excess \citep{Garufi_ea_2018}. This would perhaps be unsurprising given that the continuum and NIR emission is far more compact than the large scale gas structures. However, by this same logic we would also not necessarily expect to see a correlation between stellar accretion rate and tilt amplitude. Interestingly though, we do find marginally significant correlations with accretion rates among our cleaned sample, although our sample size is small. 

Arguably a more appropriate quantity is the warp amplitude $\psi$, which encodes within it the dynamical stability of the warp \citep{Dogan_ea_2018}. If $\psi$ is large, annuli may split. It is not necessarily clear without more detailed calculation whether a high $\psi$ would result in permanent tearing of the disc or some instability followed by reconnection of the annuli back into a continuous disc -- this may depend on the equation of state \citep{Deng_ea_2022}. Even if a disc is stable, angular momentum transport inwards would still be facilitated by damping of the warp, since the sum of the angular momentum vectors of neighbouring annuli must result in contraction when the disc returns to the plane \citep[e.g.][]{Lodata_Pringle_2006}. If we assume that what matters for accretion is a global enhancement of $\psi$ across all radii, then we can define a geometrically averaged quantity:
\begin{equation}
\label{eq:log_psi}
    \langle \log \psi \rangle_R = \frac 1 {\Delta R}\int \log \psi \, \mathrm{d}R,
\end{equation}where $\Delta R$ is the range of radii over which we integrate. This statistic has the benefit that it should not be strongly dependent on local noise, but rather a metric of whether we find high values persistently through the disc.

The star and disc properties as a function of $ \langle \log \psi \rangle_R$ are shown in Figure~\ref{fig:logpsi_mdotacc}. In this case, we find at least marginally significant correlations in all four cases for the majority of statistical tests. Despite the small sample size, we obtain $p$-values down to $p\approx 4\times 10^{-3}$ ($2.7\, \sigma$). These correlations appear compelling if not categorical; expanding the sample size remains a goal for the future.

Perhaps from an empirical point of view we should no longer be surprised; we have hypothesised that warping may contribute to scattered light spiral structures, and it has previously been shown that these spirals are correlated with NIR excess in the inner disc \citep{Garufi_ea_2018}. Nonetheless, this new correlation raises the exciting prospect that warps themselves may be responsible, in whole or part, for angular momentum transport through the outer disc. If late stage infall is a driver of the disc warping, this may connect to the recently reported correlations between local ISM density and accretion rates \citep{Winter_ea_2024c, Rogers_ea_2024, Delfini_ea_2025}. Clearly the sample size is small, and further work is needed to verify the correlation and explore the connection to infall.

\subsection{Caveats for the simple model}
\label{sec:sloshing}

\subsubsection{Warp propagation}
The first caveat for our simple model is that we have ignored the contribution to the velocity field from the propagation of the warp through the disc. We can estimate the amplitude of the non-azimuthal motions due to this propagation by noting that the period of oscillation for the warp in the bending regime is:
\begin{equation}
\label{eq:Pwarp}
    P_\mathrm{warp} \approx \frac{ R}{c_\mathrm{s}}.
\end{equation}Equation~\ref{eq:Pwarp} comes from the fact that the warp wave speed is $c_\mathrm{s}/2$ and the bending wavelength is $R/2$ \citep{Lubow_ea_2000}. Then the out-of-plane motion of the disc due to the warp propagation is:
\begin{equation}
    v_\mathrm{warp,z} \sim \beta_\mathrm{max}  R / P_\mathrm{warp},
\end{equation}where $ \beta_\mathrm{max}$ is the tilt amplitude in radians. If we assume $ \beta_\mathrm{max} = 10^\circ$, we have $v_\mathrm{warp,z} \sim 0.17 c_\mathrm{s} \sim 0.02 v_\phi$ for $H/R=0.1$. This is small compared to the expected projected contribution from the perturbed azimuthal velocity and the perturbation required to reproduce LOS variations \citep{Stadler_exoALMA}.

\subsubsection{Sloshing and breathing}

Another approximation is made in assuming throughout our calculation that all motion is in the midplane and azimuthal. This is clearly not the case, with the $^{12}$CO surface raised substantially out of the mid-plane \citep[e.g.][]{Galloway_exoALMA}. At these heights, the molecular surfaces of a warped disc may be subject to not only the motions induced by the warp propagation, but also so-called `sloshing' and `breathing' motions \citep[e.g.][]{Lodato_ea_2007}. These motions are produced by pressure gradients due to the vertical offset between annuli, which lead to epicyclic motions that exert a substantial torque on the disc \citep{Papaloizou_ea_1983}. In the limit of low tilt amplitude for inviscid discs, these can be divided into the horizontal sloshing (in $R$ and $\phi$) and the vertical breathing mode \citep{Ogilvie_ea_2013}.

While the exact impacts of each of these contributions is complex, we can make some order of magnitude estimates. Starting with the breathing motion, if the molecular emission surface is substantially above the mid-plane then we may expect a fluid element to travel a distance of order its average height $\langle z_\mathrm{surf} \rangle$ over a warp oscillation time. Then plausibly breathing modes would have velocity:
\begin{equation}
    v_\mathrm{breath} \sim \langle z_\mathrm{surf} \rangle  / P_\mathrm{warp}.
\end{equation}If we assume $\langle z_\mathrm{surf} \rangle = 2H$, for pressure scale height $H$, we have $v_\mathrm{warp,z} \sim 2 c_\mathrm{s}^2/v_\phi \sim 0.02 v_\phi$. \citet{Kimmig_ea_2024} find similar breathing motions to our estimate, $\sim 0.2 c_\mathrm{s}$. This is also a comparatively small contribution.

\citet{Lodato_ea_2007} showed that the warp produces a vertically shearing horizontal motion in the (nominal) disc plane in the form $v_{\rm slosh} \sim (\psi/\alpha)\Omega z$. The precise analysis by \citet{Dullemond_Kimmig_ea_2022} confirms this estimate to within factors of order unity and decomposed the $r$ and $\phi$ components. Taking the real part of their equations~88~and~89, appropriate in a vertically isothermal, perfectly Keplerian disc, with low viscosity (${\alpha \ll 1}$), the amplitude of the sloshing velocities is
\begin{equation}
    v_{r, \mathrm{slosh}} \approx \frac{\psi}{2 \alpha} \Omega z, \quad v_{\phi,\mathrm{slosh}} \approx \frac{\psi}{4 \alpha} \Omega z,
\end{equation}
where $\Omega$ is the angular frequency and $z$ the height above the midplane.
At one pressure scale height, ${z_\mathrm{surf}=H}$, the maximum sloshing velocity is $v_\mathrm{max} = \psi c_\mathrm{s} / (2\alpha)$, which gives $\sim 5 c_\mathrm{s}$ for ${\psi = 10^{-2}}$ and ${\alpha=10^{-3}}$.
In their simulations, \citet{Kimmig_ea_2024} indeed find radial sloshing velocities on that order of magnitude.

If these sloshing motions are indeed several times the sound speed, then they will undoubtedly contribute to the LOS velocity structures. However, we note that at altitudes $z_\mathrm{surf}\gtrsim H$ the assumptions taken to estimate the sloshing motions break down, as the vertical communication timescale becomes longer than the orbital timescale.
Shocks are likely to occur, which produce a complicated velocity field and might influence the thermal structure at the disc surfaces.
We expect both shocks and instabilities also for $\psi \gg \alpha$ \citep[see for example][]{Dogan_ea_2018}. This could lead to rapid decay of the warp, and therefore large sloshing velocities may require a persistent driving torque to be sustained.


For the purposes of this work, we note that in cases where sloshing motions are prevalent, this may be expected to break the $m=1$ symmetry to which we fit our warp model, since $m=1$ in $r$ and $\phi$ velocity components follows from axisymmetric motions on the annulus. We might then expect this to then manifest as noise in our fitting procedure. We can test this indirectly. Given sloshing motions should be vertically and radially dependent, if they strongly influence our results we may expect substantial tilt amplitude differences depending on the spatial region probed.  We explore the effects of varying molecular tracer (i.e. $z_\mathrm{surf}$) and beam size (see discussion in Section~\ref{sec:13CO_MWC758}). We find our results remain broadly unchanged. This does not prove that sloshing motions are not present, but it suggests that they might act more like noise than a strong bias. This clearly still requires investigation in future work, including the role of shock dissipation and how the emission height changes under their influence \citep{Galloway_exoALMA}, and this should be understood as a caveat of our results.

\subsubsection{Optical depth}

{\citet{Young_ea_2022} demonstrated in their numerical calculations that the optical depth of the molecular tracer can have a substantial effect on the inferred warp properties, due to the contribution to emission from different parts of the column along the LOS. This is shown particularly in their Figure 7, where the expected difference in the inferred phase angle is often tens of degrees between $^{12}$CO and $^{13}$CO (assuming factor $\sim 10$ difference in optical depth). Interestingly, we do not see a substantial difference between $^{12}$CO and $^{13}$CO, at least in the maximal tilt angle we infer, with minor systematic differences of typically $< 1^\circ$ as discussed in Section~\ref{sec:13CO_all}. Unraveling the dependence of warp structure on optical depth will certainly require an effort to produce synthetic observables for specific discs. We also highlight that our inferred tilt profiles for discs such as MWC~758 and CQ~Tau have an approximately linear twist profile with radius, which does not resemble the twist profile expected around circumbinary discs \citep[e.g.][]{Lodato_Facchini_2013}, as modeled by \citet{Young_ea_2022}. This might indicate that in some cases at least the warp is not driven by a binary. This could have further ramifications for comparisons to numerical models and synthetic observations. We leave a full physical model and radiative transfer calculations tailored to specific discs to future work, for which our warp profile fits offer a starting point.}

 \subsection{Non-uniqueness of the warping interpretation}
\label{sec:uniqueness}
 We emphasise that in the literature to date, the LOS velocity residuals have typically (although not always) been interpreted as planar axisymmetric structures \citep[e.g.][]{Stadler_exoALMA}, winds \citep[e.g.][, Benisty et al. in prep.]{Hawoth_ea_2017} spiral arms \citep[e.g.][]{Teague_ea_2022, Ren_ea_2024}, laminar flows \citep[e.g.][]{Rosenfeld_ea_2014, Zulete_ea_2024} or localised perturbations such as those due to planets \citep[e.g.][]{Pinte_ea_2018b, Pinte_ea_2020}. The warp model we discuss in this paper cannot explain any kind of feature in kinematics without azimuthal wavenumber $m=1$, although, as we show in this work, this may produce diverse structures in other tracers. Residual features after subtraction of the warp model include winds that are predominantly vertical and localised structures.

 The warp model is also degenerate with azimuthally symmetric perturbations in the radial and azimuthal component, which are both $m=1$ in their LOS component \citep[e.g.][]{Izquierdo_ea_2021}. This means that features driven by processes such as pressure support or self-gravity may also explain some of the observed structures \citep[e.g.][]{Lodato_ea_2023, Veronesi_ea_2024, Martire_ea_2024, Longarini_exoALMA}. Although these planar, axisymmetric models may need to be somewhat contrived to produce some of the spiral-like structures found in LOS kinematics, the warp interpretation is not unique. In principle, with detailed physical and radiative transfer modeling, it may be possible to distinguish the warp from other kinds of kinematic perturbation. However this is not the goal of this work, which is instead meant to offer an alternative explanation for the large scale structures seen across the exoALMA sample. We have shown that the success, simplicity and potential to explain a range of observational evidence are all arguments in favour of the warp interpretation. Nonetheless, our results should be understood as a challenge to the assumption that discs are planar, and as an upper limit on disc warping in the sense we attribute $m=1$ structures as far as possible to the warp. We do not claim that all large scale kinematic perturbations are the result of warps.

\section{Conclusions}

We have shown that much of the large scale structure in the exoALMA sample is consistent with a warped disc. Moderate amplitude warps with \( \Delta \beta \sim 0.5-2^\circ \) can reproduce the main features in several LOS kinematic residuals from a Keplerian model if these residual features have point anti-symmetry and satisfy $m=1$ periodicity on the annulus. Alongside other models, warping is not a unique interpretation for the observed LOS residuals. However, for discs where the large scale kinematic structure has $m=1$ symmetry warping is a compelling and simple model that may be understood as a benchmark with which to compare competing physical models. It also fits with growing evidence that warping and misalignment is a common occurrence among protoplanetary discs \citep{Cody_ea_2014, Garufi_ea_2018, Benisty_ea_2023}. The warp interpretation also appears to be a promising way to explain the scattered light and CO brightness temperature morphology of MWC~758, and possibly other discs with prominent spirals.

Assuming a warp structure explains a large part of observed $m=1$ kinematic substructure, we explored possible correlations of warping with departures from symmetry in the continuum, stellar accretion rates and NIR excess. {In particular, we considered both the magnitude of the tilt and the geometrically averaged warp amplitude.} We find a positive correlation only with stellar accretion for the magnitude of the tilt, but with all of the disc properties {(to varying degrees of significance) for the geometrically averaged warp amplitude.} The sample size should be increased in future work to confirm these correlations. {Nonetheless, our results} add to the growing evidence for communication between the inner and outer disc, and hint at a potential role of large scale perturbations in driving stellar accretion.

We conclude that warps and their physical drivers should be explored alongside alternative mechanisms as a plausible pathway to produce large-scale kinematic structures.

\section*{Acknowledgments}

{We thank the anonymous referee for their close reading and useful comments,} and Kees Dullemond for helpful discussion and insights into the dynamics of warped discs.
AJW has been supported by the European Union’s Horizon 2020 research and innovation programme Marie Skłodowska-Curie grant
agreement No 101104656 and by the Royal Society through a University Research Fellowship grant number URF\textbackslash R1\textbackslash 241791.
MB, JS and DF have received funding from the European Research Council (ERC) under the European Union’s Horizon 2020 research and innovation programme (PROTOPLANETS, grant agreement No. 101002188).
Support for AFI was provided by NASA through the NASA Hubble Fellowship grant No. HST-HF2-51532.001-A awarded by the Space Telescope Science Institute, which is operated by the Association of Universities for Research in Astronomy, Inc., for NASA, under contract NAS5-26555.
GL and CL have received funding from the European Union's Horizon 2020 research and innovation program under the Marie Sklodowska-Curie grant agreement No. 823823 (DUSTBUSTERS). GL acknowledges support from PRIN-MUR
20228JPA3A and from the European Union Next Generation EU,
CUP:G53D23000870006. 
GR and CK acknowledge support from the European Union (ERC Starting Grant DiscEvol, project number 101039651) and from Fondazione Cariplo, grant No. 2022-1217.
JB acknowledges support from NASA XRP grant No. 80NSSC23K1312.
NC has received funding from the European Research Council (ERC) under the European Union Horizon Europe research and innovation program (grant agreement No. 101042275, project Stellar-MADE).
PC acknowledges support by the Italian Ministero dell'Istruzione, Universit\`a e Ricerca through the grant Progetti Premiali 2012 – iALMA (CUP C52I13000140001) and by the ANID BASAL project FB210003.
SF is funded by the European Union (ERC, UNVEIL, 101076613). SF acknowledges financial contribution from PRIN-MUR 2022YP5ACE.
MF is supported by a Grant-in-Aid from the Japan Society for the Promotion of Science (KAKENHI: No. JP22H01274).
CH acknowledges support from NSF AAG grant No. A.
TH is supported by an Australian Government Research Training Program (RTP) Scholarship.
JDI acknowledges support from an STFC Ernest Rutherford Fellowship (ST/W004119/1) and a University Academic Fellowship from the University of Leeds.
CL has received funding from the UK Science and Technology research Council (STFC) via the consolidated grant ST/W000997/1.
FMe has received funding from the European Research Council (ERC) under the European Union's Horizon Europe research and innovation program (grant agreement No. 101053020, project Dust2Planets).
CP acknowledges Australian Research Council funding  via FT170100040, DP18010423, DP220103767, and DP240103290.
H-WY acknowledges support from National Science and Technology Council (NSTC) in Taiwan through grant NSTC 113-2112-M-001-035 and from the Academia Sinica Career Development Award (AS-CDA-111-M03).
TCY acknowledges support by Grant-in-Aid for JSPS Fellows JP23KJ1008.
Support for BZ was provided by The Brinson Foundation. 
Views and opinions expressed by ERC-funded scientists are however those of the author(s) only and do not necessarily reflect those of the European Union or the European Research Council. Neither the European Union nor the granting authority can be held responsible for them.

This paper makes use of the following ALMA data: ADS/JAO.ALMA\#2021.1.01123.L. ALMA is a partnership of ESO (representing its member states), NSF (USA) and NINS (Japan), together with NRC (Canada), MOST and ASIAA (Taiwan), and KASI (Republic of Korea), in cooperation with the Republic of Chile. The Joint ALMA Observatory is operated by ESO, AUI/NRAO and NAOJ. The National Radio Astronomy Observatory is a facility of the National Science Foundation operated under cooperative agreement by Associated Universities, Inc. We thank the North American ALMA Science Center (NAASC) for their generous support including providing computing facilities and financial support for student attendance at workshops and publications.

\section*{Data Availability}

{The fitting scripts used in this work are available at \url{https://github.com/ajw278/warpfitter}.  The Gaussian Process posterior samples of warp parameters for the exoALMA protoplanetary disc sample are available at \url{https://doi.org/10.5281/zenodo.15878577}. The data will be made publicly available upon publication of this article.}

\bibliography{main}{}
\bibliographystyle{aasjournal}

\appendix

\renewcommand{\thefigure}{A\arabic{figure}}
\setcounter{figure}{0}

\section{Physical warp properties}
\label{app:phys_coords}

In this appendix, we relate physical warp angles to the observational perturbations to which we fit directly. However, it must be understood that these coordinates come with an unavoidable ambiguity. Theoretical studies may choose different frames by which to define the physical warp parameters (for example, a binary orbit, inner or outer disc), {although the most physically relevant is the total angular momentum of the system.} These choices have an influence on how the profiles appear, and even on metrics extracted from the warp structure. For example, in Figure 3 of \citet{Juhasz_Facchini_2017} it can be clearly seen that the physical warp angles vary depending on the choice of reference angular momentum vector. In our work, the natural choice for the reference angular momentum unit vector is that of the unperturbed disc geometry fit by \textsc{Discminer} \citep{Izquierdo_exoALMA}. There is no perfect choice, and this consideration will become relevant for future theoretical studies aiming to explain warp structure.

\subsection{Angular momentum vector in the observed coordinates}
Given a global disc inclination \( i_0 \) and position angle \( \mathrm{PA}_0 \), the unperturbed angular momentum vector in the sky coordinate system is:
\begin{equation}
\vec{l}_0 =
\begin{pmatrix}
- \sin i_0 \sin \mathrm{PA}_0 \\
\sin i_0 \cos \mathrm{PA}_0 \\
\cos i_0
\end{pmatrix},
\end{equation}
where \( i_0 \) is the inclination (0 for face-on, \(90^\circ\) for edge-on), \( \mathrm{PA}_0 \) is the position angle, measured east of north to the red-shifted major axis of the disc, the coordinate system is defined such that \( x \) increases to the east, \( y \) increases to the north, and \( z \) is along the line of sight (toward the observer). We now introduce small, radius-dependent perturbations to the inclination and position angle:
\[
i(R) = i_0 + \delta i(R), \qquad \mathrm{PA}(R) = \mathrm{PA}_0 + \delta \mathrm{PA}(R),
\]
which imply a perturbed unit angular momentum vector of the form:
\begin{equation}
\label{eq:l_approx}
\vec{l}(R) \approx \vec{l}_0 + \delta i(R) \, \hat{\mathbf{e}}_i + \delta \mathrm{PA}(R) \sin i_0 \, \hat{\mathbf{e}}_{\mathrm{PA}},
\end{equation}
where \( \hat{\mathbf{e}}_i \) and \( \hat{\mathbf{e}}_{\mathrm{PA}} \) are local unit vectors on the sphere:
\begin{align}
\hat{\mathbf{e}}_i &= \frac{\hat{z} \times \vec{l}_0}{|\hat{z} \times \vec{l}_0|}, \\
\hat{\mathbf{e}}_{\mathrm{PA}} &= \vec{l}_0 \times \hat{\mathbf{e}}_i,
\end{align}and equation~\ref{eq:l_approx} is valid for small perturbations \(\delta i(R)\) and \(\delta \mathrm{PA}(R)\). These define the directions of increasing inclination and increasing position angle on the sphere of possible disc orientations. Note that strictly if \( i_0 = 0 \) (a face on disc), then \( \vec{l}_0 = \hat{z} \) and \( \hat{z} \times \vec{l}_0 = 0 \) -- i.e. the unit vectors become undefined when the disc is exactly face-on.

\subsection{Disc coordinates: tilt and twist}

The orientation of the warped disc at radius \( R \) can also be expressed in the {disc-aligned frame}, where the unperturbed angular momentum vector lies along the \( z \)-axis. In this frame, the orientation is defined by two angles: \( \beta(R) \), which is the {tilt} angle away from the \( z \)-axis and \( \gamma(R) \): the twist angle about the \( z \)-axis. We define the disc-aligned frame \( \{ \hat{\mathbf{e}}_{x'}, \hat{\mathbf{e}}_{y}, \hat{\mathbf{e}}_{z'} \} \) such that \( \hat{\mathbf{e}}_{z'} \) is aligned with the unperturbed angular momentum vector \( \vec{l}_0 \), which corresponds to inclination \( i_0 \) and position angle \( \mathrm{PA}_0 \). To express \( \vec{l}(R) \) in this frame, we write:
\begin{equation}
\vec{l}(R) = l_{x'}(R) \, \hat{\mathbf{e}}_{x'} + l_{y'}(R) \, \hat{\mathbf{e}}_{y'} + l_{z'}(R) \, \hat{\mathbf{e}}_{z'},
\end{equation}where
\begin{align}
l_{x'}(R) &= \sin\beta(R) \cos\gamma(R), \\
l_{y'}(R) &=\sin\beta(R) \sin\gamma(R) \, \\
l_{z'}(R) &= \cos\beta(R).
\end{align}

To compute these angles from the observed inclination \( i(R) \) and position angle \( \mathrm{PA}(R) \), we transform the angular momentum vector from the sky coordinates into the disc-aligned frame. This is done by first rotating by \( -\mathrm{PA}_0 \) about the \( z \)-axis, followed by a rotation by \( -i_0 \) about the \( x \)-axis. Re-writing the exact version of equation~\ref{eq:l_approx}, the sky-frame angular momentum vector is given by:
\begin{equation}
\vec{l}(R) =
\begin{pmatrix}
- \sin i(R) \sin \mathrm{PA}(R) \\
\sin i(R) \cos \mathrm{PA}(R) \\
\cos i(R)
\end{pmatrix}.
\end{equation}
If we apply a rotation by $-$PA$_0$ in the $z$-axis, followed by $-i_0$ in the $x$-axis, then we find the disc-frame components are:
\begin{align}
l_{x'}(R) &= - \sin i(R) \sin(\mathrm{PA}(R) - \mathrm{PA}_0), \\
l_{y'}(R) &= \cos i_0 \sin i(R) \cos(\mathrm{PA}(R) - \mathrm{PA}_0) + \sin i_0 \cos i(R), \\
l_{z'}(R) &= - \sin i_0 \sin i(R) \cos(\mathrm{PA}(R) - \mathrm{PA}_0) + \cos i_0 \cos i(R).
\end{align}
The tilt and twist angles are then
\begin{align}
\label{eq:beta_gamma_exact}
\beta(R) &= \arccos\left( l_{z'}(R) \right), \\
\gamma(R) &= \arctan2\left( l_{y'}(R), \, l_{x'}(R) \right).
\end{align}

In the limit of small perturbations \( \delta i(R), \delta \mathrm{PA}(R) \ll 1 \), we may expand these expressions to obtain:
\begin{align}
l_{x'}(R) &\approx - \sin i_0 \, \delta \mathrm{PA}(R), \\
l_{y'}(R) &\approx \delta i(R), \\
l_{z'}(R) &\approx 1 - \tfrac{1}{2} \left[ \delta i(R)^2 + \delta \mathrm{PA}(R)^2 \sin^2 i_0 \right].
\end{align}
Hence,
\begin{align}
\label{eq:beta_small}
\beta(R) &\approx \sqrt{ \delta i(R)^2 + \delta \mathrm{PA}(R)^2 \sin^2 i_0 }, \\
\label{eq:gamma_small}
\gamma(R) &\approx \arctan2\left( \delta i(R) , -\sin i_0 \, \delta \mathrm{PA}(R)  \right).
\end{align}
These expressions provide a mapping from the sky-projected disc orientation to a physical representation of the warp in terms of tilt and twist.

\subsection{Warp amplitude}

Another useful quantity relating to the geometry of the warp is defined \citep{Ogilvie_1999}:
\begin{equation}
    \vec{\psi}(R) \equiv R \frac{d\vec{l}}{dR},
\end{equation}or in our case:
\begin{equation}
\vec{\psi}(R) = R
\begin{pmatrix}
- \cos i \sin \mathrm{PA} \, \frac{\partial i}{\partial  R} - \sin i \cos \mathrm{PA} \, \frac{\partial \mathrm{PA}}{\partial  R} \\
\cos i \cos \mathrm{PA} \, \frac{\partial i}{\partial  R} - \sin i \sin \mathrm{PA} \, \frac{\partial \mathrm{PA}}{\partial  R} \\
- \sin i \, \frac{\partial i}{\partial  R}
\end{pmatrix}.
\end{equation}The magnitude of this vector is then:
\begin{equation}
\label{eq:psi_def}
    {\psi}(R) = R \sqrt{ \left( \frac{\partial i}{\partial  R} \right)^2 + \sin^2 i(R) \left( \frac{\partial \mathrm{PA}}{\partial R} \right)^2 }.
\end{equation}Although it is a measure of the rate at which the warp changes angle with $\ln R$, for the sake of this work, we will describe this as the warp amplitude for consistency with the literature. Some works have related this quantity to stability criteria for a warped disc, which depends on the effective viscosity that may be a product of the warp itself. We do not attempt a stability analysis here, but equations~\ref{eq:beta_small},~\ref{eq:gamma_small}~and~\ref{eq:psi_def} constitute a mapping from our observational coordinates into a physical warp structure.

\renewcommand{\thefigure}{B\arabic{figure}}
\setcounter{figure}{0}

\section{Constructing radiative transfer models}
\label{app:phys_warp}

\begin{figure}
    \centering
    \includegraphics[width=0.8\linewidth]{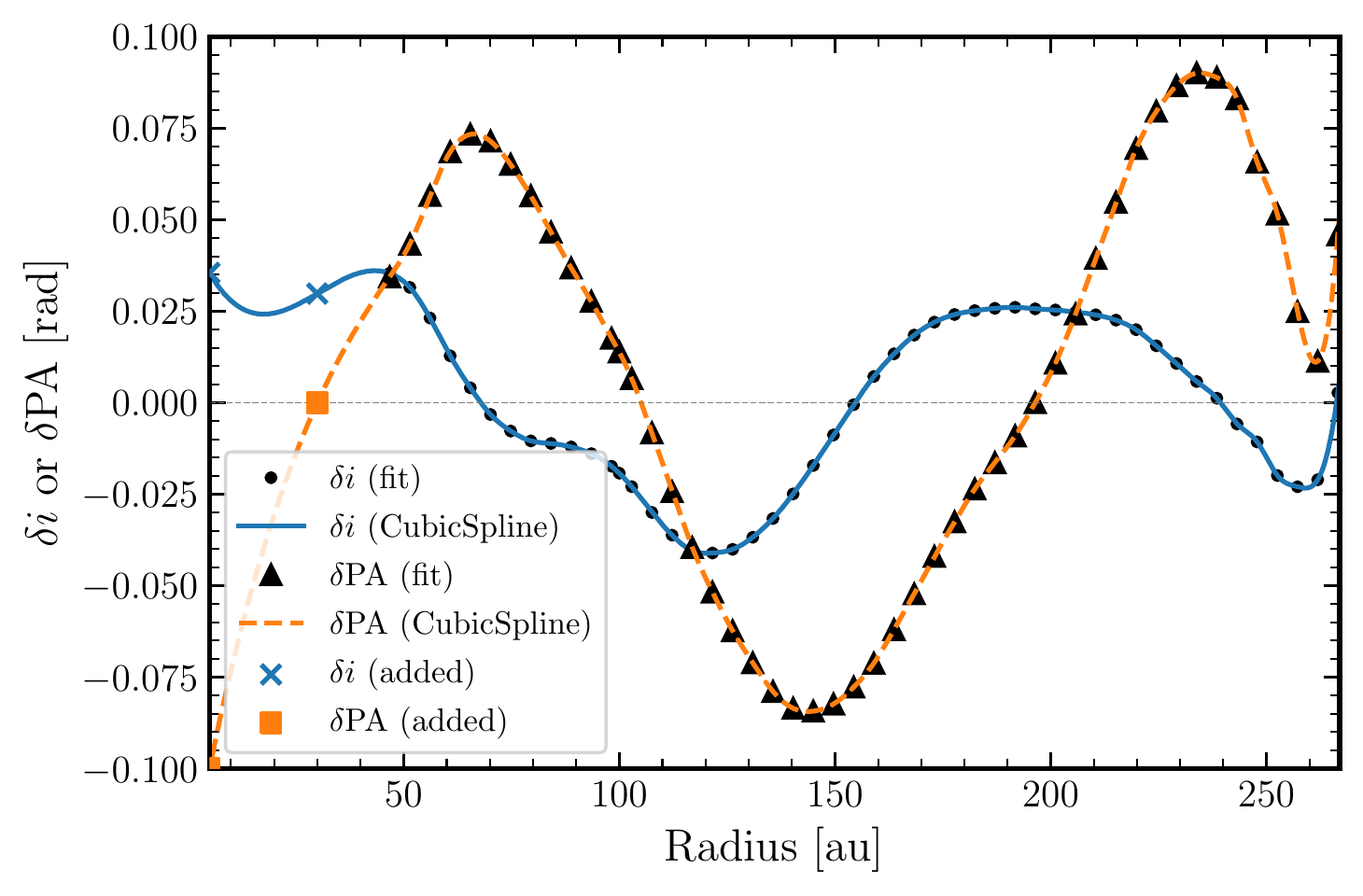}
    \caption{The extrapolation of the warp file we infer for MWC~758 into the inner disc, as required by our radiative transfer model. The two inner points were chosen arbitrarily to extrapolate the warp profile with a cubic spline. }
    \label{fig:warp_extrapolation}
\end{figure}
\begin{figure*}
    \centering
    \includegraphics[width=\linewidth]{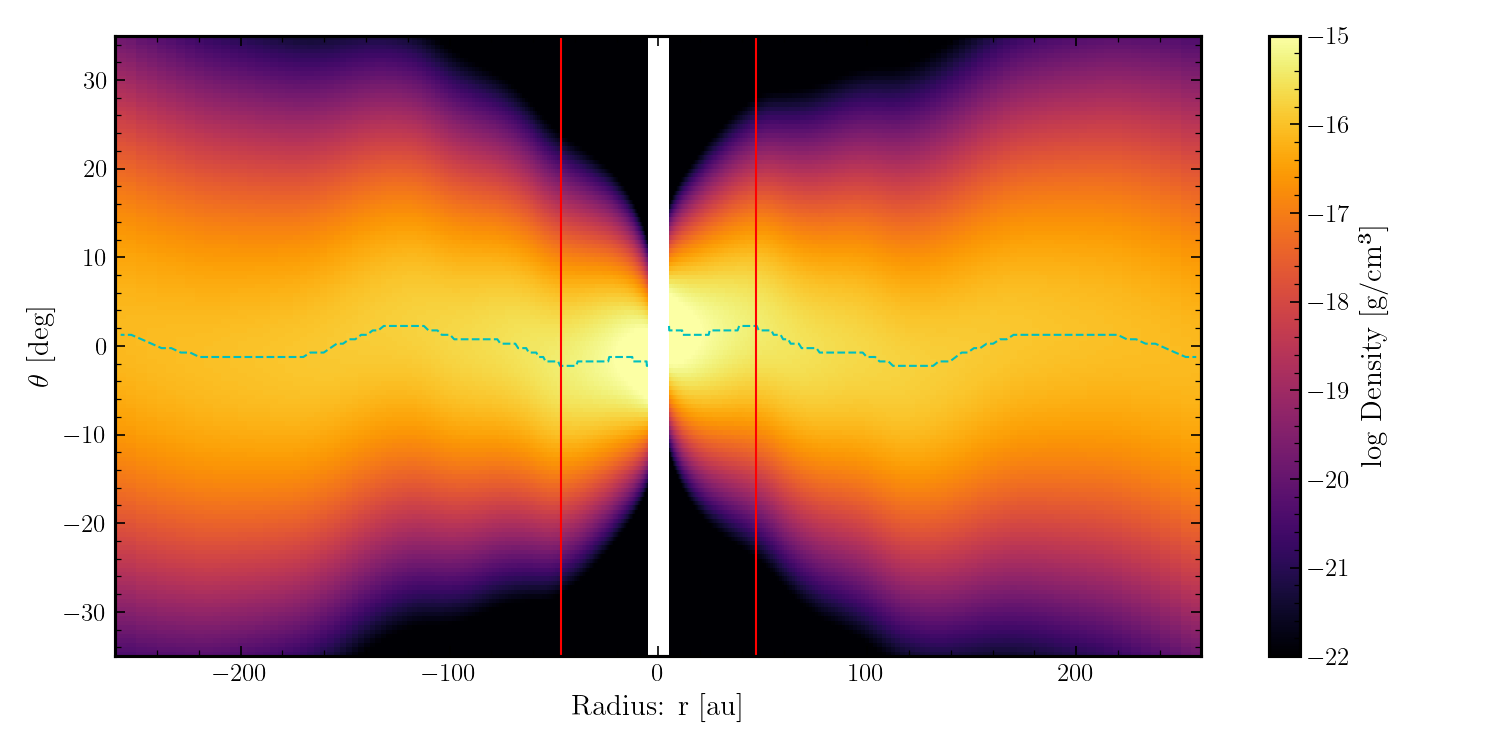}
    \caption{Slice of our model density profile used for radiative transfer calculations of scattered light. The red line shows the radius beyond which we directly use the inferred inclination and position angle profile from the residual line of sight velocity map. Inside of this we make an arbitrary cubic-spline extrapolation. The cyan dashed line traces the midplane. We use a flaring exponent of $1.03$, and a midplane density profile $\propto r^{-1}$.}
    \label{fig:model_density}
\end{figure*}

In Section~\ref{sec:scattered_light_mwc758} we discuss our application of radiative transfer calculations to model the scattered light spirals in MWC~758. As noted in that section, our warp model must be extended (extrapolated) into the inner disc in order to capture the region where the  warp casts shadows on the scattered light region. Since we are not performing a parameter space study, but a proof concept, we choose a very simple approach. We simply fix two  values of $\delta i$ and $\delta$PA at two locations interior to our kinematic warp constraint. We then perform a cubic interpolation between the radii to produce a smooth profile, as shown in Figure~\ref{fig:warp_extrapolation}. The choices of fixed points are arbitrarily chosen to produce a sensible profile, while visually attempting to estimate a warp that would produce shadows.

We show a density slice of our density profile used for our radiative transfer model in Figure~\ref{fig:model_density}. The blue line traces the midplane, as defined by the highest density at a fixed radius. The amplitude and structure of the warp remain comparable in the inner disc and outer disc. Physically, this need not be the case; there may be tearing or more complex structure in the inner disc.

\renewcommand{\thefigure}{C\arabic{figure}}
\setcounter{figure}{0}
\section{Catalogue of model fits}
\label{app:models_all}
The following figures are displayed in the style of Figure~\ref{fig:inc_pa_profiles}, but for the remainder of the exoALMA sample. We show the $^{12}$CO and then the $^{13}$CO for each disc sequentially, all for a $0.15$'' beam that is the fiducial value used in this work. The spatial ranges are determined by where the data becomes noisy and therefore do not necessarily match between different isotopes. Colour scales for the line of sight velocity are kept fixed.

\subsection{AA Tau}

AA~Tau is an example where there is already clear evidence for disc warping, with continuum rings \citep{Loomis_ea_2017} misaligned with respect to both the scattered light \citep{Cox_ea_2013} and inner disc \citep{OSullivan_ea_2005} by around $10^\circ$. As suggested by \citet{Curone_exoALMA}, this misalignment appears to cast shadows on the continuum rings. The misalignment has also been suggested as the explanation for substantial photometric variability \citep{Bouvier_ea_1999}.

AA~Tau is a complex disc to analyse kinematically due to the visible backside in $^{12}$CO line emission, being at a high inclination $i_0 \approx -59^\circ$. Figure~\ref{fig:profile_AATau} shows signatures of complex residuals that are a hazard of fitting disc properties in this case. The warp model reproduces aspects of the rotation curve, such as strong non-axisymmetric features in the inner $50$~au, particularly in the top half where the backside has not been as problematic. The outer disc exhibits symmetry ($m=2$ periodicity on the annulus), indicating that either a more complex warp model or alternative must explain these features. While the inclination profile does show some systematic trend in inclination, it is overlaid with large amplitude, small scale modulations that may be indicative of other localised structures. The nominal tilt amplitude is $6.2\pm 0.4^\circ$ assuming the structure is predominantly produced by a warp. This is slightly lower than the $\sim 10^\circ$ difference between inner and outer disc based on scattered light observations, but we are not sensitive down to the very inner disc regions.

\subsection{CQ Tau}
\citet{Bohn2022} showed with VLTI/GRAVITY and ALMA observations that CQ Tau has a substantially misaligned disc, making it another convincing case of disc warping. As shown in Figure~\ref{fig:profile_CQTau} is morphologically similar to MWC~758, and the warp model does a good job of reproducing the $m=1$ spiral-like structure. The inclination profile is similarly sinusoidal, with total tilt amplitude $3.0\pm 0.3^\circ$.

\subsection{DM~Tau}

DM~Tau is notable for being one of the few discs with directly detectable levels of turbulence via the molecular line widths of the outer disc regions, implying a high turbulent $\alpha$ \citep{Flaherty_ea_2020}. Similarly to AA Tau, DM~Tau (Figure~\ref{fig:profile_DMTau}) suffers from backside contamination and has been extracted with a double bell line profile \citep{Izquierdo_exoALMA}. At intermediate radii, a similar $m=2$ symmetry also defies the simple warp model. Unlike AA~Tau however, the warp is a more plausible explanation for the structure in the very outer disc, where the back side contaminated structure resembles $m=1$ periodicity. Outside of the inner $100$~au, the inclination profile also appears more suggestive of a large scale warp, showing a systematic linear trend with tilt amplitude $3.6\pm 0.2^\circ$.

\subsection{HD~34282}

While HD~34282 also suffers from backside contamination, it also shows a lot of $m=1$ symmetry that may be indicative of a warp. The warp model does a good job of reproducing the major features (Figure~\ref{fig:profile_HD34282}). The inclination profile is also coherent across the entire disc scale, again suggestive that a warp could be an appropriate model. However, assuming this interpretation is correct, HD~34282 exhibits a large tilt amplitude of $7.4\pm 0.3^\circ$, which is much larger than typical of discs for which backside emission is not visible.

\subsection{HD~135344B}

HD~135344B is the secondary component of a visual binary, with separation $\sim 3000$~au. It is known to have an inner disc that is misaligned compared to the outer disc \citep{Fedele_ea_2008, Grady_ea_2009, Muller_ea_2011}. It also has impressive spirals and shadows in the scattered light, as revealed by VLT/SPHERE \citep{Stolker_ea_2016}. This makes it another highly plausible kinematic-warp candidate.

As shown in Figure~\ref{fig:profile_HD135344}, the warp model does a good job of reproducing the global structure around HD~135344B, although clearly cannot reproduce the localised vortex \citep{Wölfer_exoALMA} or doppler flip ({this feature will be discussed further by} Izquierdo et al. in prep.). The inclination profile is again linear, suggestive of a large scale warp, but with moderate amplitude $1.6\pm 0.1^\circ$.

\subsection{HD 143006}

HD~143006 is another disc that is known to have a misalignment between the inner and outer disc, as evidence in the shadows observed with VLT/SPHERE \citep{Benisty_ea_2018} and VLTI/PIONIER \citep{Codron_ea_2025}, as well as inferred from the continuum ring geometry \citep{Perez_ea_2018}. \citet{Ballabio_ea_2021} had success modeling this system as an inner misaligned binary ($\lesssim 10$~au) plus an outer planet. We highlight that in this work we are here sensitive to minor warping in the outer disc and not potentially strong misalignments in the inner regions. The global kinematic structure around HD~143006 is again well produced by a warp, this time with an amplitude of only $1.1\pm 0.1^\circ$ (Figure~\ref{fig:profile_HD143006}).

\subsection{J1604}

J1604 has a misaligned inner disc with respect to the outer disc \citep{Mayama_ea_2018} as well as shadows with temporal variability \citep{Pinilla_ea_2018, Zhong_ea_2024}. {These moving shadows were successfully modeled by a \citet{Nealon_ea_2020} as a warp in a circumbinary disc.} \citet{Stadler_ea_2023} also presented kinematic evidence for a planet towards the inner edge of the dust cavity.

Here we find that the global warp again does a good job of reproducing J1604 (Figure~\ref{fig:profile_J1604}), although whether it can reproduce the structures around $150$~au depends on whether this is interpreted as a ring or tightly wound spiral \citep{Stadler_ea_2023}. Again, only a small inclination of $0.61\pm 0.04^\circ$ is required. The residuals from the warp model seem to suggest a strong wind signature in the outer disc (large negative $v_z$). If there are also substantial radial velocities arising from the wind, this may enhance the nominal warp tilt we infer outside $\sim 170$~au.

\subsection{J1615}

In scattered light observations J1615 exhibits distinctive rings that have been suggested to be an indication of truncation by an outer companion \citep{deBoer_ea_2016}. Despite the backside contamination, Figure~\ref{fig:profile_J1615} shows that J1615 is reproduced well by the warp model. The inclination profile is very linear in radius, consistent with a large scale warp with $3.6\pm 0.1^\circ$ amplitude.

\subsection{J1842}

J1842, shown in Figure~\ref{fig:profile_J1842}, is again fairly well reproduced by a warp model despite backside contamination, with total tilt amplitude is $3.6\pm 0.3^\circ$.

\subsection{J1852}

The LOS velocity structure for J1852 is shown in Figure~\ref{fig:profile_J1852}. It is consistent with a moderate tilt amplitude $1.7\pm0.2^\circ$. However, it appears very flat except in the very inner regions ($\lesssim 70$~au). Even with this inner warp, J1852 is one of the least warped discs in the exoALMA sample.

\subsection{LkCa15}

LkCa15 is another system known to have a warp in the very inner disc regions \citep{Alencar_ea_2018}, while its status as a protoplanet host remains debated \citep[e.g.][]{Currie_ea_2019, Gardener_exoALMA}. It has a nominal tilt amplitude $5.6 \pm 0.1^\circ$, but this may be overestimated because of the strong backside contribution to the residual velocity map (Figure~\ref{fig:profile_LkCa15}).

\subsection{PDS~66}

PDS~66 is notable for its apparent lack of substructure in the continuum \citep{Ribas_ea_2023, Curone_exoALMA}, although it does show a ring like structure in the scattered light \citep{Avenhaus_ea_2018}. This structure is quite symmetric however, and \citet{Ribas_ea_2023} suggest it may be the result of a shadow cast by a puffed up inner disc.  In Figure~\ref{fig:profile_PDS66} we also show that it is one of the quieter exoALMA discs. It exhibits a nominal tilt amplitude of $1.4 \pm0.1^\circ$.

\subsection{SY Cha}

\citet{Orihara_ea_2023} already suggested that the inner regions of the SY~Cha system may be warped, based on their high resolution ALMA Band 6 kinematic study. \citet{Schwarz_ea_2023} also presented a comparison of JWST/MIRI and archival \textit{Spitzer} data that showed evidence of strong inner disc variability. Our fitting procedure suggests SY Cha has the largest tilt amplitude ($10.2\pm 0.5^\circ$) of the sample, as shown in Figure~\ref{fig:profile_SYCha} despite substantial noise from the backside contribution.

However, we also see that when the warp model is extracted, we recover a systematically red inner disc region and blue outer region. As in the cases of MWC~758 and J1604, this may be the result of a small error in the systemic velocity, and an outer wind with strong negative LOS velocity residual. The  difference for SY~Cha is that if this interpretation is correct, the outer wind may be present for a substantial fraction of the disc. Interestingly, recent JWST observations revealed extended H$_2$ and [NeII] emission around SY~Cha, suggestive of a large scale wind \citep{Schwarz_ea_2025}. In this case, the residual we attribute to a warp may be strongly sub-Keplerian rotation due to the outflowing gas. The warp tilt is greatest in the hypothetical `wind' region outside of $300$~au, so the tilt amplitude for SY~Cha may be substantially overestimated due to sub-Keplerian rotation.

\subsection{V4046 Sgr}

V4046 Sgr is a binary system with a period of $2.4$~days \citep{Strempels_ea_2004}, too short to influence the outer disc. The latter is very quiet kinematically, and the tilt amplitude is $0.87\pm 0.02^\circ$.

\clearpage
\begin{figure}[ht!]
    \centering
    \includegraphics[width=\textwidth]{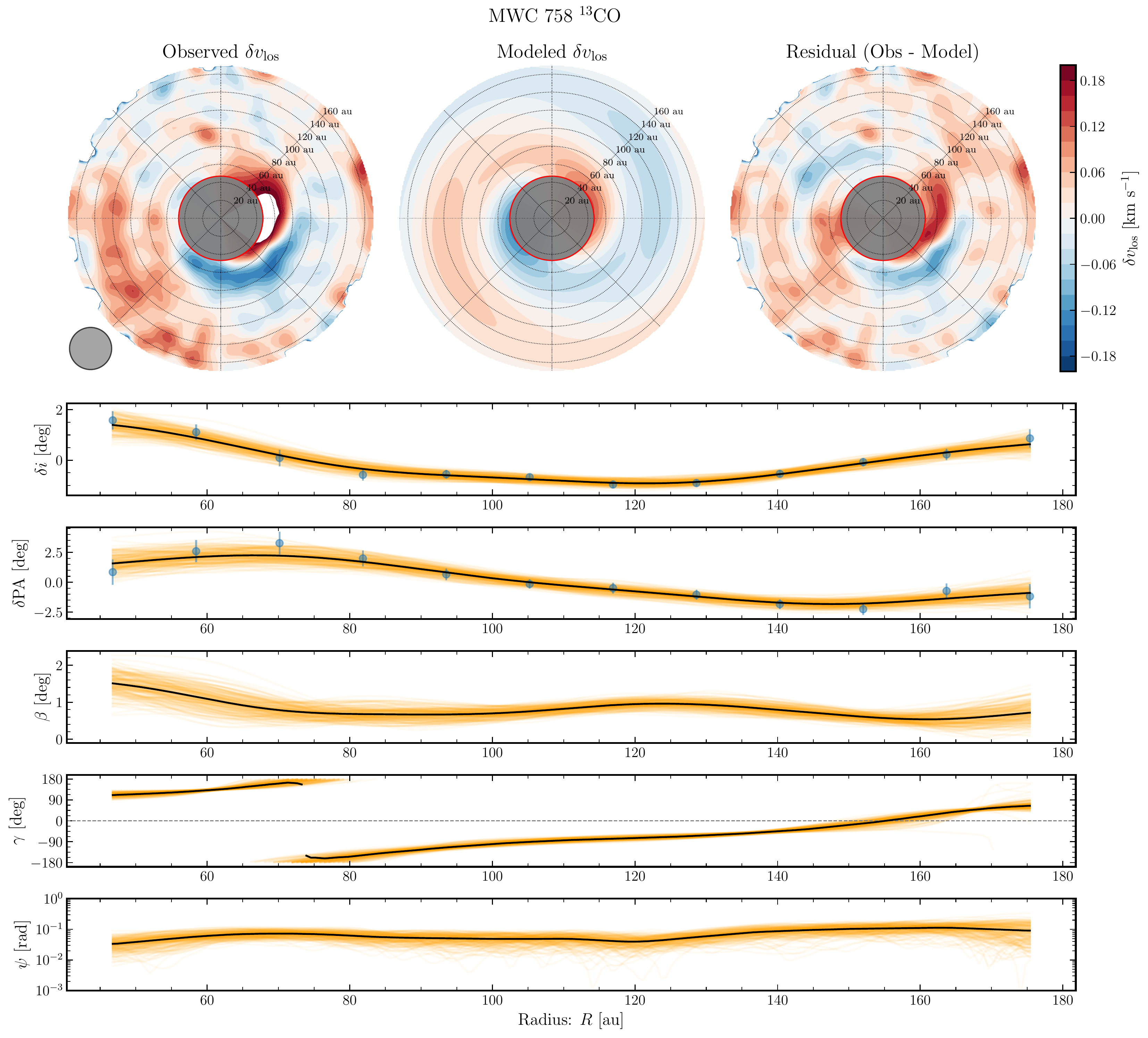}
    \caption{As in Figure~\ref{fig:inc_pa_profiles} but for $^{13}$CO.}
    \label{fig:profile_MWC758_13}
\end{figure}

\clearpage
\begin{figure}[ht!]
    \centering
    \includegraphics[width=\textwidth]{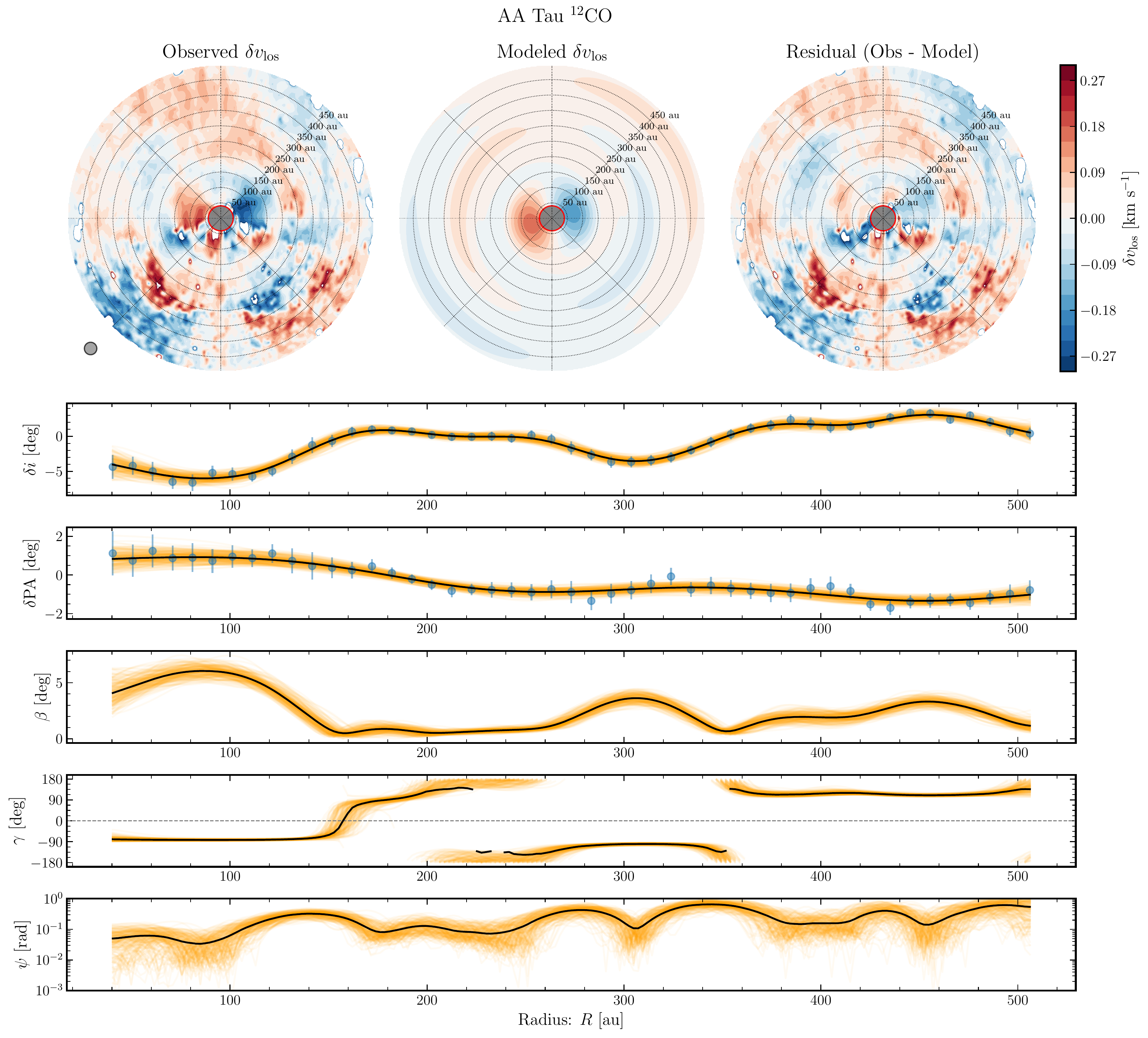}
    \caption{As in Figure~\ref{fig:inc_pa_profiles} but for AA Tau.}
    \label{fig:profile_AATau}
\end{figure}

\clearpage
\begin{figure}[ht!]
    \centering
    \includegraphics[width=\textwidth]{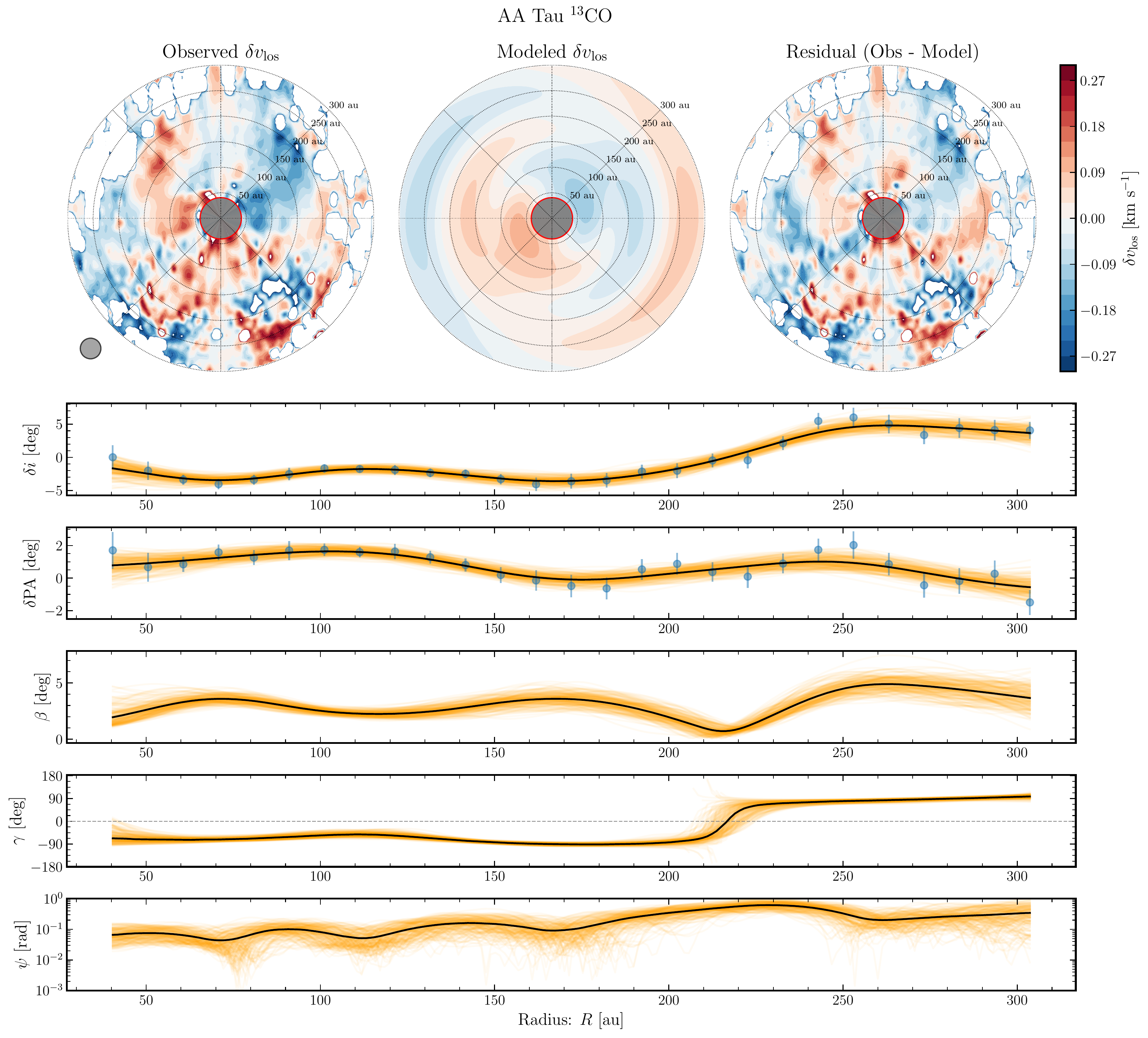}
    \caption{As in Figure~\ref{fig:profile_AATau} but for $^{13}$CO.}
    \label{fig:profile_AATau_13}
\end{figure}

\clearpage
\begin{figure}[ht!]
    \centering
    \includegraphics[width=\textwidth]{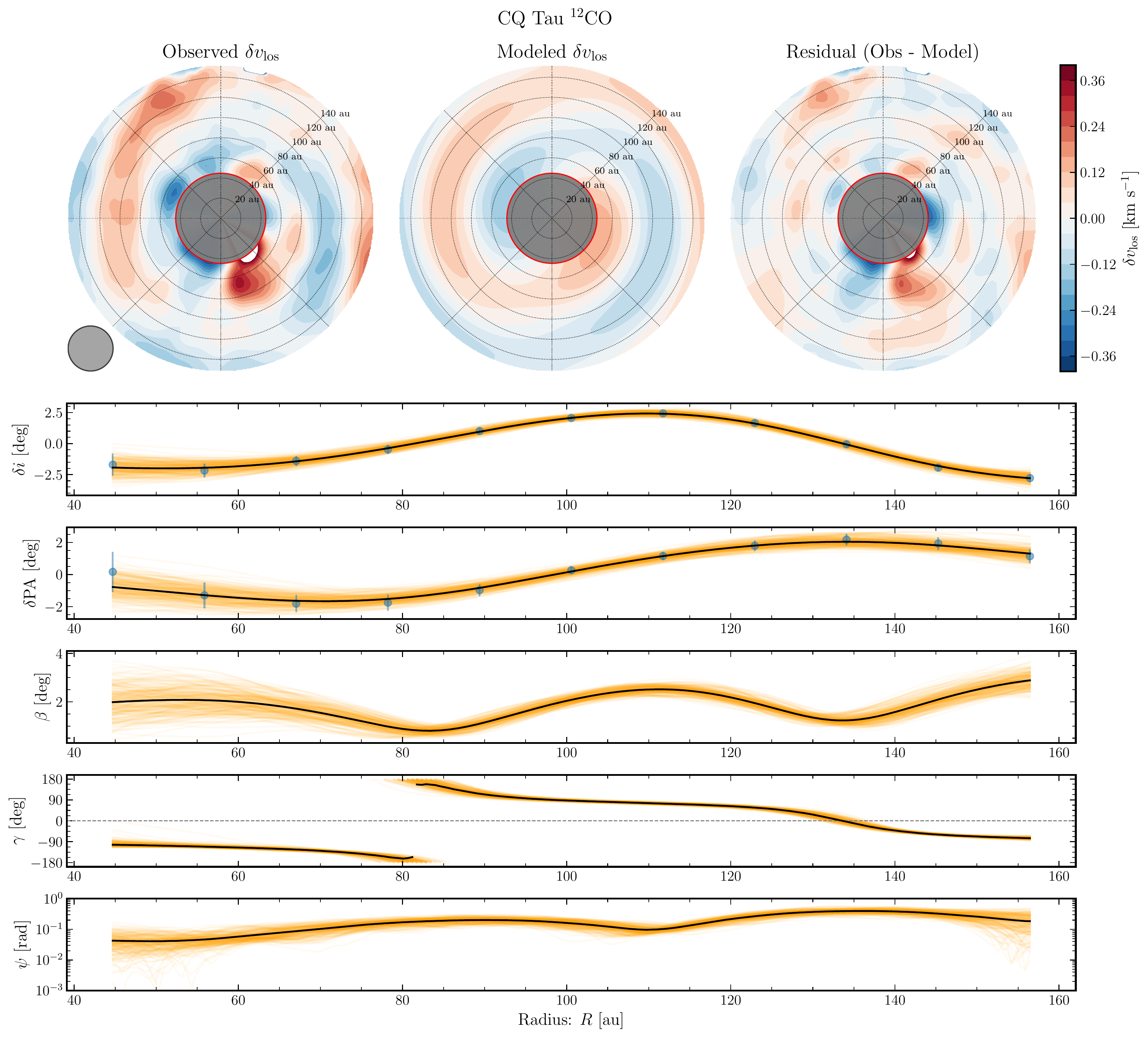}
    \caption{As in Figure~\ref{fig:inc_pa_profiles} but for CQ Tau.}
    \label{fig:profile_CQTau}
\end{figure}

\clearpage
\begin{figure}[ht!]
    \centering
    \includegraphics[width=\textwidth]{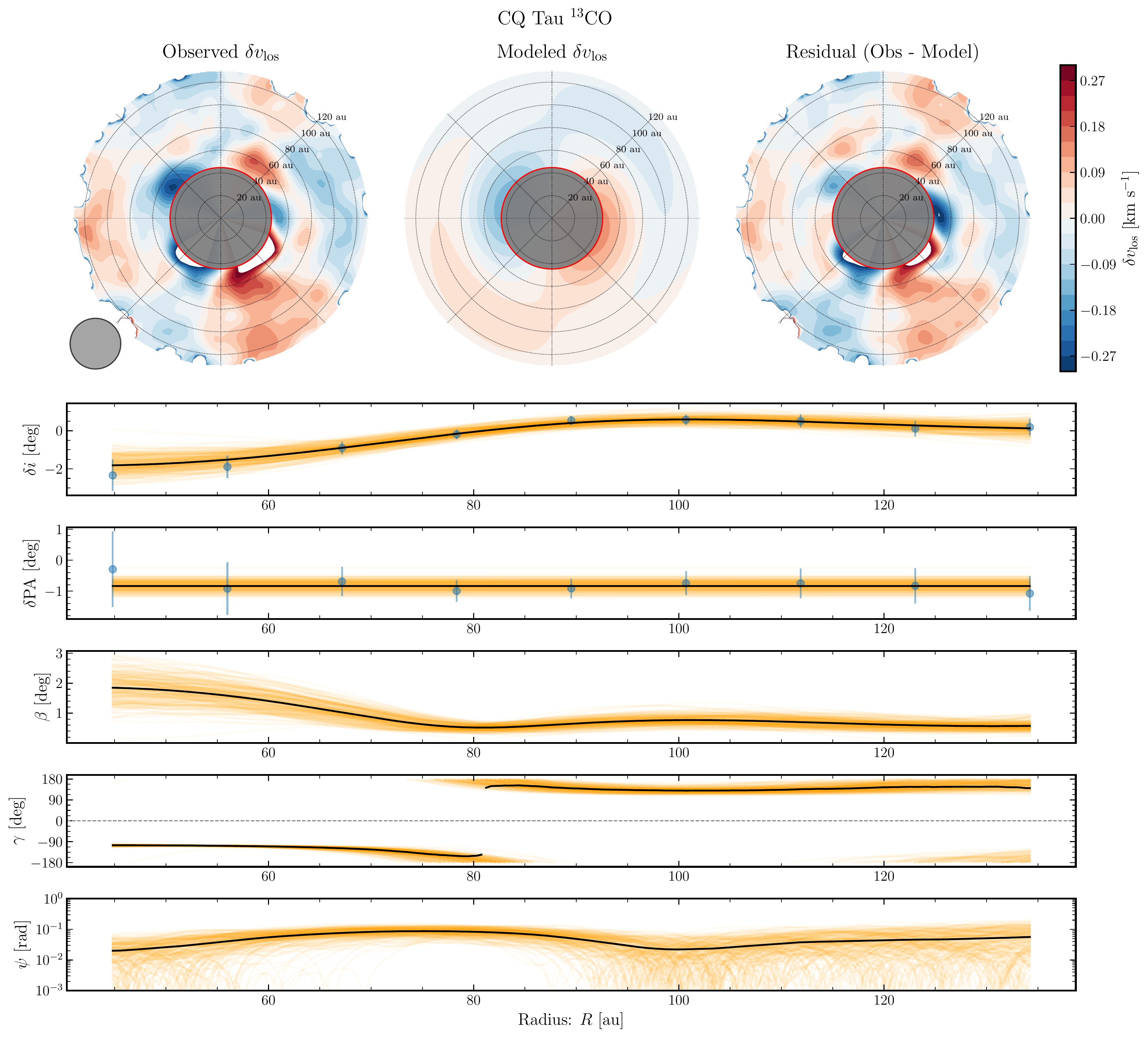}
    \caption{As in Figure~\ref{fig:profile_CQTau} but for $^{13}$CO.}
    \label{fig:profile_CQTau_13}
\end{figure}

\clearpage
\begin{figure}[ht!]
    \centering
    \includegraphics[width=\textwidth]{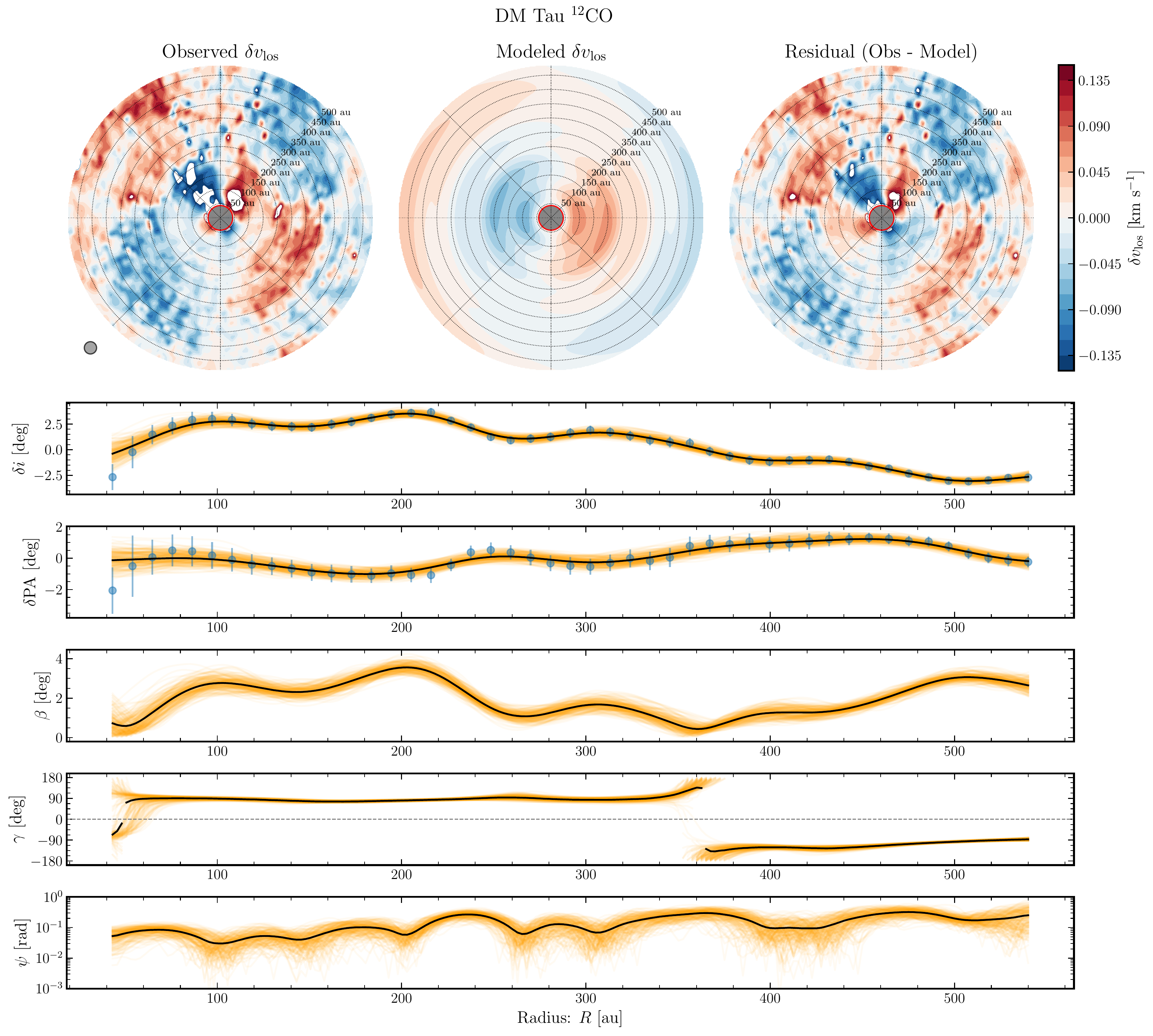}
    \caption{As in Figure~\ref{fig:inc_pa_profiles} but for DM Tau.}
    \label{fig:profile_DMTau}
\end{figure}

\clearpage
\begin{figure}[ht!]
    \centering
    \includegraphics[width=\textwidth]{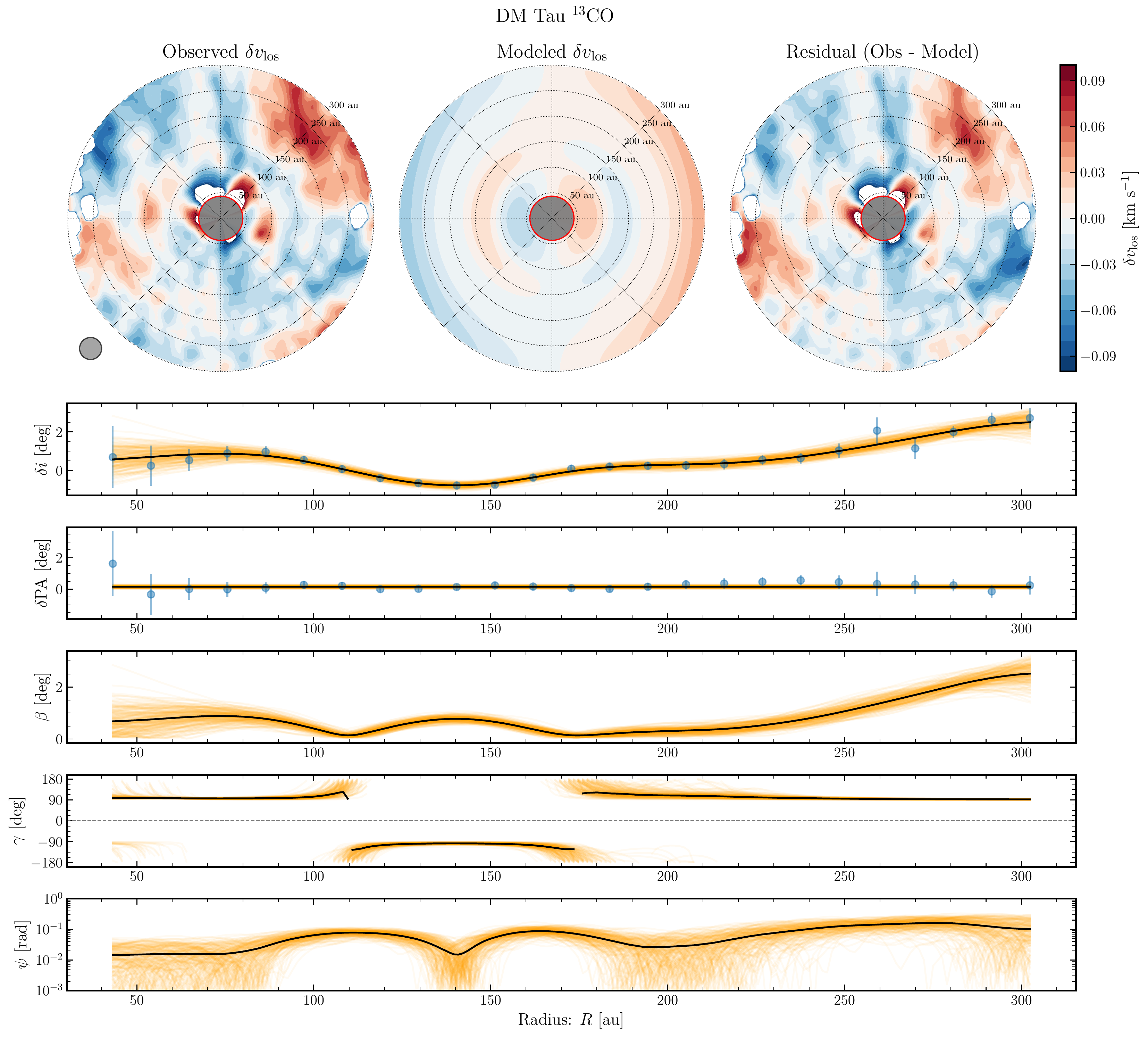}
    \caption{As in Figure~\ref{fig:profile_DMTau} but for $^{13}$CO.}
    \label{fig:profile_DMTau_13}
\end{figure}

\clearpage
\begin{figure}[ht!]
    \centering
    \includegraphics[width=\textwidth]{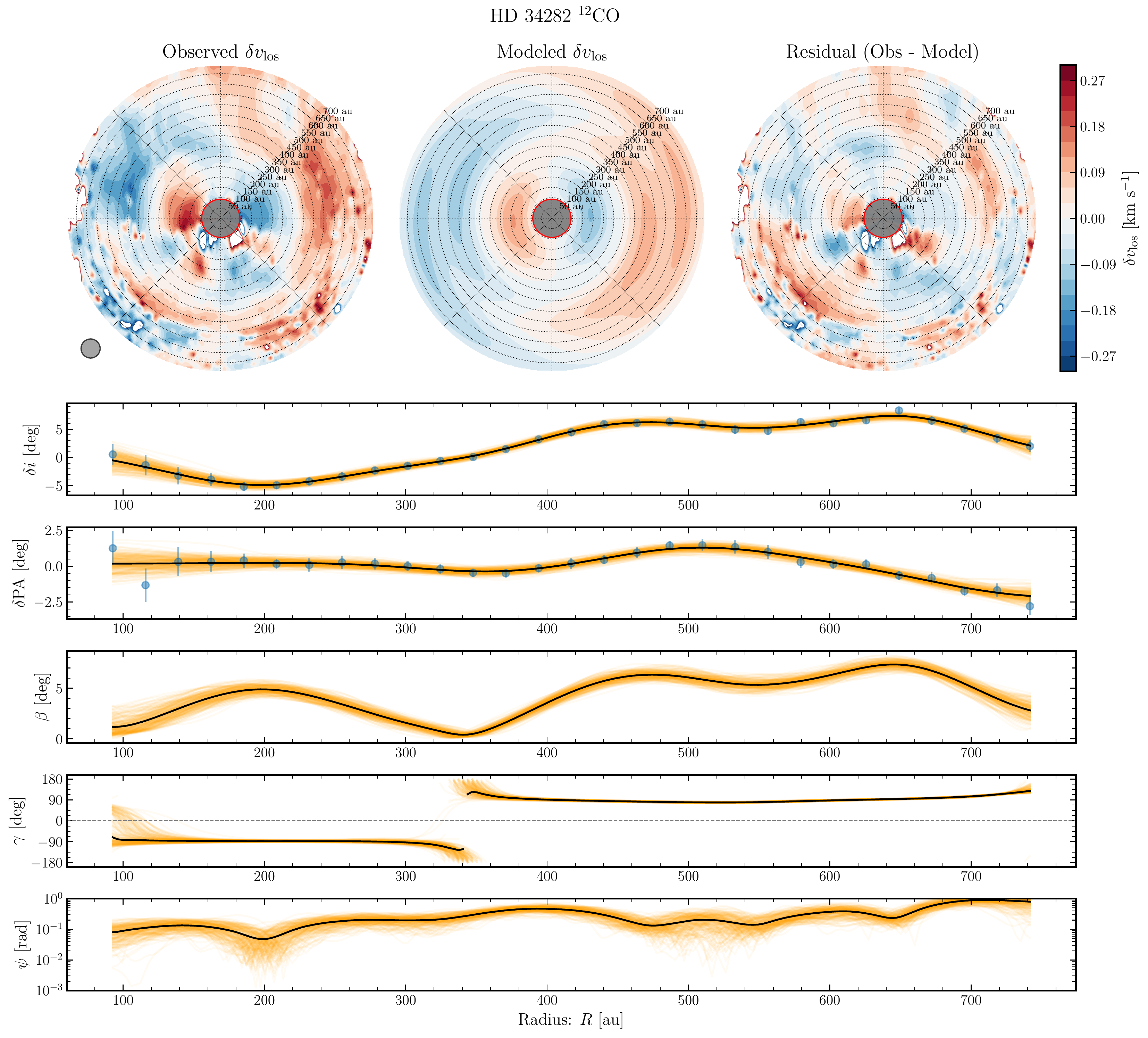}
    \caption{As in Figure~\ref{fig:inc_pa_profiles} but for HD34282.}
    \label{fig:profile_HD34282}
\end{figure}

\clearpage
\begin{figure}[ht!]
    \centering
    \includegraphics[width=\textwidth]{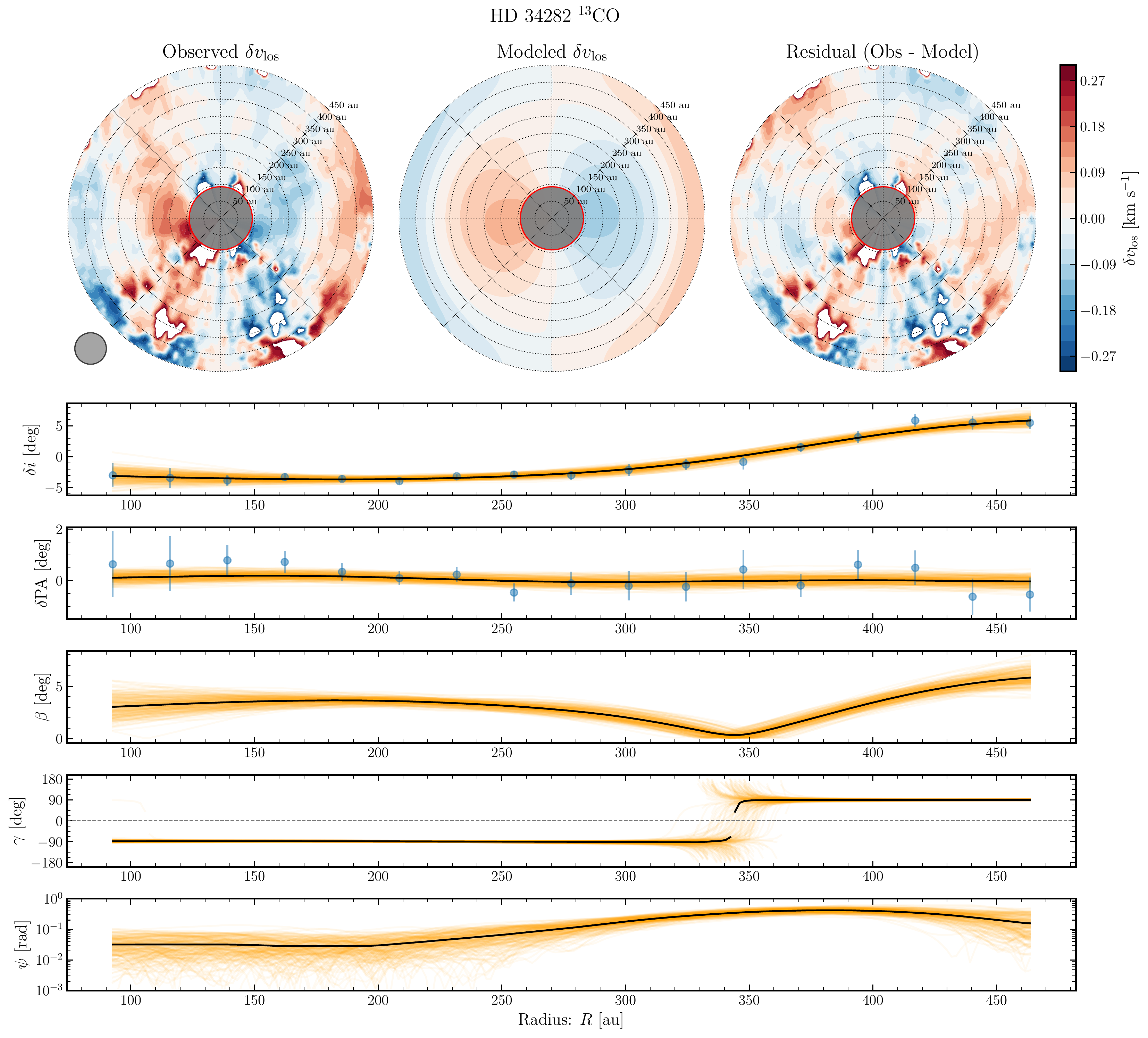}
    \caption{As in Figure~\ref{fig:profile_HD34282} but for $^{13}$CO.}
    \label{fig:profile_HD34282_13}
\end{figure}

\clearpage
\begin{figure}[ht!]
    \centering
    \includegraphics[width=\textwidth]{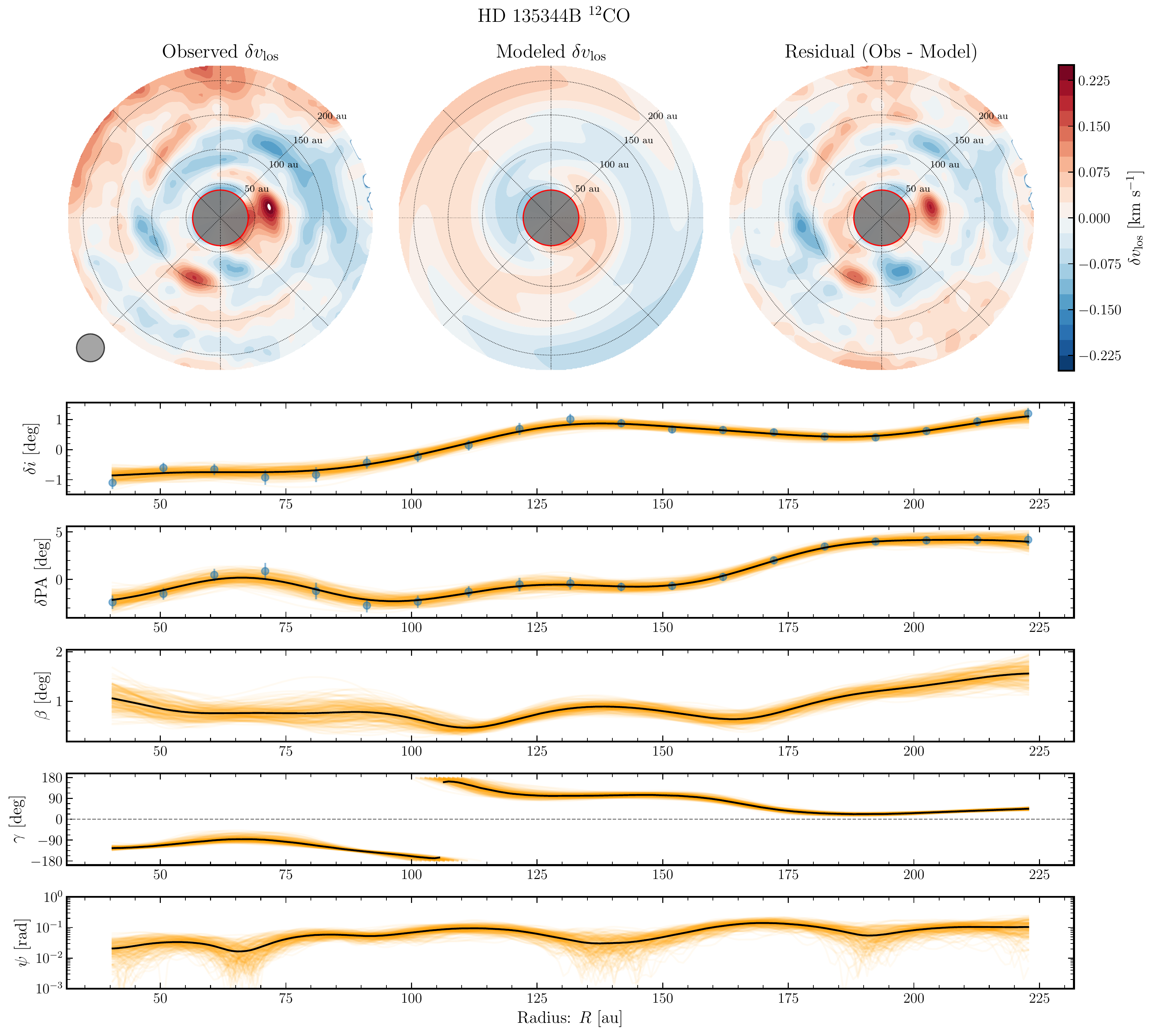}
    \caption{As in Figure~\ref{fig:inc_pa_profiles} but for HD~135344B.}
    \label{fig:profile_HD135344}
\end{figure}

\clearpage
\begin{figure}[ht!]
    \centering
    \includegraphics[width=\textwidth]{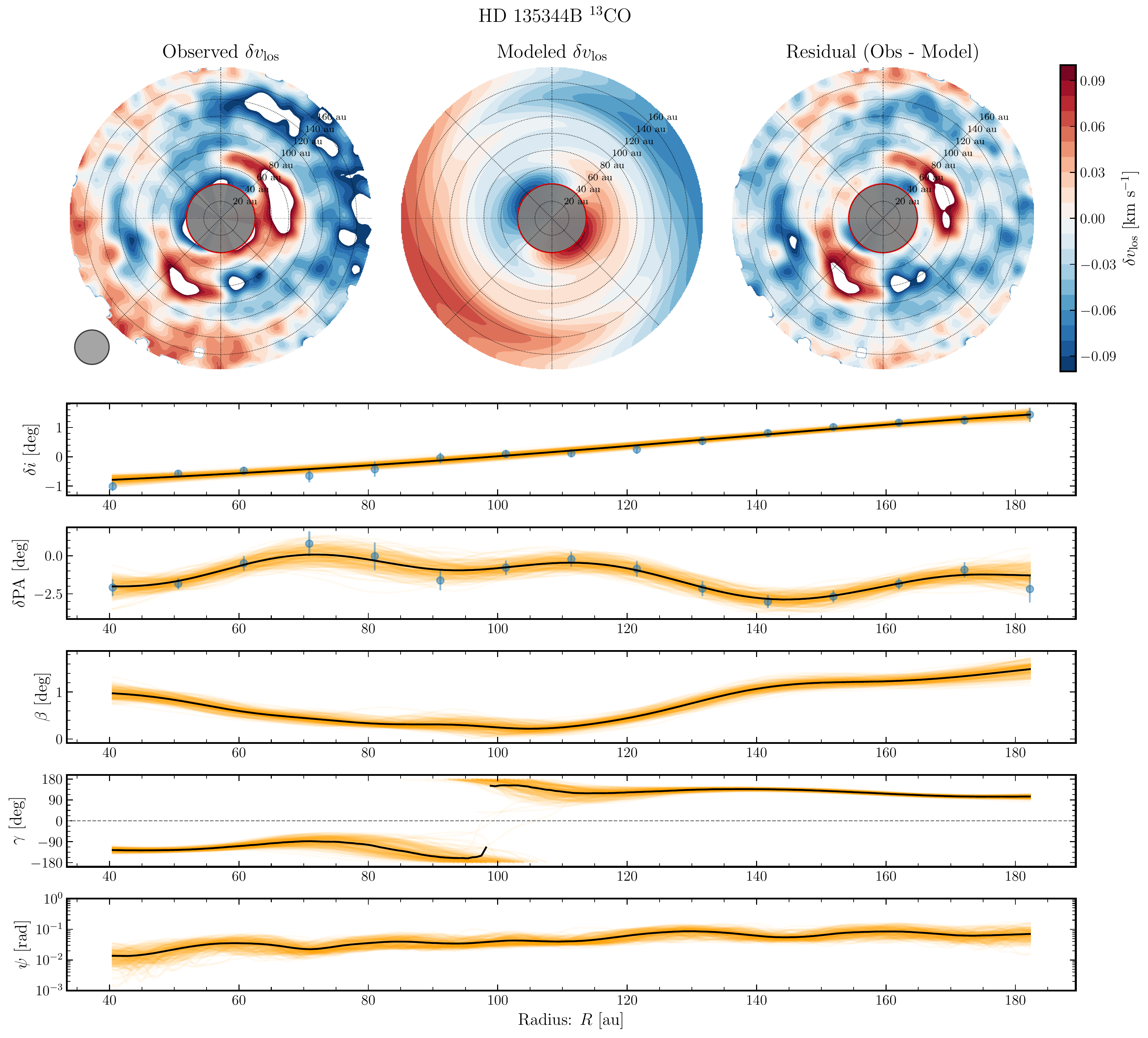}
    \caption{As in Figure~\ref{fig:profile_HD135344} but for $^{13}$CO.}
    \label{fig:profile_HD135344_13}
\end{figure}

\clearpage
\begin{figure}[ht!]
    \centering
    \includegraphics[width=\textwidth]{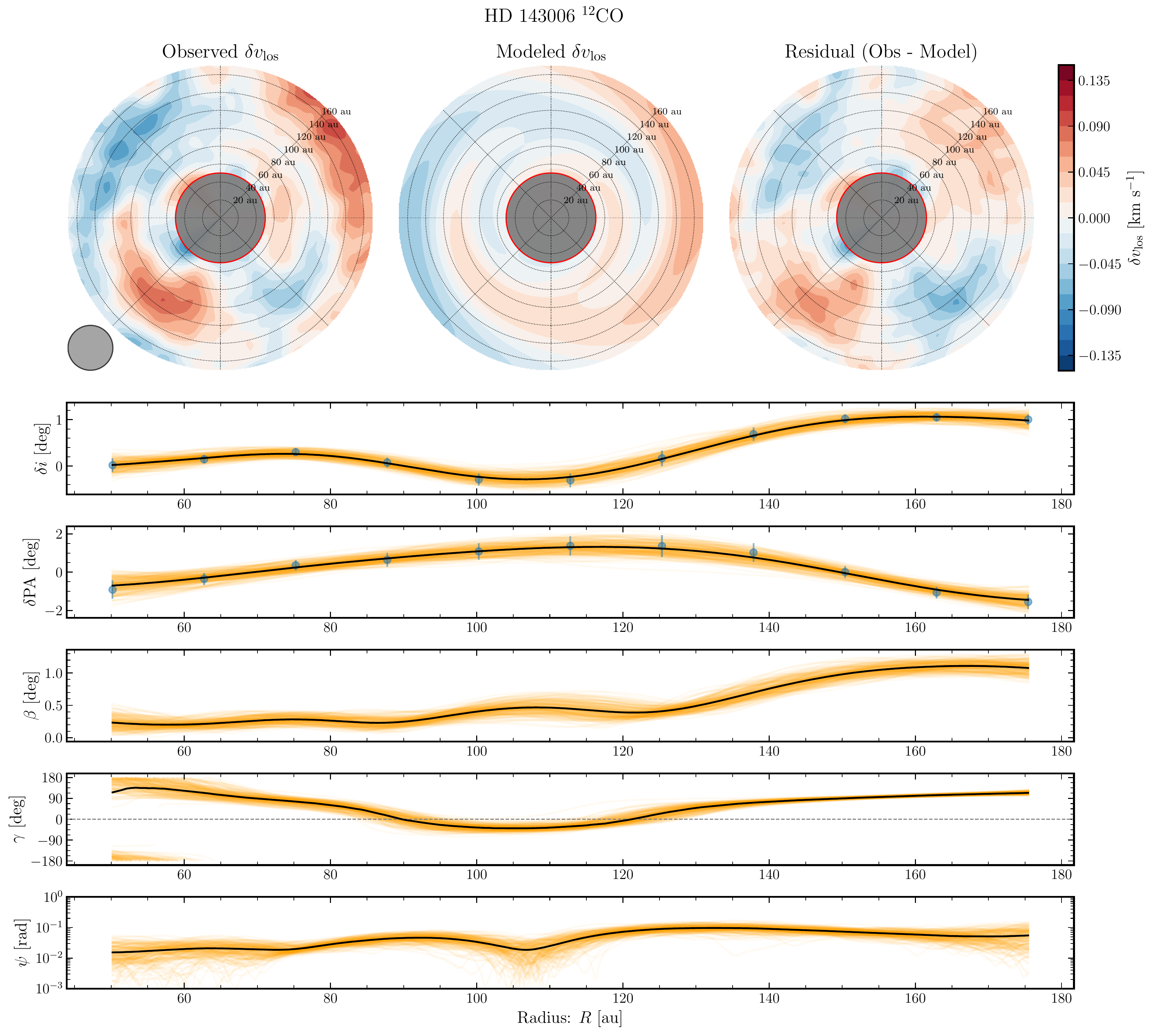}
    \caption{As in Figure~\ref{fig:inc_pa_profiles} but for HD143006.}
    \label{fig:profile_HD143006}
\end{figure}

\clearpage
\begin{figure}[ht!]
    \centering
    \includegraphics[width=\textwidth]{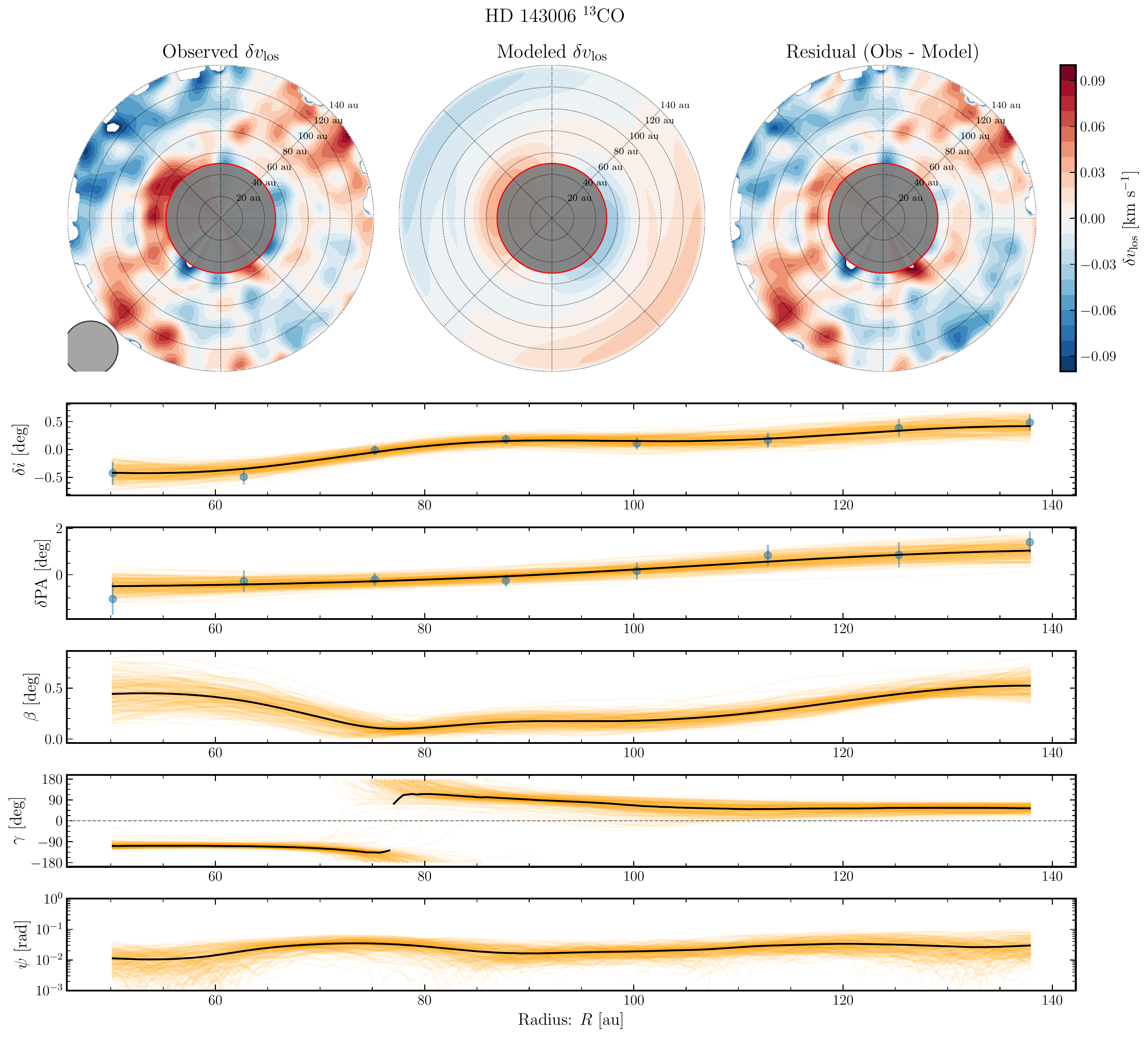}
    \caption{As in Figure~\ref{fig:profile_HD143006} but for $^{13}$CO.}
    \label{fig:profile_HD143006_13}
\end{figure}

\clearpage
\begin{figure}[ht!]
    \centering
    \includegraphics[width=\textwidth]{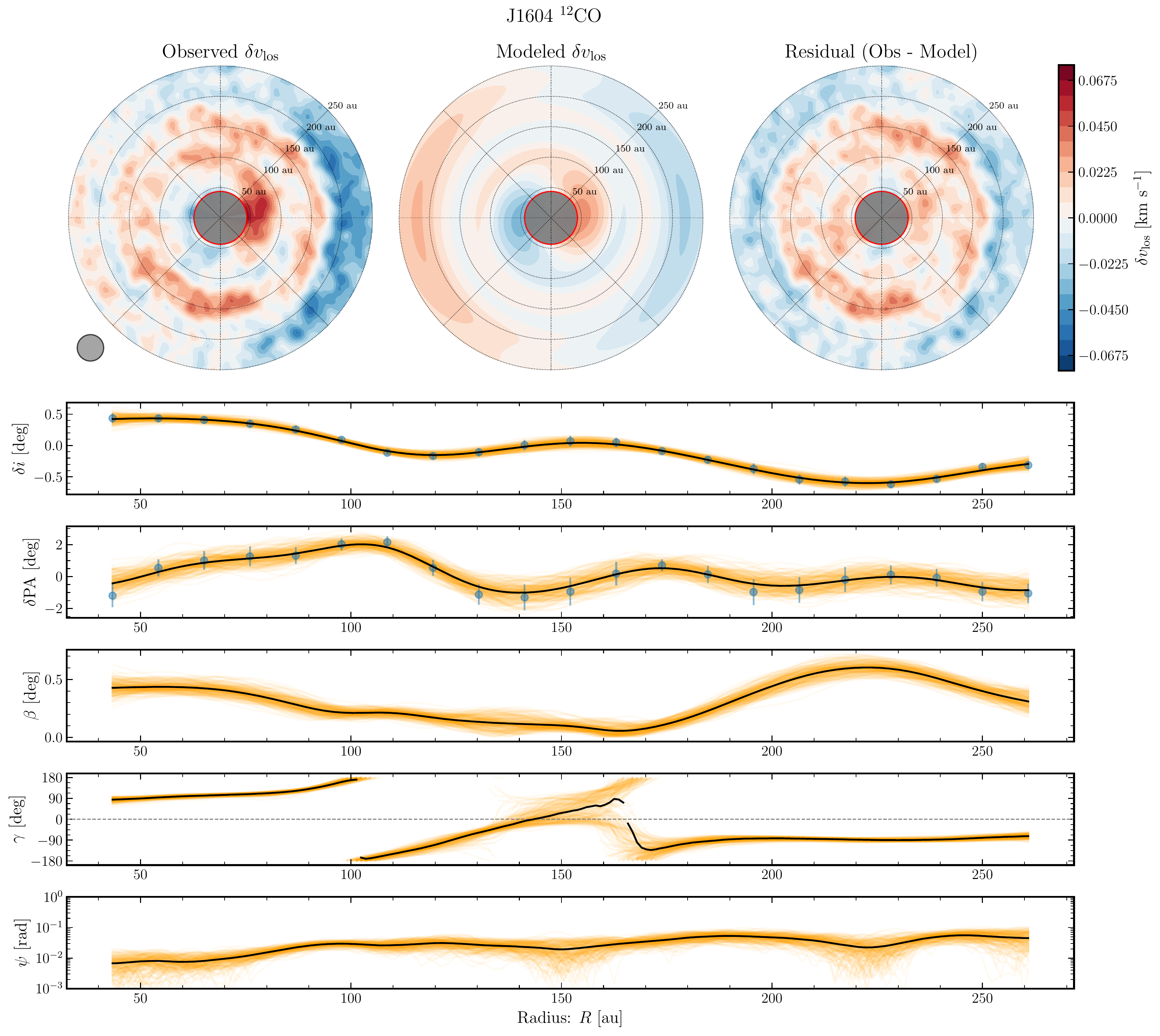}
    \caption{As in Figure~\ref{fig:inc_pa_profiles} but for J1604.}
    \label{fig:profile_J1604}
\end{figure}

\clearpage
\begin{figure}[ht!]
    \centering
    \includegraphics[width=\textwidth]{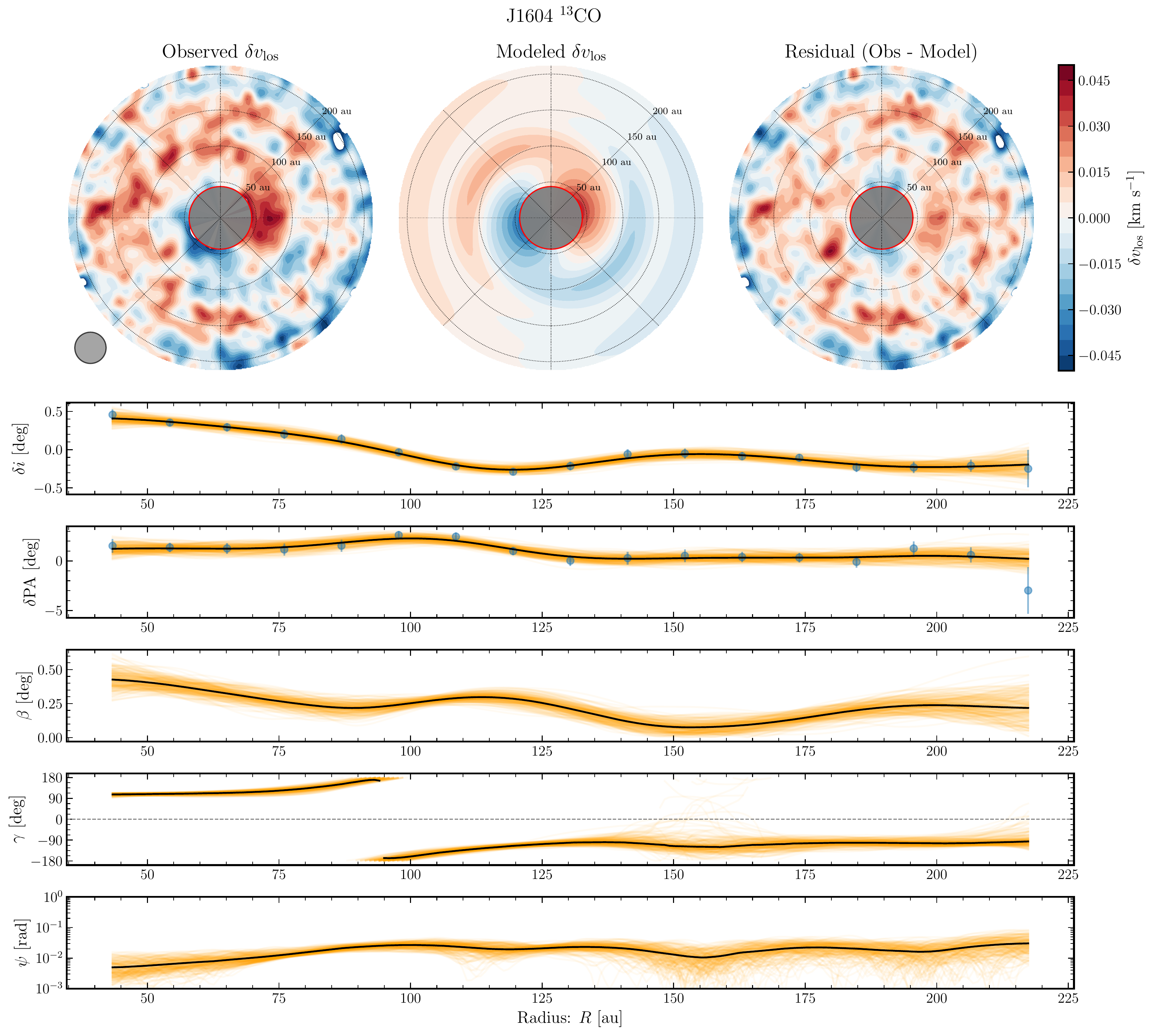}
    \caption{As in Figure~\ref{fig:profile_J1604} but for $^{13}$CO.}
    \label{fig:profile_J1604_13}
\end{figure}

\clearpage
\begin{figure}[ht!]
    \centering
    \includegraphics[width=\textwidth]{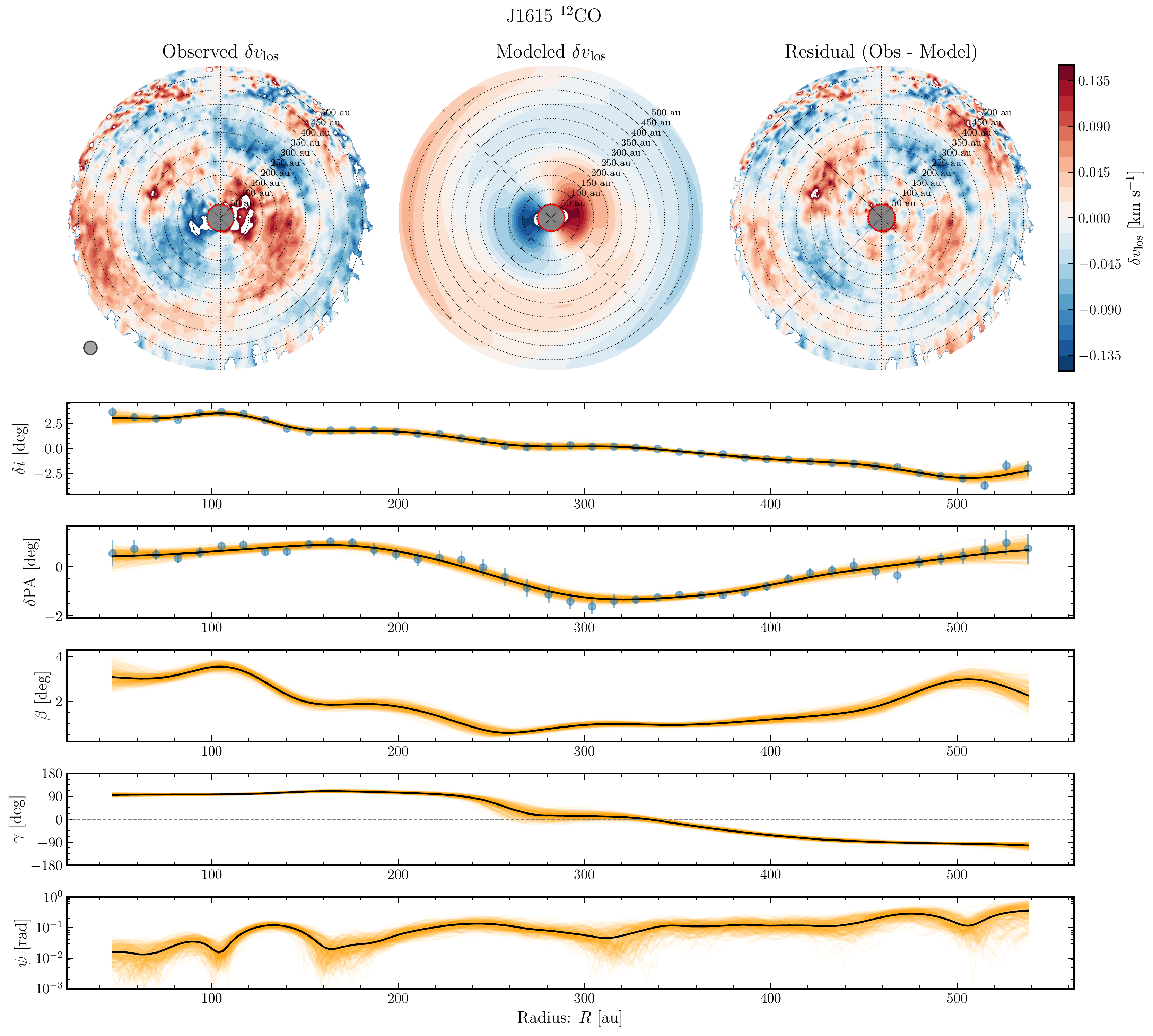}
    \caption{As in Figure~\ref{fig:inc_pa_profiles} but for J1615.}
    \label{fig:profile_J1615}
\end{figure}

\clearpage
\begin{figure}[ht!]
    \centering
    \includegraphics[width=\textwidth]{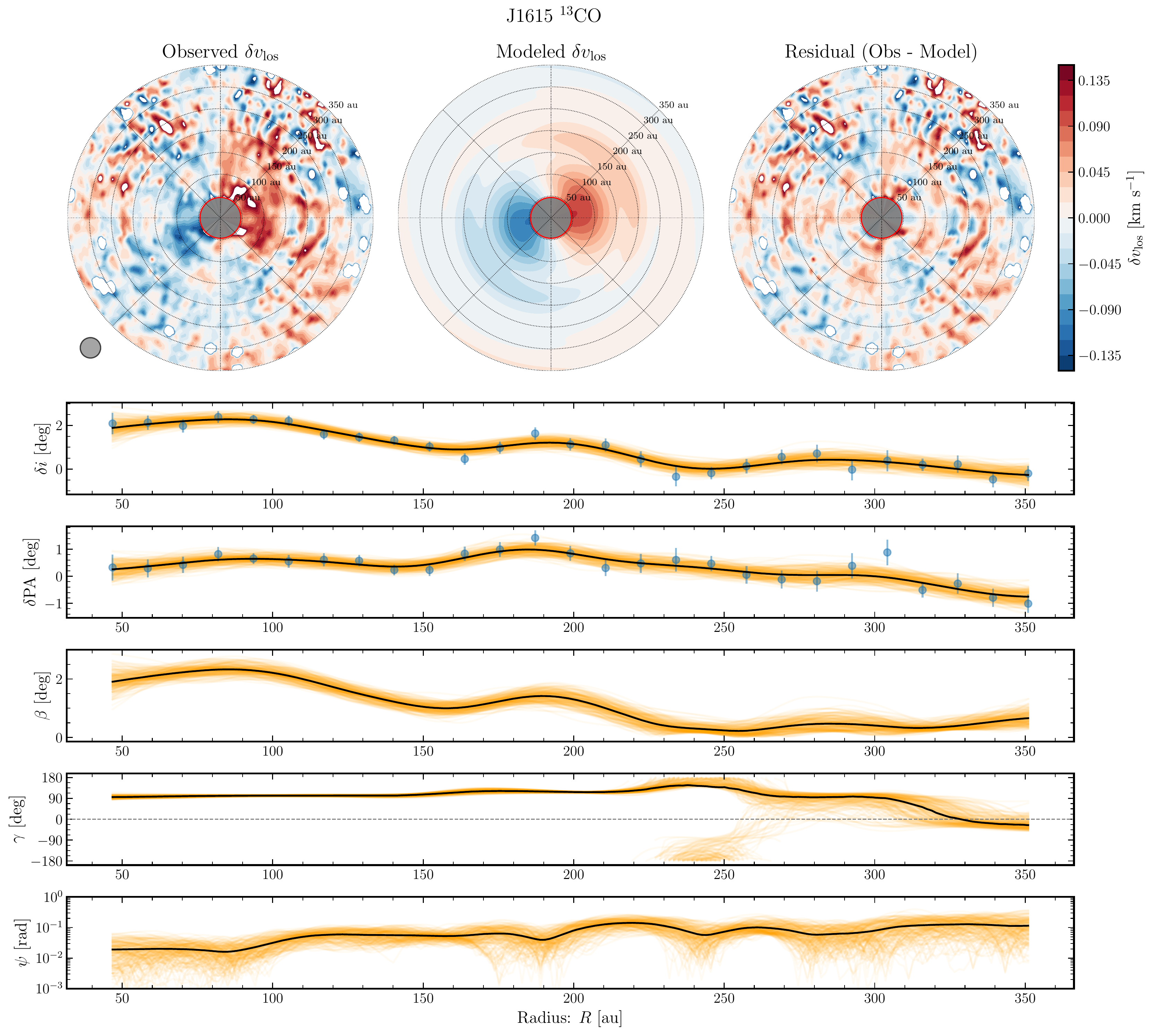}
    \caption{As in Figure~\ref{fig:profile_J1615} but for $^{13}$CO.}
    \label{fig:profile_J1615_13}
\end{figure}

\clearpage
\begin{figure}[ht!]
    \centering
    \includegraphics[width=\textwidth]{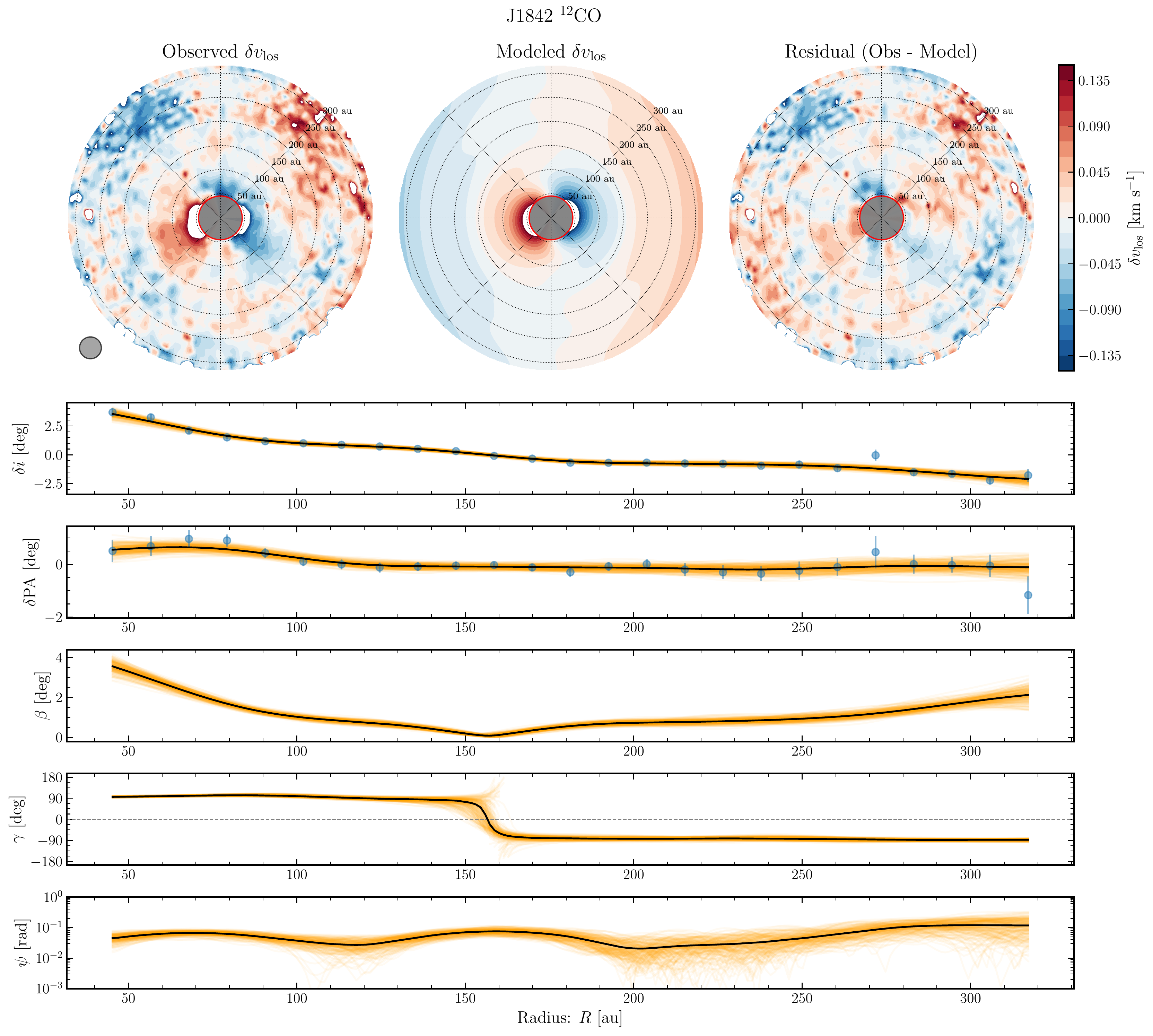}
    \caption{As in Figure~\ref{fig:inc_pa_profiles} but for J1842.}
    \label{fig:profile_J1842}
\end{figure}

\clearpage
\begin{figure}[ht!]
    \centering
    \includegraphics[width=\textwidth]{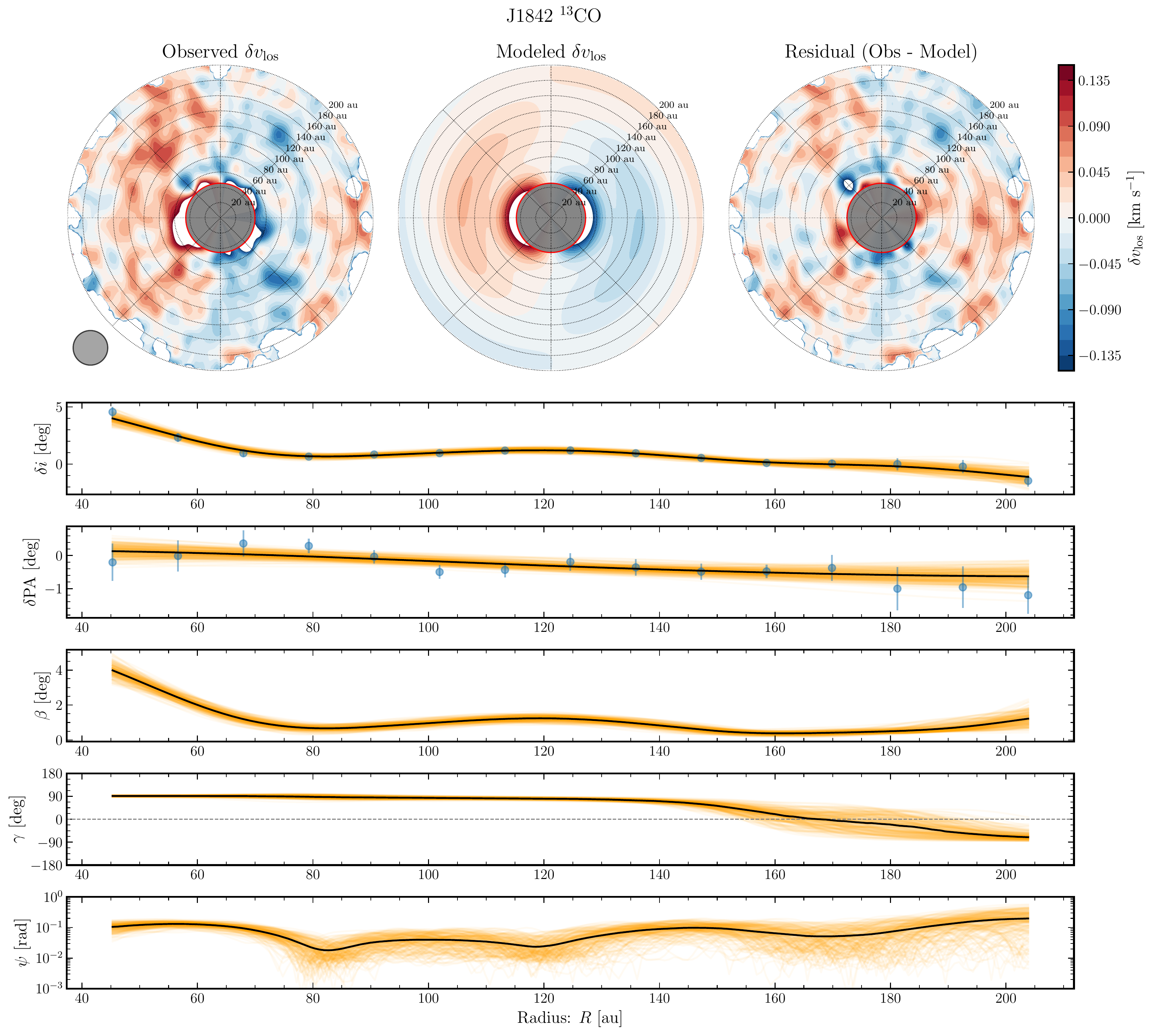}
    \caption{As in Figure~\ref{fig:profile_J1842} but for $^{13}$CO.}
    \label{fig:profile_J1842_13}
\end{figure}

\clearpage
\begin{figure}[ht!]
    \centering
    \includegraphics[width=\textwidth]{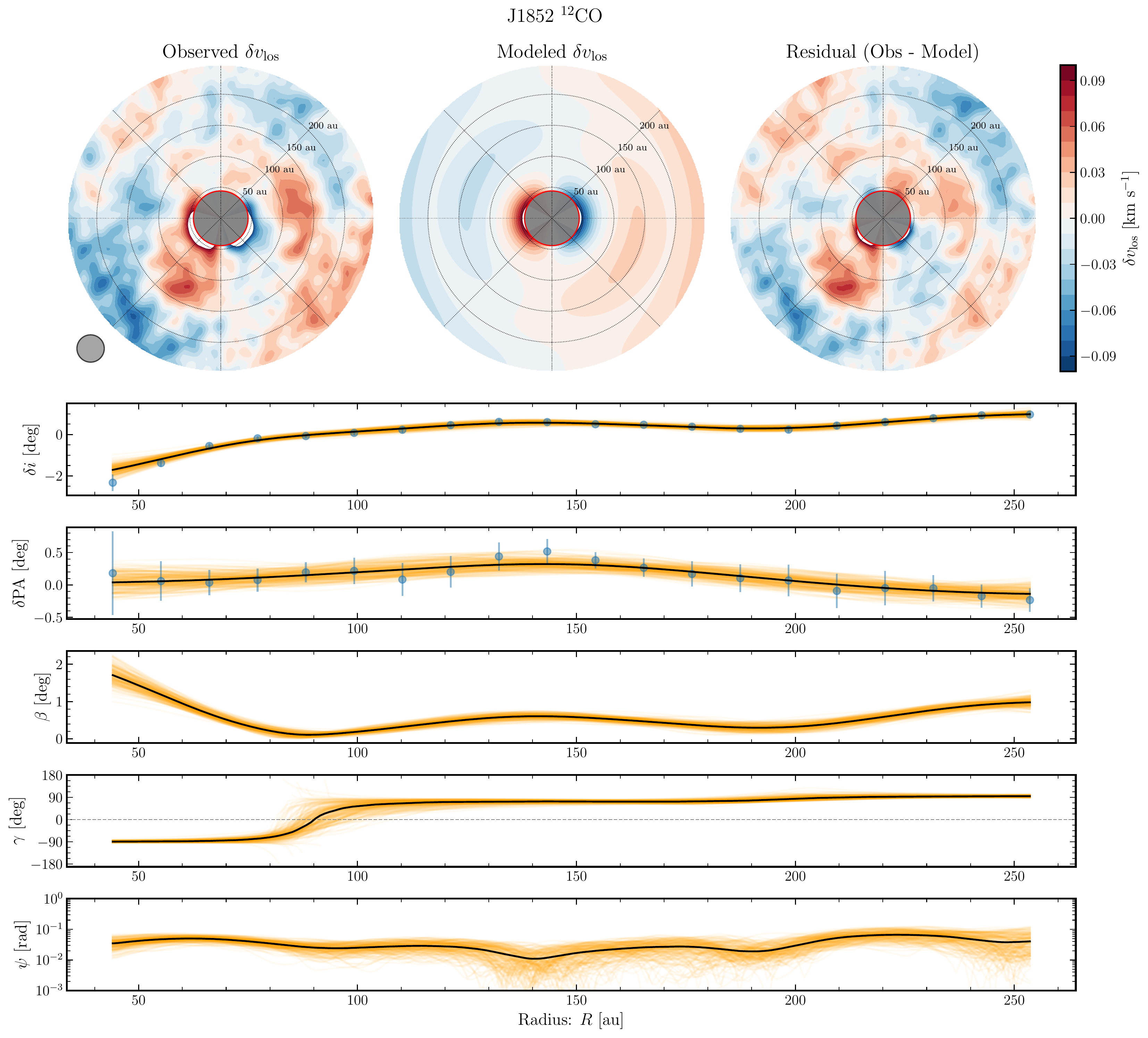}
    \caption{As in Figure~\ref{fig:inc_pa_profiles} but for J1852.}
    \label{fig:profile_J1852}
\end{figure}

\clearpage
\begin{figure}[ht!]
    \centering
    \includegraphics[width=\textwidth]{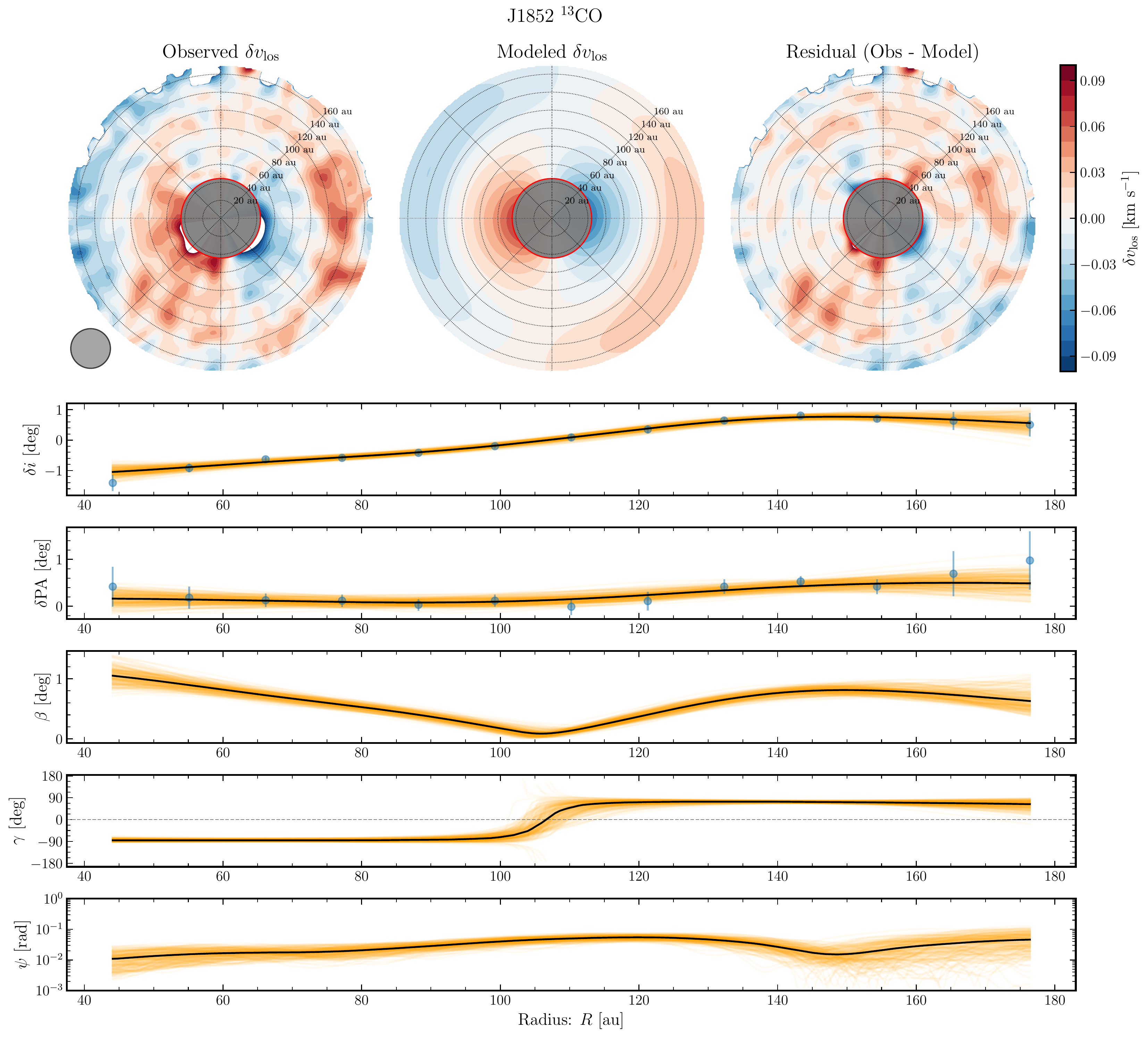}
    \caption{As in Figure~\ref{fig:profile_J1852} but for $^{13}$CO.}
    \label{fig:profile_J1852_13}
\end{figure}

\clearpage
\begin{figure}[ht!]
    \centering
    \includegraphics[width=\textwidth]{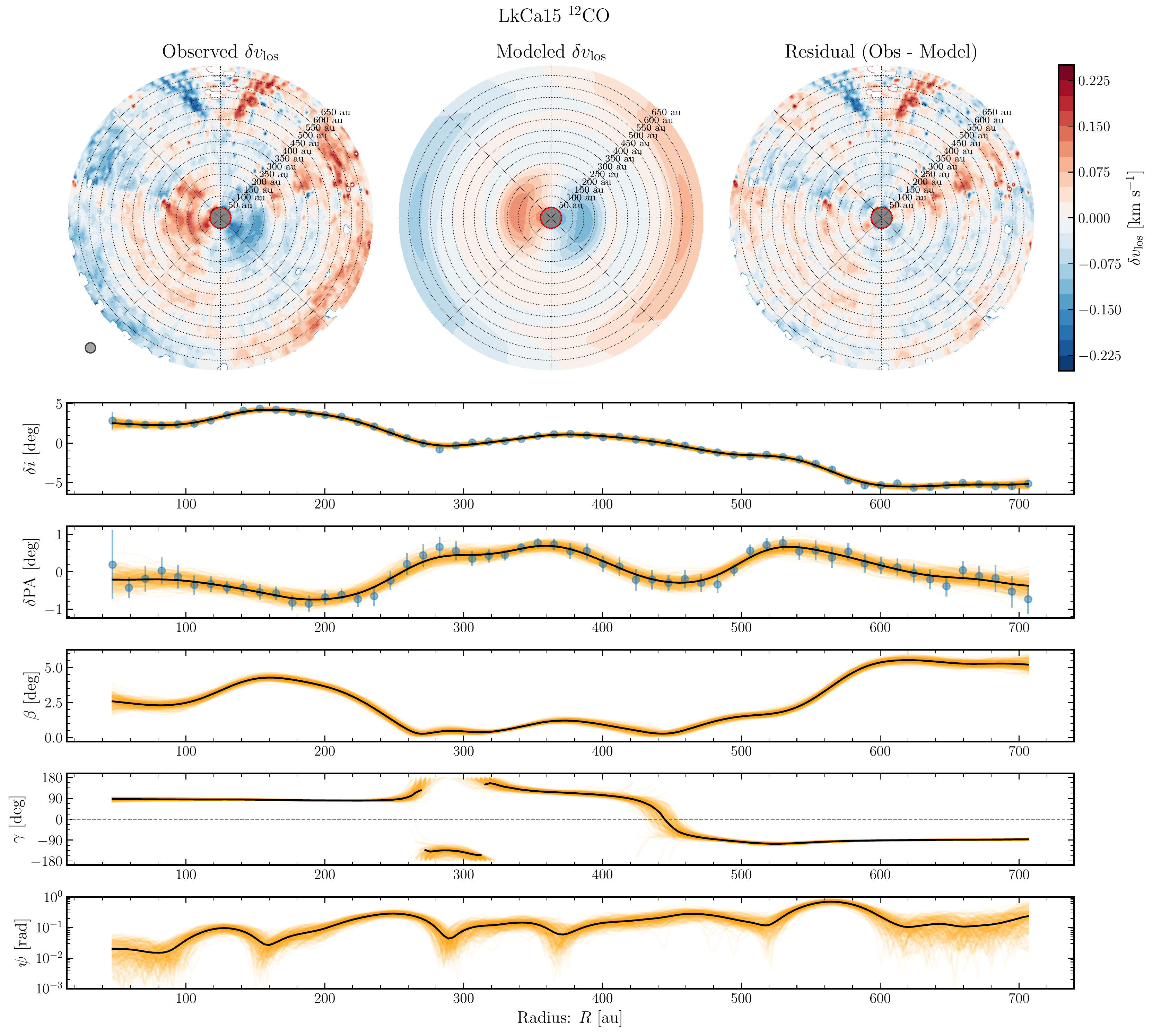}
    \caption{As in Figure~\ref{fig:inc_pa_profiles} but for LkCa15.}
    \label{fig:profile_LkCa15}
\end{figure}

\clearpage
\begin{figure}[ht!]
    \centering
    \includegraphics[width=\textwidth]{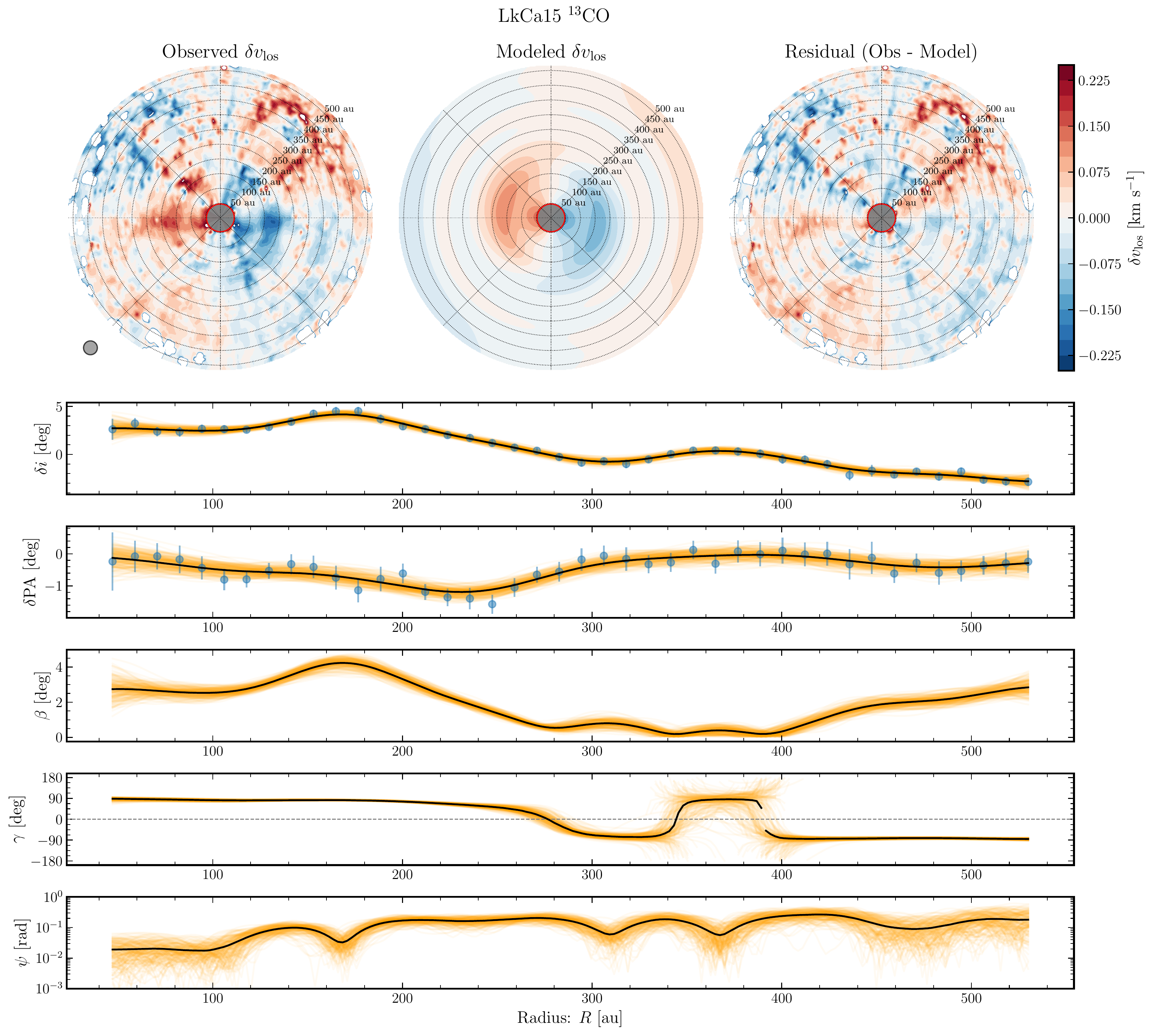}
    \caption{As in Figure~\ref{fig:profile_LkCa15} but for $^{13}$CO.}
    \label{fig:profile_LkCa15_13}
\end{figure}

\clearpage
\begin{figure}[ht!]
    \centering
    \includegraphics[width=\textwidth]{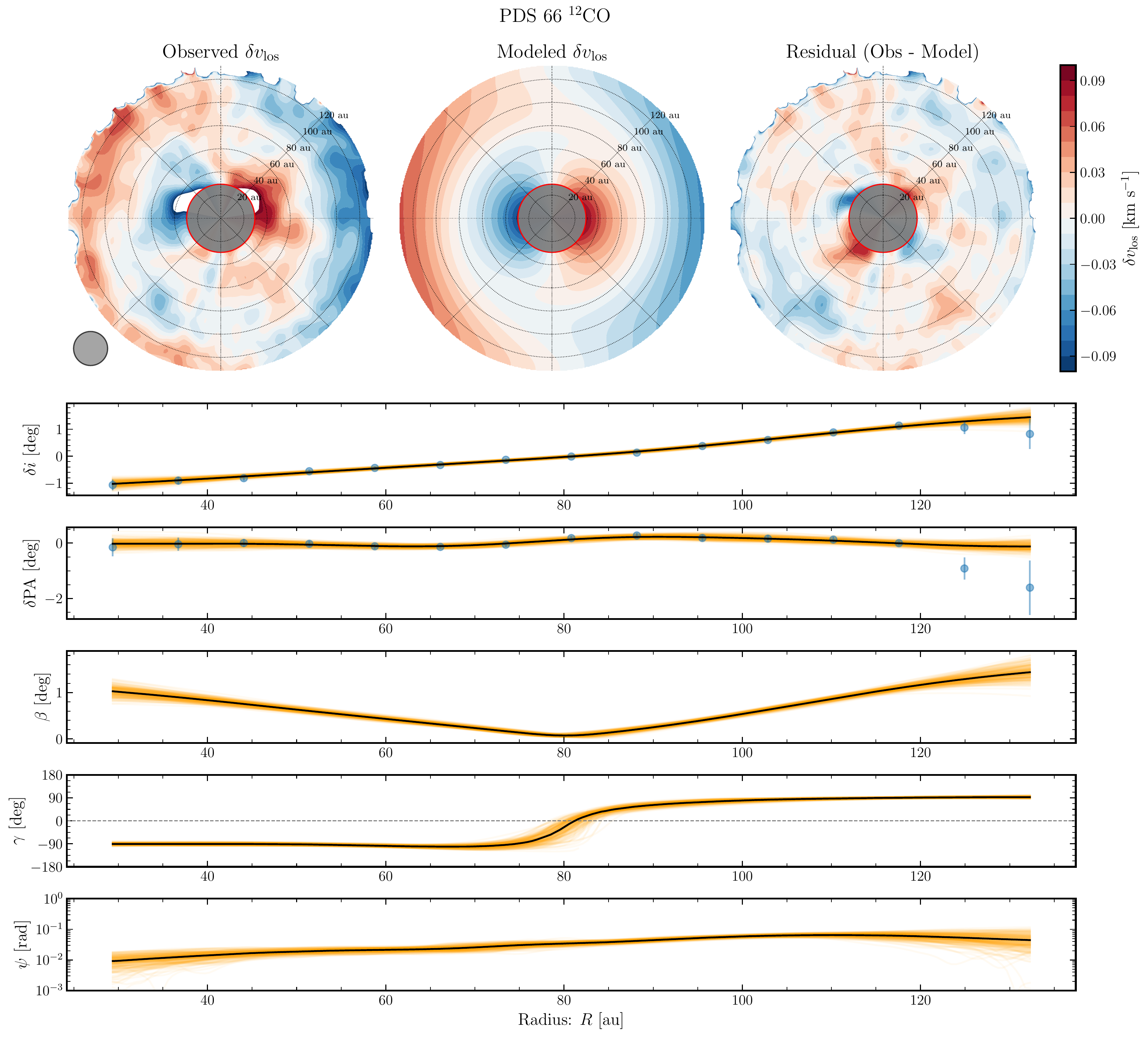}
    \caption{As in Figure~\ref{fig:inc_pa_profiles} but for PDS66.}
    \label{fig:profile_PDS66}
\end{figure}

\clearpage
\begin{figure}[ht!]
    \centering
    \includegraphics[width=\textwidth]{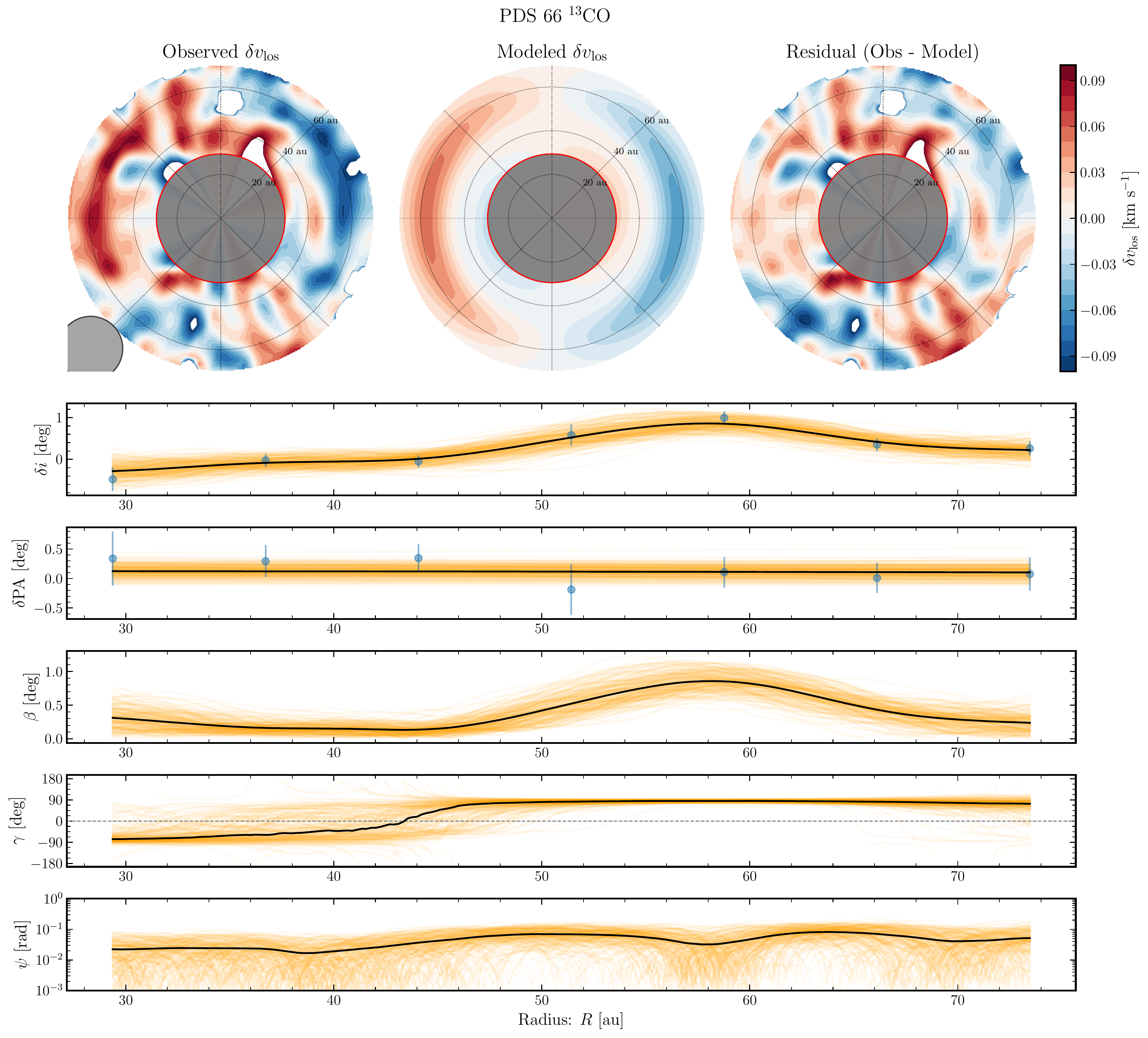}
    \caption{As in Figure~\ref{fig:profile_PDS66} but for $^{13}$CO.}
    \label{fig:profile_PDS66_13}
\end{figure}

\clearpage
\begin{figure}[ht!]
    \centering
    \includegraphics[width=\textwidth]{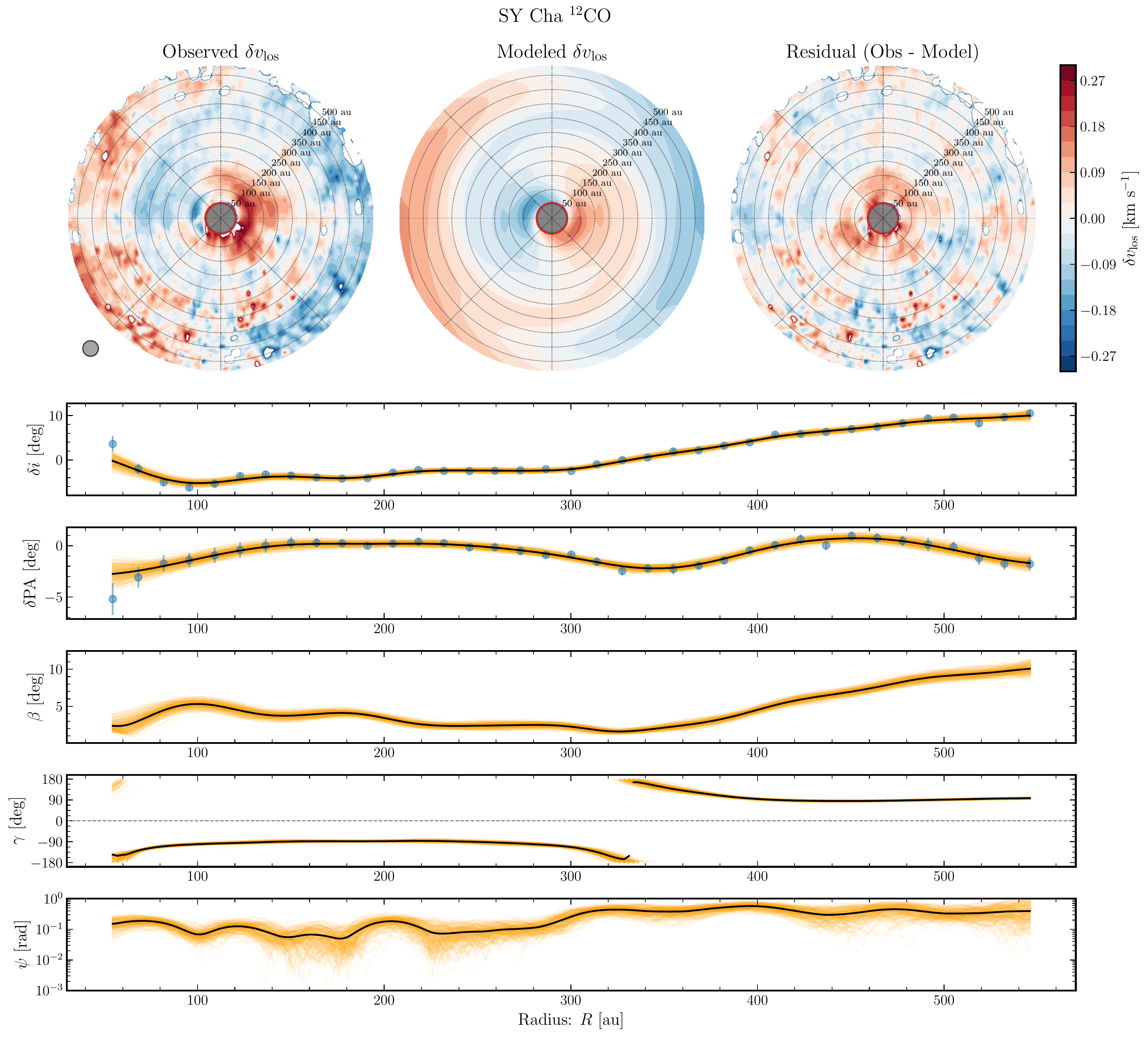}
    \caption{As in Figure~\ref{fig:inc_pa_profiles} but for SY Cha.}
    \label{fig:profile_SYCha}
\end{figure}

\clearpage
\begin{figure}[ht!]
    \centering
    \includegraphics[width=\textwidth]{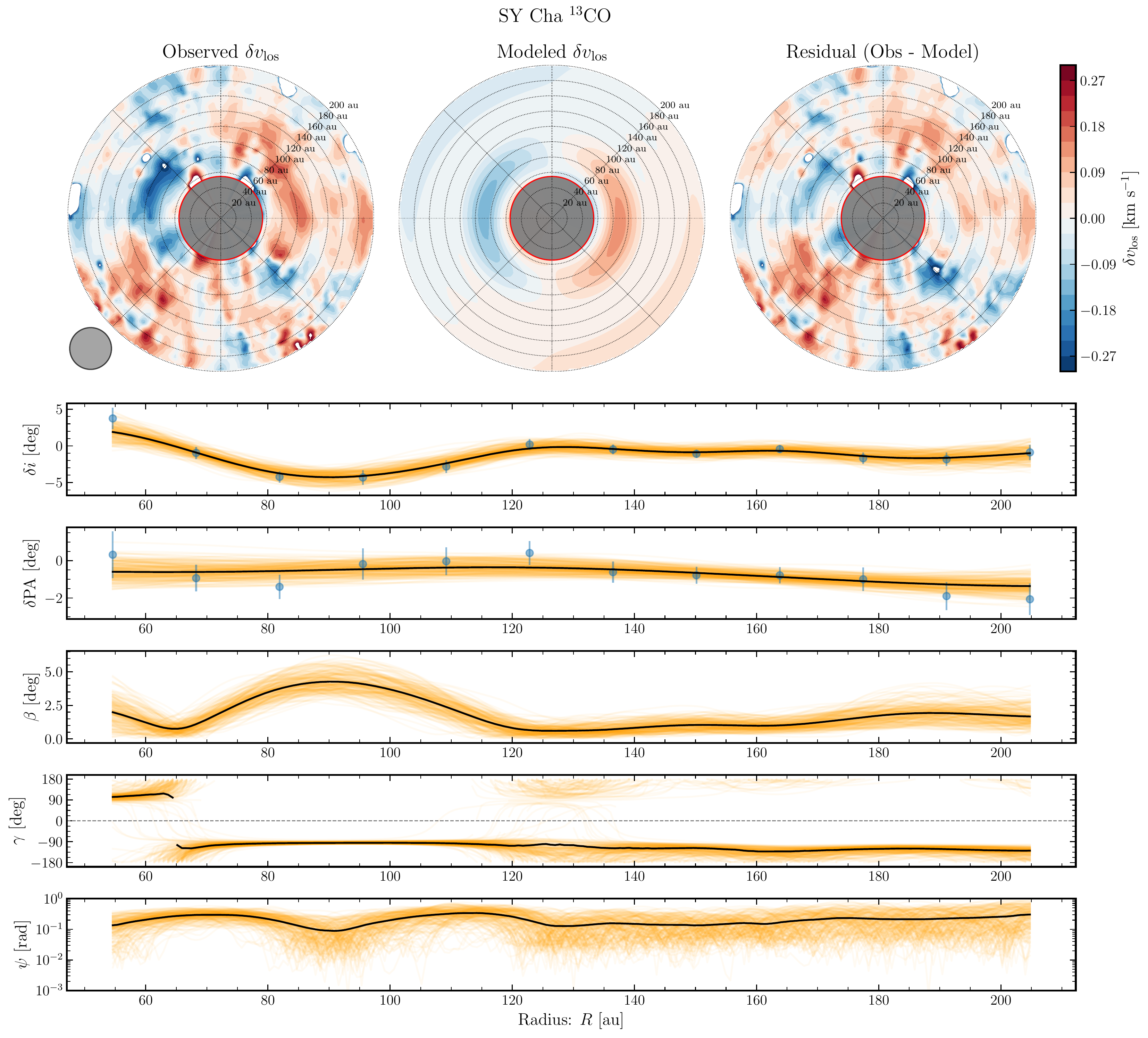}
    \caption{As in Figure~\ref{fig:profile_SYCha} but for $^{13}$CO.}
    \label{fig:profile_SYCha_13}
\end{figure}

\clearpage
\begin{figure}[ht!]
    \centering
    \includegraphics[width=\textwidth]{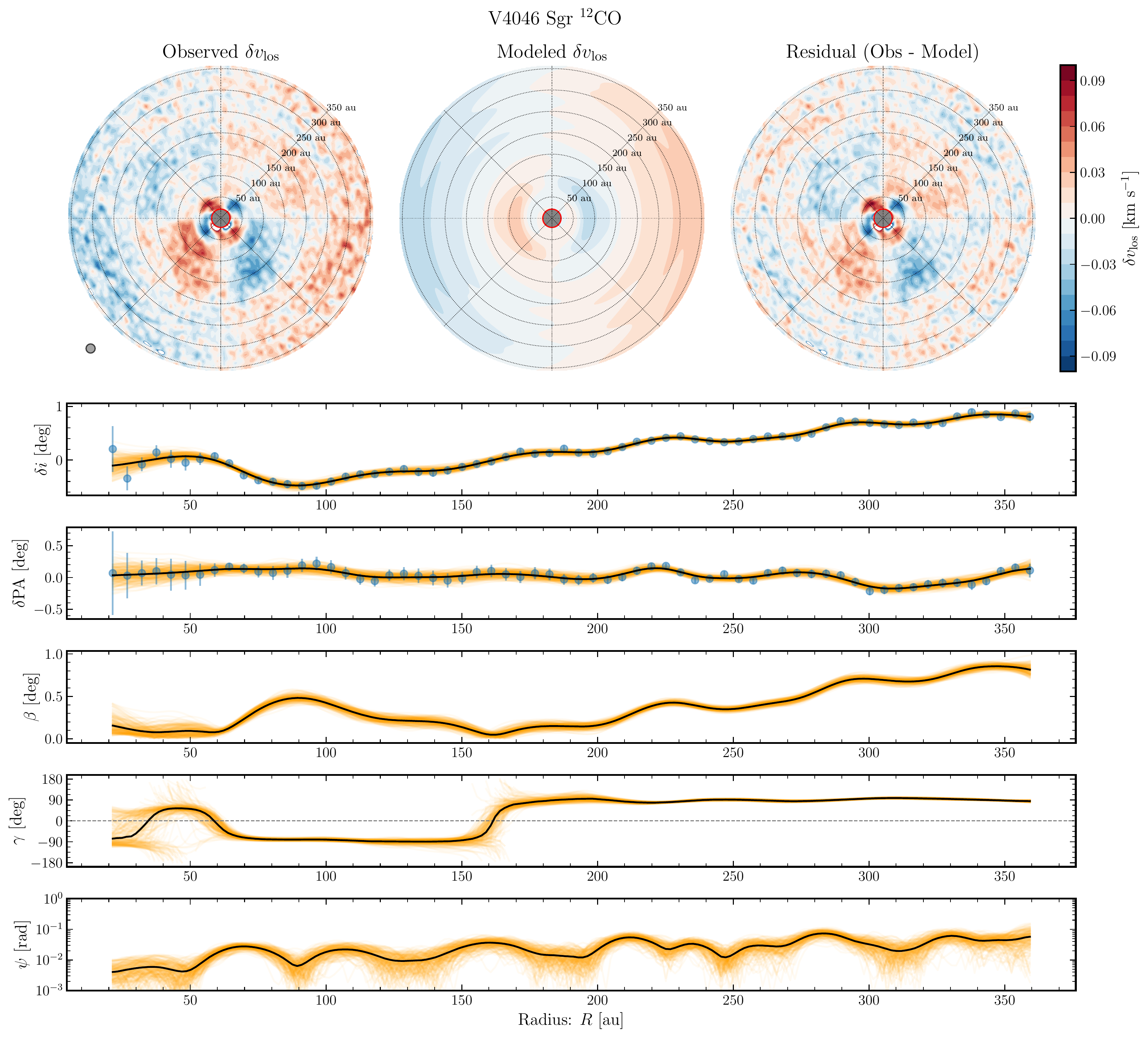}
    \caption{As in Figure~\ref{fig:inc_pa_profiles} but for V4046.}
    \label{fig:profile_V4046}
\end{figure}

\clearpage
\begin{figure}[ht!]
    \centering
    \includegraphics[width=\textwidth]{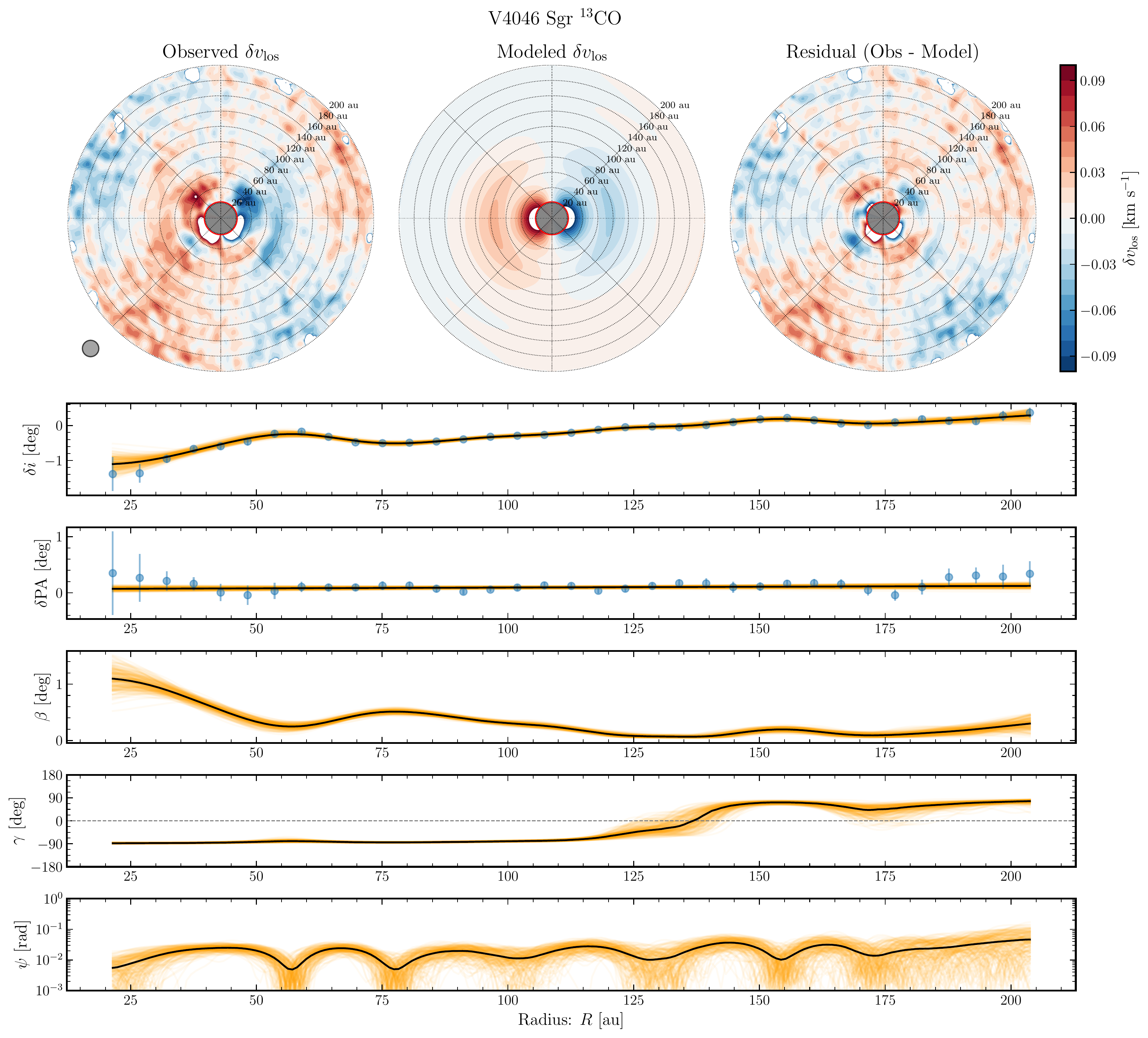}
    \caption{As in Figure~\ref{fig:profile_V4046} but for $^{13}$CO.}
    \label{fig:profile_V4046_13}
\end{figure}

\end{document}

%% file: authors.tex
\newcommand{\InstMPIA}{Max-Planck Institute for Astronomy (MPIA), Königstuhl 17, 69117 Heidelberg, Germany}
\newcommand{\InstMIT}{Department of Earth, Atmospheric, and Planetary Sciences, Massachusetts Institute of Technology, Cambridge, MA 02139, USA}
\newcommand{\InstOCA}{Université Côte d’Azur, Observatoire de la Côte d’Azur, CNRS, Laboratoire Lagrange, France}
\newcommand{\InstMilano}{Dipartimento di Fisica, Università degli Studi di Milano, Via Celoria 16, 20133 Milano, Italy}
\newcommand{\InstNAOJ}{National Astronomical Observatory of Japan, Osawa 2-21-1, Mitaka, Tokyo 181-8588, Japan}
\newcommand{\InstIPAGGrenoble}{Univ. Grenoble Alpes, CNRS, IPAG, 38000 Grenoble, France}
\newcommand{\InstMonash}{School of Physics and Astronomy, Monash University, Clayton VIC 3800, Australia}
\newcommand{\InstCfA}{Center for Astrophysics | Harvard \& Smithsonian, Cambridge, MA 02138, USA}
\newcommand{\InstFlorida}{Department of Astronomy, University of Florida, Gainesville, FL 32611, USA}
\newcommand{\InstChile}{Departamento de Astronomía, Universidad de Chile, Camino El Observatorio 1515, Las Condes, Santiago, Chile}
\newcommand{\InstStAndrewsPhysics}{School of Physics \& Astronomy, University of St. Andrews, North Haugh, St. Andrews KY16 9SS, UK}
\newcommand{\InstStAndrewsExoplanets}{Centre for Exoplanet Science, University of St. Andrews, North Haugh, St. Andrews, KY16 9SS, UK}
\newcommand{\InstRicePhysics}{Department of Physics and Astronomy, Rice University, Houston, TX 77005, USA}
\newcommand{\InstLANL}{Los Alamos National Laboratory, Los Alamos, NM 87545, USA}
\newcommand{\InstUGAphysics}{Department of Physics and Astronomy, The University of Georgia, Athens, GA 30602, USA}
\newcommand{\InstUGACSP}{Center for Simulational Physics, The University of Georgia, Athens, GA 30602, USA}
\newcommand{\InstUGAIA}{Institute for Artificial Intelligence, The University of Georgia, Athens, GA, 30602, USA}
\newcommand{\InstColumbia}{Department of Astronomy, Columbia University, 538 W. 120th Street, Pupin Hall, New York, NY, USA}
\newcommand{\InstLeeds}{School of Physics and Astronomy, University of Leeds, Leeds, UK, LS2 9JT}
\newcommand{\InstRiceSpace}{Rice Space Institute, Rice University, 6100 Main St, Houston, TX 77005, USA}
\newcommand{\InstLeiden}{Leiden Observatory, Leiden University, P.O. Box 9513, NL-2300 RA Leiden, The Netherlands}
\newcommand{\InstESO}{European Southern Observatory, Karl-Schwarzschild-Str. 2, D-85748 Garching bei München, Germany}
\newcommand{\InstNHFP}{NASA Hubble Fellowship Program Sagan Fellow}
\newcommand{\InstIbaraki}{College of Science, Ibaraki University, 2-1-1 Bunkyo, Mito, Ibaraki 310-8512, Japan}
\newcommand{\InstCambridge}{Institute of Astronomy, University of Cambridge, Madingley Road, CB3 0HA, Cambridge, UK}
\newcommand{\InstNRAO}{National Radio Astronomy Observatory, 520 Edgemont Rd., Charlottesville, VA 22903, USA}
\newcommand{\InstUNAM}{Instituto de Ciencias Físicas, Universidad Nacional Autónoma de México, Av. Universidad s/n, 62210 Cuernavaca, Mor., Mexico}
\newcommand{\InstBologna}{Alma Mater Studiorum Università di Bologna, Dipartimento di Fisica e Astronomia (DIFA), Via Gobetti 93/2, 40129 Bologna, Italy}
\newcommand{\InstArcetri}{INAF-Osservatorio Astrofisico di Arcetri, Largo E. Fermi 5, 50125 Firenze, Italy}
\newcommand{\InstASIAA}{Academia Sinica Institute of Astronomy \& Astrophysics, 11F of Astronomy-Mathematics Building, AS/NTU, No.1, Sec. 4, Roosevelt Rd, Taipei 10617, Taiwan}
\newcommand{\InstWesleyan}{Department of Astronomy, Van Vleck Observatory, Wesleyan University, 96 Foss Hill Drive, Middletown, CT 06459, USA}
\newcommand{\InstPennState}{Department of Astronomy \& Astrophysics, 525 Davey Laboratory, The Pennsylvania State University, University Park, PA 16802, USA}
\newcommand{\InstSOKENDAI}{Department of Astronomical Science, The Graduate University for Advanced Studies, SOKENDAI, 2-21-1 Osawa, Mitaka, Tokyo 181-8588, Japan}
\newcommand{\InstQMUL}{Astronomy Unit, School of Physics and Astronomy, Queen Mary University of London, London E1 4NS, UK}

\author[0000-0002-7501-9801]{Andrew J. Winter}
\affiliation{\InstQMUL}

\author[0000-0002-7695-7605]{Myriam Benisty}
\affiliation{\InstMPIA}

\author[0000-0001-8446-3026]{Andrés F. Izquierdo} 
\affiliation{\InstFlorida}
\affiliation{\InstLeiden}
\affiliation{\InstESO}
\affiliation{\InstNHFP}

\author[0000-0002-2357-7692]{Giuseppe Lodato} 
\affiliation{\InstMilano}

\author[0000-0003-1534-5186]{Richard Teague}
\affiliation{\InstMIT}

\author[0000-0001-9071-1508]{Carolin N. Kimmig} 
\affiliation{\InstMilano}


\author[0009-0000-7872-3493]{Sean M. Andrews}
\affiliation{\InstCfA}

\author[0000-0001-7258-770X]{Jaehan Bae}
\affiliation{\InstFlorida}

\author[0000-0001-6378-7873]{Marcelo Barraza-Alfaro}
\affiliation{\InstMIT}


\author[0000-0003-3713-8073]{Nicolás Cuello} 
\affiliation{\InstIPAGGrenoble}Ba

\author[0000-0003-2045-2154]{Pietro Curone} 
\affiliation{\InstChile}

\author[0000-0002-1483-8811]{Ian Czekala}
\affiliation{\InstStAndrewsPhysics}

\author[0000-0003-4689-2684]{Stefano Facchini}
\affiliation{\InstMilano}

\author[0000-0003-4679-4072]{Daniele Fasano} 
\affiliation{\InstOCA}






\author[0000-0002-8138-0425]{Cassandra Hall} 
\affiliation{\InstUGAphysics}
\affiliation{\InstUGACSP}
\affiliation{\InstUGAIA}

\author[0009-0003-7403-9207]{Caitlyn Hardiman}
\affiliation{\InstMonash}

\author[0000-0001-7641-5235]{Thomas Hilder} 
\affiliation{\InstMonash}


\author[0000-0003-1008-1142]{John D. Ilee} 
\affiliation{\InstLeeds}


\author[0000-0003-1117-9213]{Misato Fukagawa} 
\affiliation{\InstNAOJ}



\author[0000-0003-4663-0318]{Cristiano Longarini} 
\affiliation{\InstCambridge}
\affiliation{\InstMilano}



\author[0000-0002-1637-7393]{François M\'{e}nard} 
\affiliation{\InstIPAGGrenoble}

\author[0000-0003-4039-8933]{Ryuta Orihara} 
\affiliation{\InstIbaraki}

\author[0000-0001-5907-5179]{Christophe Pinte}
\affiliation{\InstIPAGGrenoble}

\author[0000-0002-4716-4235]{Daniel J. Price} 
\affiliation{\InstMonash}

\author[0000-0003-4853-5736]{Giovanni Rosotti} 
\affiliation{\InstMilano}

\author[0000-0002-0491-143X]{Jochen Stadler} 
\affiliation{\InstOCA}



\author[0000-0003-1526-7587]{David J. Wilner} 
\affiliation{\InstCfA}

\author[0000-0002-7212-2416]{Lisa Wölfer} 
\affiliation{\InstMIT}

\author[0000-0003-1412-893X]{Hsi-Wei Yen} 
\affiliation{\InstASIAA}

\author[0000-0001-8002-8473]{Tomohiro C. Yoshida} 
\affiliation{\InstNAOJ}
\affiliation{\InstSOKENDAI}

\author[0000-0001-9319-1296]{Brianna Zawadzki} 
\affiliation{\InstWesleyan}

%% file: main.bbl
\begin{thebibliography}{}
\expandafter\ifx\csname natexlab\endcsname\relax\def\natexlab#1{#1}\fi
\providecommand{\url}[1]{\href{#1}{#1}}
\providecommand{\dodoi}[1]{doi:~\href{http://doi.org/#1}{\nolinkurl{#1}}}
\providecommand{\doeprint}[1]{\href{http://ascl.net/#1}{\nolinkurl{http://ascl.net/#1}}}
\providecommand{\doarXiv}[1]{\href{https://arxiv.org/abs/#1}{\nolinkurl{https://arxiv.org/abs/#1}}}

\bibitem[{{Alencar} {et~al.}(2018){Alencar}, {Bouvier}, {Donati}, {Alecian},
  {Folsom}, {Grankin}, {Hussain}, {Hill}, {Cody}, {Carmona}, {Dougados},
  {Gregory}, {Herczeg}, {M{\'e}nard}, {Moutou}, {Malo}, {Takami}, \& {Matysse
  Collaboration}}]{Alencar_ea_2018}
{Alencar}, S.~H.~P., {Bouvier}, J., {Donati}, J.~F., {et~al.} 2018, \aap, 620,
  A195, \dodoi{10.1051/0004-6361/201834263}

\bibitem[{{Almendros-Abad} {et~al.}(2024){Almendros-Abad}, {Manara}, {Testi},
  {Natta}, {Claes}, {Mu{\v{z}}i{\'c}}, {Sanchis}, {Alcal{\'a}}, {Bayo}, \&
  {Scholz}}]{Almendros-Abad_ea_2024}
{Almendros-Abad}, V., {Manara}, C.~F., {Testi}, L., {et~al.} 2024, \aap, 685,
  A118, \dodoi{10.1051/0004-6361/202348649}

\bibitem[{{Aly} {et~al.}(2024){Aly}, {Nealon}, \& {Gonzalez}}]{Aly_ea_2024}
{Aly}, H., {Nealon}, R., \& {Gonzalez}, J.-F. 2024, \mnras, 527, 4777,
  \dodoi{10.1093/mnras/stad3494}

\bibitem[{{Ansdell} {et~al.}(2016{\natexlab{a}}){Ansdell}, {Gaidos},
  {Williams}, {Kennedy}, {Wyatt}, {LaCourse}, {Jacobs}, \&
  {Mann}}]{Ansdell_ea_2016b}
{Ansdell}, M., {Gaidos}, E., {Williams}, J.~P., {et~al.} 2016{\natexlab{a}},
  \mnras, 462, L101, \dodoi{10.1093/mnrasl/slw140}

\bibitem[{{Ansdell} {et~al.}(2016{\natexlab{b}}){Ansdell}, {Gaidos},
  {Rappaport}, {Jacobs}, {LaCourse}, {Jek}, {Mann}, {Wyatt}, {Kennedy},
  {Williams}, \& {Boyajian}}]{Ansdell_ea_2016a}
{Ansdell}, M., {Gaidos}, E., {Rappaport}, S.~A., {et~al.} 2016{\natexlab{b}},
  \apj, 816, 69, \dodoi{10.3847/0004-637X/816/2/69}

\bibitem[{{Armitage} \& {Pringle}(1997)}]{Armitage_Pringle_1997}
{Armitage}, P.~J., \& {Pringle}, J.~E. 1997, \apjl, 488, L47,
  \dodoi{10.1086/310907}

\bibitem[{{Avenhaus} {et~al.}(2018){Avenhaus}, {Quanz}, {Garufi}, {Perez},
  {Casassus}, {Pinte}, {Bertrang}, {Caceres}, {Benisty}, \&
  {Dominik}}]{Avenhaus_ea_2018}
{Avenhaus}, H., {Quanz}, S.~P., {Garufi}, A., {et~al.} 2018, \apj, 863, 44,
  \dodoi{10.3847/1538-4357/aab846}

\bibitem[{{Ballabio} {et~al.}(2021){Ballabio}, {Nealon}, {Alexander}, {Cuello},
  {Pinte}, \& {Price}}]{Ballabio_ea_2021}
{Ballabio}, G., {Nealon}, R., {Alexander}, R.~D., {et~al.} 2021, \mnras, 504,
  888, \dodoi{10.1093/mnras/stab922}

\bibitem[{{Barraza-Alfaro} {et~al.}(2025){Barraza-Alfaro}, {Flock},
  {B{\'e}thune}, {Teague}, {Bae}, {Benisty}, {Cataldi}, {Curone}, {Czekala},
  {Facchini}, {Fasano}, {Fukagawa}, {Galloway-Sprietsma}, {Garg}, {Hall},
  {Huang}, {Ilee}, {Izquierdo}, {Kanagawa}, {Koch}, {Lesur}, {Longarini},
  {Loomis}, {Orihara}, {Pinte}, {Price}, {Rosotti}, {Stadler},
  {Wafflard-Fernandez}, {Winter}, {W{\"o}lfer}, {Yen}, {Yoshida}, \&
  {Zawadzki}}]{Barraza_exoALMA}
{Barraza-Alfaro}, M., {Flock}, M., {B{\'e}thune}, W., {et~al.} 2025, \apjl,
  984, L21, \dodoi{10.3847/2041-8213/adc42d}

\bibitem[{{Bate} {et~al.}(2010){Bate}, {Lodato}, \& {Pringle}}]{Bate_ea_2010}
{Bate}, M.~R., {Lodato}, G., \& {Pringle}, J.~E. 2010, \mnras, 401, 1505,
  \dodoi{10.1111/j.1365-2966.2009.15773.x}

\bibitem[{{Begeman}(1989)}]{Begeman_ea_1989}
{Begeman}, K.~G. 1989, \aap, 223, 47

\bibitem[{{Benisty} {et~al.}(2015){Benisty}, {Juhasz}, {Boccaletti},
  {Avenhaus}, {Milli}, {Thalmann}, {Dominik}, {Pinilla}, {Buenzli}, {Pohl},
  {Beuzit}, {Birnstiel}, {de Boer}, {Bonnefoy}, {Chauvin}, {Christiaens},
  {Garufi}, {Grady}, {Henning}, {Huelamo}, {Isella}, {Langlois}, {M{\'e}nard},
  {Mouillet}, {Olofsson}, {Pantin}, {Pinte}, \& {Pueyo}}]{Benisty_es_2015}
{Benisty}, M., {Juhasz}, A., {Boccaletti}, A., {et~al.} 2015, \aap, 578, L6,
  \dodoi{10.1051/0004-6361/201526011}

\bibitem[{{Benisty} {et~al.}(2018){Benisty}, {Juh{\'a}sz}, {Facchini},
  {Pinilla}, {de Boer}, {P{\'e}rez}, {Keppler}, {Muro-Arena}, {Villenave},
  {Andrews}, {Dominik}, {Dullemond}, {Gallenne}, {Garufi}, {Ginski}, \&
  {Isella}}]{Benisty_ea_2018}
{Benisty}, M., {Juh{\'a}sz}, A., {Facchini}, S., {et~al.} 2018, \aap, 619,
  A171, \dodoi{10.1051/0004-6361/201833913}

\bibitem[{{Benisty} {et~al.}(2023){Benisty}, {Dominik}, {Follette}, {Garufi},
  {Ginski}, {Hashimoto}, {Keppler}, {Kley}, \& {Monnier}}]{Benisty_ea_2023}
{Benisty}, M., {Dominik}, C., {Follette}, K., {et~al.} 2023, in Astronomical
  Society of the Pacific Conference Series, Vol. 534, Protostars and Planets
  VII, ed. S.~{Inutsuka}, Y.~{Aikawa}, T.~{Muto}, K.~{Tomida}, \& M.~{Tamura},
  605, \dodoi{10.48550/arXiv.2203.09991}

\bibitem[{{Bohn} {et~al.}(2022){Bohn}, {Benisty}, {Perraut}, {van der Marel},
  {W{\"o}lfer}, {van Dishoeck}, {Facchini}, {Manara}, {Teague}, {Francis},
  {Berger}, {Garcia-Lopez}, {Ginski}, {Henning}, {Kenworthy}, {Kraus},
  {M{\'e}nard}, {M{\'e}rand}, \& {P{\'e}rez}}]{Bohn2022}
{Bohn}, A.~J., {Benisty}, M., {Perraut}, K., {et~al.} 2022, \aap, 658, A183,
  \dodoi{10.1051/0004-6361/202142070}

\bibitem[{{Bouvier} {et~al.}(2013){Bouvier}, {Grankin}, {Ellerbroek}, {Bouy},
  \& {Barrado}}]{Bouvier_ea_2013}
{Bouvier}, J., {Grankin}, K., {Ellerbroek}, L.~E., {Bouy}, H., \& {Barrado}, D.
  2013, \aap, 557, A77, \dodoi{10.1051/0004-6361/201321389}

\bibitem[{{Bouvier} {et~al.}(1999){Bouvier}, {Chelli}, {Allain}, {Carrasco},
  {Costero}, {Cruz-Gonzalez}, {Dougados}, {Fern{\'a}ndez}, {Mart{\'\i}n},
  {M{\'e}nard}, {Mennessier}, {Mujica}, {Recillas}, {Salas}, {Schmidt}, \&
  {Wichmann}}]{Bouvier_ea_1999}
{Bouvier}, J., {Chelli}, A., {Allain}, S., {et~al.} 1999, \aap, 349, 619

\bibitem[{{Casassus}(2022)}]{Casassus_2022}
{Casassus}, S. 2022, {ConeRot: Velocity perturbations extractor}, Astrophysics
  Source Code Library, record ascl:2207.027

\bibitem[{{Casassus} \& {P{\'e}rez}(2019)}]{Casassus_Perez_2019}
{Casassus}, S., \& {P{\'e}rez}, S. 2019, \apjl, 883, L41,
  \dodoi{10.3847/2041-8213/ab4425}

\bibitem[{{Casassus} {et~al.}(2015){Casassus}, {Marino}, {P{\'e}rez}, {Roman},
  {Dunhill}, {Armitage}, {Cuadra}, {Wootten}, {van der Plas}, {Cieza}, {Moral},
  {Christiaens}, \& {Montesinos}}]{Casassus_ea_2015}
{Casassus}, S., {Marino}, S., {P{\'e}rez}, S., {et~al.} 2015, \apj, 811, 92,
  \dodoi{10.1088/0004-637X/811/2/92}

\bibitem[{{Casassus} {et~al.}(2018){Casassus}, {Avenhaus}, {P{\'e}rez},
  {Navarro}, {C{\'a}rcamo}, {Marino}, {Cieza}, {Quanz}, {Alarc{\'o}n}, {Zurlo},
  {Osses}, {Rannou}, {Rom{\'a}n}, \& {Barraza}}]{Casassus_ea_2018}
{Casassus}, S., {Avenhaus}, H., {P{\'e}rez}, S., {et~al.} 2018, \mnras, 477,
  5104, \dodoi{10.1093/mnras/sty894}

\bibitem[{{Codron} {et~al.}(2025){Codron}, {Kraus}, {Monnier}, {Marino},
  {Davies}, {Anugu}, {Gardner}, {Ibrahim}, {Lanthermann}, \& {Le
  Bouquin}}]{Codron_ea_2025}
{Codron}, I., {Kraus}, S., {Monnier}, J.~D., {et~al.} 2025, arXiv e-prints,
  arXiv:2506.19668.
\newblock \doarXiv{2506.19668}

\bibitem[{{Cody} {et~al.}(2014){Cody}, {Stauffer}, {Baglin}, {Micela},
  {Rebull}, {Flaccomio}, {Morales-Calder{\'o}n}, {Aigrain}, {Bouvier},
  {Hillenbrand}, {Gutermuth}, {Song}, {Turner}, {Alencar}, {Zwintz},
  {Plavchan}, {Carpenter}, {Findeisen}, {Carey}, {Terebey}, {Hartmann},
  {Calvet}, {Teixeira}, {Vrba}, {Wolk}, {Covey}, {Poppenhaeger}, {G{\"u}nther},
  {Forbrich}, {Whitney}, {Affer}, {Herbst}, {Hora}, {Barrado}, {Holtzman},
  {Marchis}, {Wood}, {Medeiros Guimar{\~a}es}, {Lillo Box}, {Gillen},
  {McQuillan}, {Espaillat}, {Allen}, {D'Alessio}, \& {Favata}}]{Cody_ea_2014}
{Cody}, A.~M., {Stauffer}, J., {Baglin}, A., {et~al.} 2014, \aj, 147, 82,
  \dodoi{10.1088/0004-6256/147/4/82}

\bibitem[{{Cox} {et~al.}(2013){Cox}, {Grady}, {Hammel}, {Hornbeck}, {Russell},
  {Sitko}, \& {Woodgate}}]{Cox_ea_2013}
{Cox}, A.~W., {Grady}, C.~A., {Hammel}, H.~B., {et~al.} 2013, \apj, 762, 40,
  \dodoi{10.1088/0004-637X/762/1/40}

\bibitem[{{Cuello} {et~al.}(2023){Cuello}, {M{\'e}nard}, \&
  {Price}}]{Cuello_ea_2023}
{Cuello}, N., {M{\'e}nard}, F., \& {Price}, D.~J. 2023, European Physical
  Journal Plus, 138, 11, \dodoi{10.1140/epjp/s13360-022-03602-w}

\bibitem[{{Curone} {et~al.}(2025){Curone}, {Facchini}, {Andrews}, {Testi},
  {Benisty}, {Czekala}, {Huang}, {Ilee}, {Isella}, {Lodato}, {Loomis},
  {Stadler}, {Winter}, {Bae}, {Barraza-Alfaro}, {Cataldi}, {Cuello}, {Fasano},
  {Flock}, {Fukagawa}, {Galloway-Sprietsma}, {Garg}, {Hall}, {Izquierdo},
  {Kanagawa}, {Lesur}, {Longarini}, {Menard}, {Orihara}, {Pinte}, {Price},
  {Rosotti}, {Teague}, {Wafflard-Fernandez}, {Wilner}, {W{\"o}lfer}, {Yen},
  {Yoshida}, \& {Zawadzki}}]{Curone_exoALMA}
{Curone}, P., {Facchini}, S., {Andrews}, S.~M., {et~al.} 2025, \apjl, 984, L9,
  \dodoi{10.3847/2041-8213/adc438}

\bibitem[{{Currie} {et~al.}(2019){Currie}, {Marois}, {Cieza}, {Mulders},
  {Lawson}, {Caceres}, {Rodriguez-Ruiz}, {Wisniewski}, {Guyon}, {Brandt},
  {Kasdin}, {Groff}, {Lozi}, {Chilcote}, {Hodapp}, {Jovanovic}, {Martinache},
  {Skaf}, {Lyra}, {Tamura}, {Asensio-Torres}, {Dong}, {Grady}, {Gerard},
  {Fukagawa}, {Hand}, {Hayashi}, {Henning}, {Kudo}, {Kuzuhara}, {Kwon},
  {McElwain}, \& {Uyama}}]{Currie_ea_2019}
{Currie}, T., {Marois}, C., {Cieza}, L., {et~al.} 2019, \apjl, 877, L3,
  \dodoi{10.3847/2041-8213/ab1b42}

\bibitem[{{de Boer} {et~al.}(2016){de Boer}, {Salter}, {Benisty}, {Vigan},
  {Boccaletti}, {Pinilla}, {Ginski}, {Juhasz}, {Maire}, {Messina}, {Desidera},
  {Cheetham}, {Girard}, {Wahhaj}, {Langlois}, {Bonnefoy}, {Beuzit}, {Buenzli},
  {Chauvin}, {Dominik}, {Feldt}, {Gratton}, {Hagelberg}, {Isella}, {Janson},
  {Keller}, {Lagrange}, {Lannier}, {Menard}, {Mesa}, {Mouillet}, {Mugrauer},
  {Peretti}, {Perrot}, {Sissa}, {Snik}, {Vogt}, {Zurlo}, \& {SPHERE
  Consortium}}]{deBoer_ea_2016}
{de Boer}, J., {Salter}, G., {Benisty}, M., {et~al.} 2016, \aap, 595, A114,
  \dodoi{10.1051/0004-6361/201629267}

\bibitem[{{Debes} {et~al.}(2023){Debes}, {Nealon}, {Alexander}, {Weinberger},
  {Wolff}, {Hines}, {Kastner}, {Jang-Condell}, {Pinte}, {Plavchan}, \&
  {Pueyo}}]{Debes_ea_2023}
{Debes}, J., {Nealon}, R., {Alexander}, R., {et~al.} 2023, \apj, 948, 36,
  \dodoi{10.3847/1538-4357/acbdf1}

\bibitem[{{Debes} {et~al.}(2016){Debes}, {Jang-Condell}, \&
  {Schneider}}]{Debes_ea_2016}
{Debes}, J.~H., {Jang-Condell}, H., \& {Schneider}, G. 2016, \apjl, 819, L1,
  \dodoi{10.3847/2041-8205/819/1/L1}

\bibitem[{{Delfini} {et~al.}(2025){Delfini}, {Vioque}, {Ribas}, \&
  {Hodgkin}}]{Delfini_ea_2025}
{Delfini}, L., {Vioque}, M., {Ribas}, {\'A}., \& {Hodgkin}, S. 2025, arXiv
  e-prints, arXiv:2505.04699, \dodoi{10.48550/arXiv.2505.04699}

\bibitem[{{Deng} \& {Ogilvie}(2022)}]{Deng_ea_2022}
{Deng}, H., \& {Ogilvie}, G.~I. 2022, \mnras, 512, 6078,
  \dodoi{10.1093/mnras/stac858}

\bibitem[{{Donati} {et~al.}(2011){Donati}, {Gregory}, {Montmerle}, {Maggio},
  {Argiroffi}, {Sacco}, {Hussain}, {Kastner}, {Alencar}, {Audard}, {Bouvier},
  {Damiani}, {G{\"u}del}, {Huenemoerder}, \& {Wade}}]{Donati_ea_2011}
{Donati}, J.~F., {Gregory}, S.~G., {Montmerle}, T., {et~al.} 2011, \mnras, 417,
  1747, \dodoi{10.1111/j.1365-2966.2011.19366.x}

\bibitem[{{Donehew} \& {Brittain}(2011)}]{Donehew_Brittain_2011}
{Donehew}, B., \& {Brittain}, S. 2011, \aj, 141, 46,
  \dodoi{10.1088/0004-6256/141/2/46}

\bibitem[{{Dorschner} {et~al.}(1995){Dorschner}, {Begemann}, {Henning},
  {Jaeger}, \& {Mutschke}}]{Dorschner_ea_1995}
{Dorschner}, J., {Begemann}, B., {Henning}, T., {Jaeger}, C., \& {Mutschke}, H.
  1995, \aap, 300, 503

\bibitem[{{Do{\v{g}}an} {et~al.}(2018){Do{\v{g}}an}, {Nixon}, {King}, \&
  {Pringle}}]{Dogan_ea_2018}
{Do{\v{g}}an}, S., {Nixon}, C.~J., {King}, A.~R., \& {Pringle}, J.~E. 2018,
  \mnras, 476, 1519, \dodoi{10.1093/mnras/sty155}

\bibitem[{{Do{\v{g}}an} {et~al.}(2023){Do{\v{g}}an}, {Nixon}, {King},
  {Pringle}, \& {Price}}]{Dogan_ea_2023}
{Do{\v{g}}an}, S., {Nixon}, C.~J., {King}, A.~R., {Pringle}, J.~E., \& {Price},
  D. 2023, in IAU Symposium, Vol. 362, The Predictive Power of Computational
  Astrophysics as a Discover Tool, ed. D.~{Bisikalo}, D.~{Wiebe}, \&
  C.~{Boily}, 177--183, \dodoi{10.1017/S1743921322001387}

\bibitem[{{Dullemond} {et~al.}(2012){Dullemond}, {Juhasz}, {Pohl}, {Sereshti},
  {Shetty}, {Peters}, {Commercon}, \& {Flock}}]{Dullemond_ea_2012}
{Dullemond}, C.~P., {Juhasz}, A., {Pohl}, A., {et~al.} 2012, {RADMC-3D: A
  multi-purpose radiative transfer tool}, Astrophysics Source Code Library,
  record ascl:1202.015

\bibitem[{{Dullemond} {et~al.}(2022){Dullemond}, {Kimmig}, \&
  {Zanazzi}}]{Dullemond_Kimmig_ea_2022}
{Dullemond}, C.~P., {Kimmig}, C.~N., \& {Zanazzi}, J.~J. 2022, \mnras, 511,
  2925, \dodoi{10.1093/mnras/stab2791}

\bibitem[{{Facchini} {et~al.}(2018){Facchini}, {Juh{\'a}sz}, \&
  {Lodato}}]{Facchini_ea_2018}
{Facchini}, S., {Juh{\'a}sz}, A., \& {Lodato}, G. 2018, \mnras, 473, 4459,
  \dodoi{10.1093/mnras/stx2523}

\bibitem[{{Facchini} {et~al.}(2013){Facchini}, {Lodato}, \&
  {Price}}]{Facchini_ea_2013}
{Facchini}, S., {Lodato}, G., \& {Price}, D.~J. 2013, \mnras, 433, 2142,
  \dodoi{10.1093/mnras/stt877}

\bibitem[{{Fairlamb} {et~al.}(2015){Fairlamb}, {Oudmaijer}, {Mendigut{\'\i}a},
  {Ilee}, \& {van den Ancker}}]{Fairlamb_ea_2015}
{Fairlamb}, J.~R., {Oudmaijer}, R.~D., {Mendigut{\'\i}a}, I., {Ilee}, J.~D., \&
  {van den Ancker}, M.~E. 2015, \mnras, 453, 976, \dodoi{10.1093/mnras/stv1576}

\bibitem[{{Fedele} {et~al.}(2008){Fedele}, {van den Ancker}, {Acke}, {van der
  Plas}, {van Boekel}, {Wittkowski}, {Henning}, {Bouwman}, {Meeus}, \&
  {Rafanelli}}]{Fedele_ea_2008}
{Fedele}, D., {van den Ancker}, M.~E., {Acke}, B., {et~al.} 2008, \aap, 491,
  809, \dodoi{10.1051/0004-6361:200810126}

\bibitem[{{Flaherty} {et~al.}(2020){Flaherty}, {Hughes}, {Simon}, {Qi}, {Bai},
  {Bulatek}, {Andrews}, {Wilner}, \& {K{\'o}sp{\'a}l}}]{Flaherty_ea_2020}
{Flaherty}, K., {Hughes}, A.~M., {Simon}, J.~B., {et~al.} 2020, \apj, 895, 109,
  \dodoi{10.3847/1538-4357/ab8cc5}

\bibitem[{{Foucart} \& {Lai}(2011)}]{Foucart_Lai_2011}
{Foucart}, F., \& {Lai}, D. 2011, \mnras, 412, 2799,
  \dodoi{10.1111/j.1365-2966.2010.18176.x}

\bibitem[{{Galloway-Sprietsma} {et~al.}(2025){Galloway-Sprietsma}, {Bae},
  {Izquierdo}, {Stadler}, {Longarini}, {Teague}, {Andrews}, {Winter},
  {Benisty}, {Facchini}, {Rosotti}, {Zawadzki}, {Pinte}, {Fasano},
  {Barraza-Alfaro}, {Cataldi}, {Cuello}, {Curone}, {Czekala}, {Flock},
  {Fukagawa}, {Gardner}, {Garg}, {Hall}, {Huang}, {Ilee}, {Kanagawa}, {Lesur},
  {Lodato}, {Loomis}, {Menard}, {Orihara}, {Price}, {Wafflard-Fernandez},
  {Wilner}, {W{\"o}lfer}, {Yen}, \& {Yoshida}}]{Galloway_exoALMA}
{Galloway-Sprietsma}, M., {Bae}, J., {Izquierdo}, A.~F., {et~al.} 2025, \apjl,
  984, L10, \dodoi{10.3847/2041-8213/adc437}

\bibitem[{{Gardner} {et~al.}(2025){Gardner}, {Isella}, {Li}, {Li}, {Bae},
  {Barraza-Alfaro}, {Benisty}, {Cataldi}, {Curone}, {Eisner}, {Facchini},
  {Fasano}, {Flock}, {Follette}, {Fukagawa}, {Galloway-Sprietsma}, {Garg},
  {Hall}, {Huang}, {Ilee}, {Ireland}, {Izquierdo}, {Johns-Krull}, {Kanagawa},
  {Kraus}, {Lesur}, {Liu}, {Longarini}, {Loomis}, {Menard}, {Orihara}, {Pinte},
  {Price}, {Ricci}, {Rosotti}, {Sallum}, {Stadler}, {Teague},
  {Wafflard-Fernandez}, {Wilner}, {Winter}, {W{\"o}lfer}, {Yen}, {Yoshida},
  {Zawadzki}, \& {Zhu}}]{Gardener_exoALMA}
{Gardner}, C.~H., {Isella}, A., {Li}, H., {et~al.} 2025, \apjl, 984, L16,
  \dodoi{10.3847/2041-8213/adc432}

\bibitem[{{Garufi} {et~al.}(2018){Garufi}, {Benisty}, {Pinilla}, {Tazzari},
  {Dominik}, {Ginski}, {Henning}, {Kral}, {Langlois}, {M{\'e}nard}, {Stolker},
  {Szulagyi}, {Villenave}, \& {van der Plas}}]{Garufi_ea_2018}
{Garufi}, A., {Benisty}, M., {Pinilla}, P., {et~al.} 2018, \aap, 620, A94,
  \dodoi{10.1051/0004-6361/201833872}

\bibitem[{{Garufi} {et~al.}(2024){Garufi}, {Ginski}, {van Holstein}, {Benisty},
  {Manara}, {P{\'e}rez}, {Pinilla}, {Ribas}, {Weber}, {Williams}, {Cieza},
  {Dominik}, {Facchini}, {Huang}, {Zurlo}, {Bae}, {Hagelberg}, {Henning},
  {Hogerheijde}, {Janson}, {M{\'e}nard}, {Messina}, {Meyer}, {Pinte}, {Quanz},
  {Rigliaco}, {Roccatagliata}, {Schmid}, {Szul{\'a}gyi}, {van Boekel},
  {Wahhaj}, {Antichi}, {Baruffolo}, \& {Moulin}}]{Garufi_ea_2024}
{Garufi}, A., {Ginski}, C., {van Holstein}, R.~G., {et~al.} 2024, \aap, 685,
  A53, \dodoi{10.1051/0004-6361/202347586}

\bibitem[{{Ginski} {et~al.}(2021){Ginski}, {Facchini}, {Huang}, {Benisty},
  {Vaendel}, {Stapper}, {Dominik}, {Bae}, {M{\'e}nard}, {Muro-Arena},
  {Hogerheijde}, {McClure}, {van Holstein}, {Birnstiel}, {Boehler}, {Bohn},
  {Flock}, {Mamajek}, {Manara}, {Pinilla}, {Pinte}, \&
  {Ribas}}]{Ginski_ea_2021}
{Ginski}, C., {Facchini}, S., {Huang}, J., {et~al.} 2021, \apjl, 908, L25,
  \dodoi{10.3847/2041-8213/abdf57}

\bibitem[{{Grady} {et~al.}(2009){Grady}, {Schneider}, {Sitko}, {Williger},
  {Hamaguchi}, {Brittain}, {Ablordeppey}, {Apai}, {Beerman}, {Carpenter},
  {Collins}, {Fukagawa}, {Hammel}, {Henning}, {Hines}, {Kimes}, {Lynch},
  {M{\'e}nard}, {Pearson}, {Russell}, {Silverstone}, {Smith}, {Troutman},
  {Wilner}, {Woodgate}, \& {Clampin}}]{Grady_ea_2009}
{Grady}, C.~A., {Schneider}, G., {Sitko}, M.~L., {et~al.} 2009, \apj, 699,
  1822, \dodoi{10.1088/0004-637X/699/2/1822}

\bibitem[{Harris {et~al.}(2020)Harris, Millman, van~der Walt, Gommers,
  Virtanen, Cournapeau, Wieser, Taylor, Berg, Smith, Kern, Picus, Hoyer, van
  Kerkwijk, Brett, Haldane, del R{\'{i}}o, Wiebe, Peterson,
  G{\'{e}}rard-Marchant, Sheppard, Reddy, Weckesser, Abbasi, Gohlke, \&
  Oliphant}]{Numpy_citation}
Harris, C.~R., Millman, K.~J., van~der Walt, S.~J., {et~al.} 2020, Nature, 585,
  357, \dodoi{10.1038/s41586-020-2649-2}

\bibitem[{{Haworth} {et~al.}(2017){Haworth}, {Facchini}, {Clarke}, \&
  {Cleeves}}]{Hawoth_ea_2017}
{Haworth}, T.~J., {Facchini}, S., {Clarke}, C.~J., \& {Cleeves}, L.~I. 2017,
  \mnras, 468, L108, \dodoi{10.1093/mnrasl/slx037}

\bibitem[{{Held} \& {Ogilvie}(2024)}]{Held_ea_2024}
{Held}, L.~E., \& {Ogilvie}, G.~I. 2024, \mnras, 535, 3108,
  \dodoi{10.1093/mnras/stae2487}

\bibitem[{{Hu{\'e}lamo} {et~al.}(2018){Hu{\'e}lamo}, {Chauvin}, {Schmid},
  {Quanz}, {Whelan}, {Lillo-Box}, {Barrado}, {Montesinos}, {Alcal{\'a}},
  {Benisty}, {de Gregorio-Monsalvo}, {Mendigut{\'\i}a}, {Bouy}, {Mer{\'\i}n},
  {de Boer}, {Garufi}, \& {Pantin}}]{Huelamo_ea_2018}
{Hu{\'e}lamo}, N., {Chauvin}, G., {Schmid}, H.~M., {et~al.} 2018, \aap, 613,
  L5, \dodoi{10.1051/0004-6361/201832874}

\bibitem[{{Ingleby} {et~al.}(2013){Ingleby}, {Calvet}, {Herczeg}, {Blaty},
  {Walter}, {Ardila}, {Alexander}, {Edwards}, {Espaillat}, {Gregory},
  {Hillenbrand}, \& {Brown}}]{Ingleby_ea_2013}
{Ingleby}, L., {Calvet}, N., {Herczeg}, G., {et~al.} 2013, \apj, 767, 112,
  \dodoi{10.1088/0004-637X/767/2/112}

\bibitem[{{Izquierdo} {et~al.}(2021){Izquierdo}, {Testi}, {Facchini},
  {Rosotti}, \& {van Dishoeck}}]{Izquierdo_ea_2021}
{Izquierdo}, A.~F., {Testi}, L., {Facchini}, S., {Rosotti}, G.~P., \& {van
  Dishoeck}, E.~F. 2021, \aap, 650, A179, \dodoi{10.1051/0004-6361/202140779}

\bibitem[{{Izquierdo} {et~al.}(2025){Izquierdo}, {Stadler},
  {Galloway-Sprietsma}, {Benisty}, {Pinte}, {Bae}, {Teague}, {Facchini},
  {W{\"o}lfer}, {Longarini}, {Curone}, {Andrews}, {Barraza-Alfaro}, {Cataldi},
  {Cuello}, {Czekala}, {Fasano}, {Flock}, {Fukagawa}, {Garg}, {Hall},
  {Hammond}, {Hilder}, {Huang}, {Ilee}, {Isella}, {Kanagawa}, {Lesur},
  {Lodato}, {Loomis}, {Orihara}, {Price}, {Rosotti}, {Testi}, {Yen},
  {Wafflard-Fernandez}, {Wilner}, {Winter}, {Yoshida}, \&
  {Zawadzki}}]{Izquierdo_exoALMA}
{Izquierdo}, A.~F., {Stadler}, J., {Galloway-Sprietsma}, M., {et~al.} 2025,
  \apjl, 984, L8, \dodoi{10.3847/2041-8213/adc439}

\bibitem[{{Jaeger} {et~al.}(1994){Jaeger}, {Mutschke}, {Begemann}, {Dorschner},
  \& {Henning}}]{Jaeger_ea_1994}
{Jaeger}, C., {Mutschke}, H., {Begemann}, B., {Dorschner}, J., \& {Henning}, T.
  1994, \aap, 292, 641

\bibitem[{{Juh{\'a}sz} \& {Facchini}(2017)}]{Juhasz_Facchini_2017}
{Juh{\'a}sz}, A., \& {Facchini}, S. 2017, \mnras, 466, 4053,
  \dodoi{10.1093/mnras/stw3389}

\bibitem[{{Kimmig} \& {Dullemond}(2024)}]{Kimmig_ea_2024}
{Kimmig}, C.~N., \& {Dullemond}, C.~P. 2024, \aap, 689, A45,
  \dodoi{10.1051/0004-6361/202348660}

\bibitem[{{Kimmig} \& {Villenave}(2025)}]{Kimmig_ea_2025}
{Kimmig}, C.~N., \& {Villenave}, M. 2025, arXiv e-prints, arXiv:2504.05399,
  \dodoi{10.48550/arXiv.2504.05399}

\bibitem[{{Kraus} {et~al.}(2020){Kraus}, {Kreplin}, {Young}, {Bate}, {Monnier},
  {Harries}, {Avenhaus}, {Kluska}, {Laws}, {Rich}, {Willson}, {Aarnio},
  {Adams}, {Andrews}, {Anugu}, {Bae}, {ten Brummelaar}, {Calvet}, {Cur{\'e}},
  {Davies}, {Ennis}, {Espaillat}, {Gardner}, {Hartmann}, {Hinkley}, {Labdon},
  {Lanthermann}, {LeBouquin}, {Schaefer}, {Setterholm}, {Wilner}, \&
  {Zhu}}]{Kraus_ea_2020}
{Kraus}, S., {Kreplin}, A., {Young}, A.~K., {et~al.} 2020, Science, 369, 1233,
  \dodoi{10.1126/science.aba4633}

\bibitem[{{Kuffmeier} {et~al.}(2023){Kuffmeier}, {Jensen}, \&
  {Haugb{\o}lle}}]{Kuffmeier_ea_2023}
{Kuffmeier}, M., {Jensen}, S.~S., \& {Haugb{\o}lle}, T. 2023, European Physical
  Journal Plus, 138, 272, \dodoi{10.1140/epjp/s13360-023-03880-y}

\bibitem[{{Kuffmeier} {et~al.}(2024){Kuffmeier}, {Pineda}, {Segura-Cox}, \&
  {Haugb{\o}lle}}]{Kuffmeier_ea_2024}
{Kuffmeier}, M., {Pineda}, J.~E., {Segura-Cox}, D., \& {Haugb{\o}lle}, T. 2024,
  \aap, 690, A297, \dodoi{10.1051/0004-6361/202450410}

\bibitem[{{Lai}(1999)}]{Lai_1999}
{Lai}, D. 1999, \apj, 524, 1030, \dodoi{10.1086/307850}

\bibitem[{{Lazareff} {et~al.}(2017){Lazareff}, {Berger}, {Kluska}, {Le
  Bouquin}, {Benisty}, {Malbet}, {Koen}, {Pinte}, {Thi}, {Absil}, {Baron},
  {Delboulb{\'e}}, {Duvert}, {Isella}, {Jocou}, {Juhasz}, {Kraus}, {Lachaume},
  {M{\'e}nard}, {Millan-Gabet}, {Monnier}, {Moulin}, {Perraut}, {Rochat},
  {Soulez}, {Tallon}, {Thi{\'e}baut}, {Traub}, \& {Zins}}]{Lazareff_ea_2017}
{Lazareff}, B., {Berger}, J.~P., {Kluska}, J., {et~al.} 2017, \aap, 599, A85,
  \dodoi{10.1051/0004-6361/201629305}

\bibitem[{{Lodato} \& {Facchini}(2013)}]{Lodato_Facchini_2013}
{Lodato}, G., \& {Facchini}, S. 2013, \mnras, 433, 2157,
  \dodoi{10.1093/mnras/stt878}

\bibitem[{{Lodato} \& {Price}(2010)}]{Lodato_ea_2010}
{Lodato}, G., \& {Price}, D.~J. 2010, \mnras, 405, 1212,
  \dodoi{10.1111/j.1365-2966.2010.16526.x}

\bibitem[{{Lodato} \& {Pringle}(2006)}]{Lodata_Pringle_2006}
{Lodato}, G., \& {Pringle}, J.~E. 2006, \mnras, 368, 1196,
  \dodoi{10.1111/j.1365-2966.2006.10194.x}

\bibitem[{{Lodato} \& {Pringle}(2007)}]{Lodato_ea_2007}
---. 2007, \mnras, 381, 1287, \dodoi{10.1111/j.1365-2966.2007.12332.x}

\bibitem[{{Lodato} {et~al.}(2023){Lodato}, {Rampinelli}, {Viscardi},
  {Longarini}, {Izquierdo}, {Paneque-Carre{\~n}o}, {Testi}, {Facchini},
  {Miotello}, {Veronesi}, \& {Hall}}]{Lodato_ea_2023}
{Lodato}, G., {Rampinelli}, L., {Viscardi}, E., {et~al.} 2023, \mnras, 518,
  4481, \dodoi{10.1093/mnras/stac3223}

\bibitem[{{Longarini} {et~al.}(2021){Longarini}, {Lodato}, {Toci}, \&
  {Aly}}]{Longarini_ea_2021a}
{Longarini}, C., {Lodato}, G., {Toci}, C., \& {Aly}, H. 2021, \mnras, 503,
  4930, \dodoi{10.1093/mnras/stab843}

\bibitem[{{Longarini} {et~al.}(2025){Longarini}, {Lodato}, {Rosotti},
  {Andrews}, {Winter}, {Stadler}, {Izquierdo}, {Galloway-Sprietsma},
  {Facchini}, {Curone}, {Benisty}, {Teague}, {Bae}, {Barraza-Alfaro},
  {Cataldi}, {Czekala}, {Cuello}, {Fasano}, {Flock}, {Fukagawa}, {Garg},
  {Hall}, {Hammond}, {Hardiman}, {Hilder}, {Huang}, {Ilee}, {Isella},
  {Kanagawa}, {Lesur}, {Loomis}, {M{\'e}nard}, {Orihara}, {Pinte}, {Price},
  {Testi}, {Fernandez}, {W{\"o}lfer}, {Yen}, {Yoshida}, \&
  {Zawadzki}}]{Longarini_exoALMA}
{Longarini}, C., {Lodato}, G., {Rosotti}, G., {et~al.} 2025, \apjl, 984, L17,
  \dodoi{10.3847/2041-8213/adc431}

\bibitem[{{Loomis} {et~al.}(2017){Loomis}, {{\"O}berg}, {Andrews}, \&
  {MacGregor}}]{Loomis_ea_2017}
{Loomis}, R.~A., {{\"O}berg}, K.~I., {Andrews}, S.~M., \& {MacGregor}, M.~A.
  2017, \apj, 840, 23, \dodoi{10.3847/1538-4357/aa6c63}

\bibitem[{{Loomis} {et~al.}(2025){Loomis}, {Facchini}, {Benisty}, {Curone},
  {Ilee}, {Cataldi}, {Yen}, {Teague}, {Pinte}, {Huang}, {Garg}, {Orihara},
  {Czekala}, {Zawadzki}, {Andrews}, {Wilner}, {Bae}, {Barraza-Alfaro},
  {Fasano}, {Flock}, {Fukagawa}, {Galloway-Sprietsma}, {Izquierdo}, {Kanagawa},
  {Lesur}, {Longarini}, {Menard}, {Price}, {Rosotti}, {Stadler},
  {Wafflard-Fernandez}, {W{\"o}lfer}, \& {Yoshida}}]{Loomis_exoALMA}
{Loomis}, R.~A., {Facchini}, S., {Benisty}, M., {et~al.} 2025, \apjl, 984, L7,
  \dodoi{10.3847/2041-8213/adc43a}

\bibitem[{{Lubow} \& {Ogilvie}(2000)}]{Lubow_ea_2000}
{Lubow}, S.~H., \& {Ogilvie}, G.~I. 2000, \apj, 538, 326,
  \dodoi{10.1086/309101}

\bibitem[{{Manara} {et~al.}(2023){Manara}, {Ansdell}, {Rosotti}, {Hughes},
  {Armitage}, {Lodato}, \& {Williams}}]{Manara_ea_2023}
{Manara}, C.~F., {Ansdell}, M., {Rosotti}, G.~P., {et~al.} 2023, in
  Astronomical Society of the Pacific Conference Series, Vol. 534, Protostars
  and Planets VII, ed. S.~{Inutsuka}, Y.~{Aikawa}, T.~{Muto}, K.~{Tomida}, \&
  M.~{Tamura}, 539, \dodoi{10.48550/arXiv.2203.09930}

\bibitem[{{Manara} {et~al.}(2014){Manara}, {Testi}, {Natta}, {Rosotti},
  {Benisty}, {Ercolano}, \& {Ricci}}]{Manara_ea_2014}
{Manara}, C.~F., {Testi}, L., {Natta}, A., {et~al.} 2014, \aap, 568, A18,
  \dodoi{10.1051/0004-6361/201323318}

\bibitem[{{Manara} {et~al.}(2017){Manara}, {Testi}, {Herczeg}, {Pascucci},
  {Alcal{\'a}}, {Natta}, {Antoniucci}, {Fedele}, {Mulders}, {Henning},
  {Mohanty}, {Prusti}, \& {Rigliaco}}]{Manara_ea_2017}
{Manara}, C.~F., {Testi}, L., {Herczeg}, G.~J., {et~al.} 2017, \aap, 604, A127,
  \dodoi{10.1051/0004-6361/201630147}

\bibitem[{{Martire} {et~al.}(2024){Martire}, {Longarini}, {Lodato}, {Rosotti},
  {Winter}, {Facchini}, {Hardiman}, {Benisty}, {Stadler}, {Izquierdo}, \&
  {Testi}}]{Martire_ea_2024}
{Martire}, P., {Longarini}, C., {Lodato}, G., {et~al.} 2024, \aap, 686, A9,
  \dodoi{10.1051/0004-6361/202348546}

\bibitem[{{Mayama} {et~al.}(2018){Mayama}, {Akiyama}, {Pani{\'c}}, {Miley},
  {Tsukagoshi}, {Muto}, {Dong}, {de Leon}, {Mizuki}, {Oh}, {Hashimoto}, {Sai},
  {Currie}, {Takami}, {Grady}, {Hayashi}, {Tamura}, \&
  {Inutsuka}}]{Mayama_ea_2018}
{Mayama}, S., {Akiyama}, E., {Pani{\'c}}, O., {et~al.} 2018, \apjl, 868, L3,
  \dodoi{10.3847/2041-8213/aae88b}

\bibitem[{{M{\"u}ller} {et~al.}(2011){M{\"u}ller}, {van den Ancker},
  {Launhardt}, {Pott}, {Fedele}, \& {Henning}}]{Muller_ea_2011}
{M{\"u}ller}, A., {van den Ancker}, M.~E., {Launhardt}, R., {et~al.} 2011,
  \aap, 530, A85, \dodoi{10.1051/0004-6361/201116732}

\bibitem[{{Muro-Arena} {et~al.}(2020){Muro-Arena}, {Benisty}, {Ginski},
  {Dominik}, {Facchini}, {Villenave}, {van Boekel}, {Chauvin}, {Garufi},
  {Henning}, {Janson}, {Keppler}, {Matter}, {M{\'e}nard}, {Stolker}, {Zurlo},
  {Blanchard}, {Maurel}, {Moeller-Nilsson}, {Petit}, {Roux}, {Sevin}, \&
  {Wildi}}]{Muro-Arena_ea_2020}
{Muro-Arena}, G.~A., {Benisty}, M., {Ginski}, C., {et~al.} 2020, \aap, 635,
  A121, \dodoi{10.1051/0004-6361/201936509}

\bibitem[{{Nealon} {et~al.}(2018){Nealon}, {Dipierro}, {Alexander}, {Martin},
  \& {Nixon}}]{Nealon_ea_2018}
{Nealon}, R., {Dipierro}, G., {Alexander}, R., {Martin}, R.~G., \& {Nixon}, C.
  2018, \mnras, 481, 20, \dodoi{10.1093/mnras/sty2267}

\bibitem[{{Nealon} {et~al.}(2019){Nealon}, {Pinte}, {Alexander}, {Mentiplay},
  \& {Dipierro}}]{Nealon_ea_2019}
{Nealon}, R., {Pinte}, C., {Alexander}, R., {Mentiplay}, D., \& {Dipierro}, G.
  2019, \mnras, 484, 4951, \dodoi{10.1093/mnras/stz346}

\bibitem[{{Nealon} {et~al.}(2020){Nealon}, {Price}, \&
  {Pinte}}]{Nealon_ea_2020}
{Nealon}, R., {Price}, D.~J., \& {Pinte}, C. 2020, \mnras, 493, L143,
  \dodoi{10.1093/mnrasl/slaa026}

\bibitem[{{Nixon} {et~al.}(2013){Nixon}, {King}, \& {Price}}]{Nixon_ea_2013}
{Nixon}, C., {King}, A., \& {Price}, D. 2013, \mnras, 434, 1946,
  \dodoi{10.1093/mnras/stt1136}

\bibitem[{{Ogilvie}(1999)}]{Ogilvie_1999}
{Ogilvie}, G.~I. 1999, \mnras, 304, 557,
  \dodoi{10.1046/j.1365-8711.1999.02340.x}

\bibitem[{{Ogilvie} \& {Latter}(2013)}]{Ogilvie_ea_2013}
{Ogilvie}, G.~I., \& {Latter}, H.~N. 2013, \mnras, 433, 2403,
  \dodoi{10.1093/mnras/stt916}

\bibitem[{{Orihara} \& {Momose}(2025)}]{Orihara_ea_2025}
{Orihara}, R., \& {Momose}, M. 2025, arXiv e-prints, arXiv:2505.06044,
  \dodoi{10.48550/arXiv.2505.06044}

\bibitem[{{Orihara} {et~al.}(2023){Orihara}, {Momose}, {Muto}, {Hashimoto},
  {Liu}, {Tsukagoshi}, {Kudo}, {Takahashi}, {Yang}, {Hasegawa}, {Dong},
  {Konishi}, \& {Akiyama}}]{Orihara_ea_2023}
{Orihara}, R., {Momose}, M., {Muto}, T., {et~al.} 2023, \pasj, 75, 424,
  \dodoi{10.1093/pasj/psad009}

\bibitem[{{O'Sullivan} {et~al.}(2005){O'Sullivan}, {Truss}, {Walker}, {Wood},
  {Matthews}, {Whitney}, \& {Bjorkman}}]{OSullivan_ea_2005}
{O'Sullivan}, M., {Truss}, M., {Walker}, C., {et~al.} 2005, \mnras, 358, 632,
  \dodoi{10.1111/j.1365-2966.2005.08805.x}

\bibitem[{{Paardekooper} \& {Ogilvie}(2019)}]{Paardekooper_Ogilvie_2019}
{Paardekooper}, S.-J., \& {Ogilvie}, G.~I. 2019, \mnras, 483, 3738,
  \dodoi{10.1093/mnras/sty3349}

\bibitem[{{Papaloizou} \& {Pringle}(1983)}]{Papaloizou_ea_1983}
{Papaloizou}, J.~C.~B., \& {Pringle}, J.~E. 1983, \mnras, 202, 1181,
  \dodoi{10.1093/mnras/202.4.1181}

\bibitem[{Pedregosa {et~al.}(2011)Pedregosa, Varoquaux, Gramfort, Michel,
  Thirion, Grisel, Blondel, Prettenhofer, Weiss, Dubourg, Vanderplas, Passos,
  Cournapeau, Brucher, Perrot, \& Duchesnay}]{scikit-learn_citation}
Pedregosa, F., Varoquaux, G., Gramfort, A., {et~al.} 2011, Journal of Machine
  Learning Research, 12, 2825

\bibitem[{{P{\'e}rez} {et~al.}(2018){P{\'e}rez}, {Benisty}, {Andrews},
  {Isella}, {Dullemond}, {Huang}, {Kurtovic}, {Guzm{\'a}n}, {Zhu}, {Birnstiel},
  {Zhang}, {Carpenter}, {Wilner}, {Ricci}, {Bai}, {Weaver}, \&
  {{\"O}berg}}]{Perez_ea_2018}
{P{\'e}rez}, L.~M., {Benisty}, M., {Andrews}, S.~M., {et~al.} 2018, \apjl, 869,
  L50, \dodoi{10.3847/2041-8213/aaf745}

\bibitem[{{Pineda} {et~al.}(2014){Pineda}, {Quanz}, {Meru}, {Mulders}, {Meyer},
  {Pani{\'c}}, \& {Avenhaus}}]{Pineda_ea_2014}
{Pineda}, J.~E., {Quanz}, S.~P., {Meru}, F., {et~al.} 2014, \apjl, 788, L34,
  \dodoi{10.1088/2041-8205/788/2/L34}

\bibitem[{{Pinilla} {et~al.}(2018){Pinilla}, {Benisty}, {de Boer}, {Manara},
  {Bouvier}, {Dominik}, {Ginski}, {Loomis}, \& {Sicilia
  Aguilar}}]{Pinilla_ea_2018}
{Pinilla}, P., {Benisty}, M., {de Boer}, J., {et~al.} 2018, \apj, 868, 85,
  \dodoi{10.3847/1538-4357/aae824}

\bibitem[{{Pinte} {et~al.}(2018){Pinte}, {Price}, {M{\'e}nard}, {Duch{\^e}ne},
  {Dent}, {Hill}, {de Gregorio-Monsalvo}, {Hales}, \&
  {Mentiplay}}]{Pinte_ea_2018b}
{Pinte}, C., {Price}, D.~J., {M{\'e}nard}, F., {et~al.} 2018, \apjl, 860, L13,
  \dodoi{10.3847/2041-8213/aac6dc}

\bibitem[{{Pinte} {et~al.}(2020){Pinte}, {Price}, {M{\'e}nard}, {Duch{\^e}ne},
  {Christiaens}, {Andrews}, {Huang}, {Hill}, {van der Plas}, {Perez}, {Isella},
  {Boehler}, {Dent}, {Mentiplay}, \& {Loomis}}]{Pinte_ea_2020}
---. 2020, \apjl, 890, L9, \dodoi{10.3847/2041-8213/ab6dda}

\bibitem[{{Pinte} {et~al.}(2025){Pinte}, {Ilee}, {Huang}, {Benisty},
  {Facchini}, {Fukagawa}, {Teague}, {Bae}, {Barraza-Alfaro}, {Cataldi},
  {Cuello}, {Curone}, {Czekala}, {Fasano}, {Flock}, {Galloway-Sprietsma},
  {Garg}, {Hall}, {Hammond}, {Hardiman}, {Hilder}, {Izquierdo}, {Kanagawa},
  {Lesur}, {Lodato}, {Longarini}, {Loomis}, {Masset}, {Menard}, {Orihara},
  {Price}, {Rosotti}, {Stadler}, {Yen}, {Wafflard-Fernandez}, {Wilner},
  {Winter}, {W{\"o}lfer}, {Yoshida}, \& {Zawadzki}}]{Pinte_exoALMA}
{Pinte}, C., {Ilee}, J.~D., {Huang}, J., {et~al.} 2025, \apjl, 984, L15,
  \dodoi{10.3847/2041-8213/adc433}

\bibitem[{{Price} {et~al.}(2018){Price}, {Cuello}, {Pinte}, {Mentiplay},
  {Casassus}, {Christiaens}, {Kennedy}, {Cuadra}, {Sebastian Perez}, {Marino},
  {Armitage}, {Zurlo}, {Juhasz}, {Ragusa}, {Laibe}, \&
  {Lodato}}]{Price_ea_2018}
{Price}, D.~J., {Cuello}, N., {Pinte}, C., {et~al.} 2018, \mnras, 477, 1270,
  \dodoi{10.1093/mnras/sty647}

\bibitem[{{Pringle}(1996)}]{Pringle96}
{Pringle}, J.~E. 1996, \mnras, 281, 357, \dodoi{10.1093/mnras/281.1.357}

\bibitem[{{Ren} {et~al.}(2023){Ren}, {Benisty}, {Ginski}, {Tazaki}, {Wallack},
  {Milli}, {Garufi}, {Bae}, {Facchini}, {M{\'e}nard}, {Pinilla}, {Swastik},
  {Teague}, \& {Wahhaj}}]{Ren_ea_2023}
{Ren}, B.~B., {Benisty}, M., {Ginski}, C., {et~al.} 2023, \aap, 680, A114,
  \dodoi{10.1051/0004-6361/202347353}

\bibitem[{{Ren} {et~al.}(2024){Ren}, {Xie}, {Benisty}, {Dong}, {Bae},
  {Stolker}, {van Holstein}, {Debes}, {Garufi}, {Ginski}, \&
  {Kraus}}]{Ren_ea_2024}
{Ren}, B.~B., {Xie}, C., {Benisty}, M., {et~al.} 2024, \aap, 681, L2,
  \dodoi{10.1051/0004-6361/202348114}

\bibitem[{{Ribas} {et~al.}(2023){Ribas}, {Mac{\'\i}as}, {Weber}, {P{\'e}rez},
  {Cuello}, {Dong}, {Aguayo}, {C{\'a}ceres}, {Carpenter}, {Dent}, {de
  Gregorio-Monsalvo}, {Duch{\^e}ne}, {Espaillat}, {Riviere-Marichalar}, \&
  {Villenave}}]{Ribas_ea_2023}
{Ribas}, {\'A}., {Mac{\'\i}as}, E., {Weber}, P., {et~al.} 2023, \aap, 673, A77,
  \dodoi{10.1051/0004-6361/202245637}

\bibitem[{{Rigliaco} {et~al.}(2015){Rigliaco}, {Pascucci}, {Duchene},
  {Edwards}, {Ardila}, {Grady}, {Mendigut{\'\i}a}, {Montesinos}, {Mulders},
  {Najita}, {Carpenter}, {Furlan}, {Gorti}, {Meijerink}, \&
  {Meyer}}]{Rigliaco_ea_2015}
{Rigliaco}, E., {Pascucci}, I., {Duchene}, G., {et~al.} 2015, \apj, 801, 31,
  \dodoi{10.1088/0004-637X/801/1/31}

\bibitem[{{Rogers} {et~al.}(2024){Rogers}, {de Marchi}, \&
  {Brandl}}]{Rogers_ea_2024}
{Rogers}, C., {de Marchi}, G., \& {Brandl}, B. 2024, arXiv e-prints,
  arXiv:2412.05650, \dodoi{10.48550/arXiv.2412.05650}

\bibitem[{{Romanova} {et~al.}(2021){Romanova}, {Koldoba}, {Ustyugova},
  {Blinova}, {Lai}, \& {Lovelace}}]{Romanova_ea_2021}
{Romanova}, M.~M., {Koldoba}, A.~V., {Ustyugova}, G.~V., {et~al.} 2021, \mnras,
  506, 372, \dodoi{10.1093/mnras/stab1724}

\bibitem[{{Rosenfeld} {et~al.}(2014){Rosenfeld}, {Chiang}, \&
  {Andrews}}]{Rosenfeld_ea_2014}
{Rosenfeld}, K.~A., {Chiang}, E., \& {Andrews}, S.~M. 2014, \apj, 782, 62,
  \dodoi{10.1088/0004-637X/782/2/62}

\bibitem[{{Rosotti} {et~al.}(2014){Rosotti}, {Dale}, {de Juan Ovelar},
  {Hubber}, {Kruijssen}, {Ercolano}, \& {Walch}}]{Rosotti_ea_2014}
{Rosotti}, G.~P., {Dale}, J.~E., {de Juan Ovelar}, M., {et~al.} 2014, \mnras,
  441, 2094, \dodoi{10.1093/mnras/stu679}

\bibitem[{{Sakai} {et~al.}(2014){Sakai}, {Oya}, {Sakai}, {Watanabe}, {Hirota},
  {Ceccarelli}, {Kahane}, {Lopez-Sepulcre}, {Lefloch}, {Vastel}, {Bottinelli},
  {Caux}, {Coutens}, {Aikawa}, {Takakuwa}, {Ohashi}, {Yen}, \&
  {Yamamoto}}]{Sakai_ea_2014}
{Sakai}, N., {Oya}, Y., {Sakai}, T., {et~al.} 2014, \apjl, 791, L38,
  \dodoi{10.1088/2041-8205/791/2/L38}

\bibitem[{{Schwarz} {et~al.}(2024){Schwarz}, {Henning}, {Christiaens},
  {Gasman}, {Samland}, {Perotti}, {Jang}, {Grant}, {Tabone},
  {Morales-Calder{\'o}n}, {Kamp}, {van Dishoeck}, {G{\"u}del}, {Lagage},
  {Barrado}, {Caratti o Garatti}, {Glauser}, {Ray}, {Vandenbussche}, {Waters},
  {Arabhavi}, {Kanwar}, {Olofsson}, {Rodgers-Lee}, {Schreiber}, \&
  {Temmink}}]{Schwarz_ea_2023}
{Schwarz}, K.~R., {Henning}, T., {Christiaens}, V., {et~al.} 2024, \apj, 962,
  8, \dodoi{10.3847/1538-4357/ad1393}

\bibitem[{{Schwarz} {et~al.}(2025){Schwarz}, {Samland}, {Olofsson}, {Henning},
  {Sellek}, {G{\"u}del}, {Tabone}, {Kamp}, {Lagage}, {van Dishoeck}, {Caratti o
  Garatti}, {Glauser}, {Ray}, {Arabhavi}, {Christiaens}, {Franceschi},
  {Gasman}, {Grant}, {Kanwar}, {Kaeufer}, {Kurtovic}, {Perotti}, {Temmink}, \&
  {Vlasblom}}]{Schwarz_ea_2025}
{Schwarz}, K.~R., {Samland}, M., {Olofsson}, G., {et~al.} 2025, \apj, 980, 148,
  \dodoi{10.3847/1538-4357/adaa79}

\bibitem[{{Shuai} {et~al.}(2022){Shuai}, {Ren}, {Dong}, {Zhou}, {Pueyo}, {De
  Rosa}, {Fang}, \& {Mawet}}]{Shuai_ea_2022}
{Shuai}, L., {Ren}, B.~B., {Dong}, R., {et~al.} 2022, \apjs, 263, 31,
  \dodoi{10.3847/1538-4365/ac98fd}

\bibitem[{{Sicilia-Aguilar} {et~al.}(2020){Sicilia-Aguilar}, {Manara}, {de
  Boer}, {Benisty}, {Pinilla}, \& {Bouvier}}]{Sicilia-Aguilar_ea_2020}
{Sicilia-Aguilar}, A., {Manara}, C.~F., {de Boer}, J., {et~al.} 2020, \aap,
  633, A37, \dodoi{10.1051/0004-6361/201936565}

\bibitem[{{Sitko} {et~al.}(2012){Sitko}, {Day}, {Kimes}, {Beerman}, {Martus},
  {Lynch}, {Russell}, {Grady}, {Schneider}, {Lisse}, {Nuth}, {Cur{\'e}},
  {Henden}, {Kraus}, {Motta}, {Tamura}, {Hornbeck}, {Williger}, \&
  {Fugazza}}]{Sitko_ea_2012}
{Sitko}, M.~L., {Day}, A.~N., {Kimes}, R.~L., {et~al.} 2012, \apj, 745, 29,
  \dodoi{10.1088/0004-637X/745/1/29}

\bibitem[{{Stadler} {et~al.}(2023){Stadler}, {Benisty}, {Izquierdo},
  {Facchini}, {Teague}, {Kurtovic}, {Pinilla}, {Bae}, {Ansdell}, {Loomis},
  {Mayama}, {Perez}, \& {Testi}}]{Stadler_ea_2023}
{Stadler}, J., {Benisty}, M., {Izquierdo}, A., {et~al.} 2023, \aap, 670, L1,
  \dodoi{10.1051/0004-6361/202245381}

\bibitem[{{Stadler} {et~al.}(2025){Stadler}, {Benisty}, {Winter}, {Izquierdo},
  {Longarini}, {Galloway-Sprietsma}, {Curone}, {Andrews}, {Bae}, {Facchini},
  {Rosotti}, {Teague}, {Barraza-Alfaro}, {Cataldi}, {Cuello}, {Czekala},
  {Fasano}, {Flock}, {Fukagawa}, {Garg}, {Hall}, {Hammond}, {Hilder}, {Huang},
  {Ilee}, {Kanagawa}, {Lesur}, {Lodato}, {Loomis}, {Menard}, {Orihara},
  {Pinte}, {Price}, {Yen}, {Wafflard-Fernandez}, {Wilner}, {W{\"o}lfer},
  {Yoshida}, \& {Zawadzki}}]{Stadler_exoALMA}
{Stadler}, J., {Benisty}, M., {Winter}, A.~J., {et~al.} 2025, \apjl, 984, L11,
  \dodoi{10.3847/2041-8213/adb152}

\bibitem[{{Stauffer} {et~al.}(2015){Stauffer}, {Cody}, {McGinnis}, {Rebull},
  {Hillenbrand}, {Turner}, {Carpenter}, {Plavchan}, {Carey}, {Terebey},
  {Morales-Calder{\'o}n}, {Alencar}, {Bouvier}, {Venuti}, {Hartmann}, {Calvet},
  {Micela}, {Flaccomio}, {Song}, {Gutermuth}, {Barrado}, {Vrba}, {Covey},
  {Padgett}, {Herbst}, {Gillen}, {Lyra}, {Medeiros Guimaraes}, {Bouy}, \&
  {Favata}}]{Stauffer_ea_2015}
{Stauffer}, J., {Cody}, A.~M., {McGinnis}, P., {et~al.} 2015, \aj, 149, 130,
  \dodoi{10.1088/0004-6256/149/4/130}

\bibitem[{{Stempels} \& {Gahm}(2004)}]{Strempels_ea_2004}
{Stempels}, H.~C., \& {Gahm}, G.~F. 2004, \aap, 421, 1159,
  \dodoi{10.1051/0004-6361:20034502}

\bibitem[{{Stolker} {et~al.}(2016){Stolker}, {Dominik}, {Avenhaus}, {Min}, {de
  Boer}, {Ginski}, {Schmid}, {Juhasz}, {Bazzon}, {Waters}, {Garufi},
  {Augereau}, {Benisty}, {Boccaletti}, {Henning}, {Langlois}, {Maire},
  {M{\'e}nard}, {Meyer}, {Pinte}, {Quanz}, {Thalmann}, {Beuzit}, {Carbillet},
  {Costille}, {Dohlen}, {Feldt}, {Gisler}, {Mouillet}, {Pavlov}, {Perret},
  {Petit}, {Pragt}, {Rochat}, {Roelfsema}, {Salasnich}, {Soenke}, \&
  {Wildi}}]{Stolker_ea_2016}
{Stolker}, T., {Dominik}, C., {Avenhaus}, H., {et~al.} 2016, \aap, 595, A113,
  \dodoi{10.1051/0004-6361/201528039}

\bibitem[{{Teague} {et~al.}(2022){Teague}, {Bae}, {Andrews}, {Benisty},
  {Bergin}, {Facchini}, {Huang}, {Longarini}, \& {Wilner}}]{Teague_ea_2022}
{Teague}, R., {Bae}, J., {Andrews}, S.~M., {et~al.} 2022, \apj, 936, 163,
  \dodoi{10.3847/1538-4357/ac88ca}

\bibitem[{{Teague} {et~al.}(2025){Teague}, {Benisty}, {Facchini}, {Fukagawa},
  {Pinte}, {Andrews}, {Bae}, {Barraza-Alfaro}, {Cataldi}, {Cuello}, {Curone},
  {Czekala}, {Fasano}, {Flock}, {Galloway-Sprietsma}, {Garg}, {Hall},
  {Hammond}, {Hilder}, {Huang}, {Ilee}, {Izquierdo}, {Kanagawa}, {Lesur},
  {Lodato}, {Longarini}, {Loomis}, {Masset}, {Menard}, {Orihara}, {Price},
  {Rosotti}, {Stadler}, {Testi}, {Yen}, {Wafflard-Fernandez}, {Wilner},
  {Winter}, {W{\"o}lfer}, {Yoshida}, \& {Zawadzki}}]{Teague_exoALMA}
{Teague}, R., {Benisty}, M., {Facchini}, S., {et~al.} 2025, \apjl, 984, L6,
  \dodoi{10.3847/2041-8213/adc43b}

\bibitem[{{Veronesi} {et~al.}(2024){Veronesi}, {Longarini}, {Lodato}, {Laibe},
  {Hall}, {Facchini}, \& {Testi}}]{Veronesi_ea_2024}
{Veronesi}, B., {Longarini}, C., {Lodato}, G., {et~al.} 2024, arXiv e-prints,
  arXiv:2405.15944, \dodoi{10.48550/arXiv.2405.15944}

\bibitem[{{Villenave} {et~al.}(2023){Villenave}, {Podio}, {Duch{\^e}ne},
  {Stapelfeldt}, {Melis}, {Carrasco-Gonzalez}, {Le Gouellec}, {M{\'e}nard}, {de
  Simone}, {Chandler}, {Garufi}, {Pinte}, {Bianchi}, \&
  {Codella}}]{Villenave_ea_2023}
{Villenave}, M., {Podio}, L., {Duch{\^e}ne}, G., {et~al.} 2023, \apj, 946, 70,
  \dodoi{10.3847/1538-4357/acb92e}

\bibitem[{{Watson} \& {Stapelfeldt}(2007)}]{Watson_ea_2007}
{Watson}, A.~M., \& {Stapelfeldt}, K.~R. 2007, \aj, 133, 845,
  \dodoi{10.1086/510455}

\bibitem[{{Winter} {et~al.}(2024{\natexlab{a}}){Winter}, {Benisty}, {Manara},
  \& {Gupta}}]{Winter_ea_2024c}
{Winter}, A.~J., {Benisty}, M., {Manara}, C.~F., \& {Gupta}, A.
  2024{\natexlab{a}}, \aap, 691, A169, \dodoi{10.1051/0004-6361/202452120}

\bibitem[{{Winter} {et~al.}(2024{\natexlab{b}}){Winter}, {Benisty}, {Shuai},
  {D{\^u}chene}, {Cuello}, {Anania}, {Cadiou}, \& {Joncour}}]{Winter_ea_2024b}
{Winter}, A.~J., {Benisty}, M., {Shuai}, L., {et~al.} 2024{\natexlab{b}}, \aap,
  691, A43, \dodoi{10.1051/0004-6361/202450842}

\bibitem[{{W{\"o}lfer} {et~al.}(2025){W{\"o}lfer}, {Barraza-Alfaro}, {Teague},
  {Curone}, {Benisty}, {Fukagawa}, {Bae}, {Cataldi}, {Czekala}, {Facchini},
  {Fasano}, {Flock}, {Galloway-Sprietsma}, {Garg}, {Hall}, {Huang}, {Ilee},
  {Izquierdo}, {Kanagawa}, {Lesur}, {Longarini}, {Loomis}, {Menard}, {Nath},
  {Orihara}, {Pinte}, {Price}, {Rosotti}, {Stadler}, {Wafflard-Fernandez},
  {Winter}, {Yen}, {Yoshida}, \& {Zawadzki}}]{Wölfer_exoALMA}
{W{\"o}lfer}, L., {Barraza-Alfaro}, M., {Teague}, R., {et~al.} 2025, \apjl,
  984, L22, \dodoi{10.3847/2041-8213/adc42c}

\bibitem[{{Young} {et~al.}(2022){Young}, {Alexander}, {Rosotti}, \&
  {Pinte}}]{Young_ea_2022}
{Young}, A.~K., {Alexander}, R., {Rosotti}, G., \& {Pinte}, C. 2022, \mnras,
  513, 487, \dodoi{10.1093/mnras/stac840}

\bibitem[{{Young} {et~al.}(2021){Young}, {Alexander}, {Walsh}, {Nealon},
  {Booth}, \& {Pinte}}]{Young_ea_2021}
{Young}, A.~K., {Alexander}, R., {Walsh}, C., {et~al.} 2021, \mnras, 505, 4821,
  \dodoi{10.1093/mnras/stab1675}

\bibitem[{{Young} {et~al.}(2023){Young}, {Stevenson}, {Nixon}, \&
  {Rice}}]{Young_ea_2023}
{Young}, A.~K., {Stevenson}, S., {Nixon}, C.~J., \& {Rice}, K. 2023, \mnras,
  525, 2616, \dodoi{10.1093/mnras/stad2451}

\bibitem[{{Zagaria} {et~al.}(2025){Zagaria}, {Jiang}, {Cataldi}, {Facchini},
  {Benisty}, {Aikawa}, {Andrews}, {Bae}, {Barraza-Alfaro}, {Curone}, {Czekala},
  {Fasano}, {Hall}, {Hammond}, {Huang}, {Ilee}, {Izquierdo}, {Lawrence},
  {Lodato}, {M{\'e}nard}, {Pinte}, {Rosotti}, {Stadler}, {Teague}, {Testi},
  {Wilner}, {Winter}, \& {Yoshida}}]{Zagaria_ea_2025}
{Zagaria}, F., {Jiang}, H., {Cataldi}, G., {et~al.} 2025, arXiv e-prints,
  arXiv:2506.16481, \dodoi{10.48550/arXiv.2506.16481}

\bibitem[{{Zawadzki} {et~al.}(2025){Zawadzki}, {Czekala}, {Galloway-Sprietsma},
  {Bae}, {Barraza-Alfaro}, {Benisty}, {Cataldi}, {Curone}, {Facchini},
  {Fasano}, {Flock}, {Fukagawa}, {Garg}, {Hall}, {Hilder}, {Huang}, {Ilee},
  {Isella}, {Izquierdo}, {Kanagawa}, {Lesur}, {Longarini}, {Loomis}, {Orihara},
  {Pinte}, {Price}, {Rosotti}, {Stadler}, {Teague}, {Yen},
  {Wafflard-Fernandez}, {Wilner}, {Winter}, {W{\"o}lfer}, \&
  {Yoshida}}]{Zawadzki_exoALMA}
{Zawadzki}, B., {Czekala}, I., {Galloway-Sprietsma}, M., {et~al.} 2025, \apjl,
  984, L14, \dodoi{10.3847/2041-8213/adc434}

\bibitem[{{Zhang} \& {Zhu}(2024)}]{Zhang_Zhu_2024}
{Zhang}, S., \& {Zhu}, Z. 2024, \apjl, 974, L38,
  \dodoi{10.3847/2041-8213/ad815f}

\bibitem[{{Zhong} {et~al.}(2024){Zhong}, {Ren}, {Ma}, {Xie}, {Ma}, {Wallack},
  {Mawet}, \& {Ruane}}]{Zhong_ea_2024}
{Zhong}, H., {Ren}, B.~B., {Ma}, B., {et~al.} 2024, \aap, 684, A168,
  \dodoi{10.1051/0004-6361/202348874}

\bibitem[{{Zhu}(2019)}]{Zhu_ea_2019}
{Zhu}, Z. 2019, \mnras, 483, 4221, \dodoi{10.1093/mnras/sty3358}

\bibitem[{{Ziampras} {et~al.}(2025){Ziampras}, {Dullemond}, {Birnstiel},
  {Benisty}, \& {Nelson}}]{Ziampras_ea_2025}
{Ziampras}, A., {Dullemond}, C.~P., {Birnstiel}, T., {Benisty}, M., \&
  {Nelson}, R.~P. 2025, \mnras, 540, 1185, \dodoi{10.1093/mnras/staf785}

\bibitem[{{Zuleta} {et~al.}(2024){Zuleta}, {Birnstiel}, \&
  {Teague}}]{Zulete_ea_2024}
{Zuleta}, A., {Birnstiel}, T., \& {Teague}, R. 2024, \aap, 692, A56,
  \dodoi{10.1051/0004-6361/202451145}

\end{thebibliography}
